\documentclass[12pt,prd,onecolumn,showpacs,amsmath,amssymb,aps,floats,floatfix,nofootinbib]{revtex4-1}
\usepackage[colorlinks=true,urlcolor=blue,anchorcolor=blue,citecolor=blue,filecolor=blue,linkcolor=blue,menucolor=blue,pagecolor=blue,linktocpage=true]{hyperref} 

\usepackage[inline]{enumitem}
\usepackage[multidot]{grffile} 
\usepackage{dcolumn}
\usepackage{bm}
\usepackage{amsmath}
\usepackage{amsfonts}
\usepackage{amssymb}
\usepackage{dsfont}
\usepackage{color}
\usepackage{latexsym}
\usepackage{slashed} 
\usepackage{indentfirst}
\usepackage{mathrsfs}
\usepackage{multirow}
\usepackage{epsfig,psfrag}
\usepackage{subfigure}
\usepackage{mathtools}
\usepackage{setspace} 
\usepackage[utf8]{inputenc} 
\usepackage[scientific-notation=true]{siunitx} 
\usepackage{verbatim}

\graphicspath{{fig/}}
\allowdisplaybreaks 

\newcommand{\uvec}{\boldsymbol}
\newcommand{\ud}{\mathrm{d}}

\makeatother

\begin{document}
\title{Nucleon relativistic polarization and magnetization distributions}
\author{Yi Chen}
\affiliation{Interdisciplinary Center for Theoretical Study and Department of Modern Physics, University of Science and Technology of China, Hefei, Anhui 230026, China}
\affiliation{Shanghai Institute of Applied Physics, Chinese Academy of Sciences, Shanghai 201800, China}
\affiliation{University of Chinese Academy of Sciences, Beijing 100049, China}
\author{C\'edric Lorc\'e}~\email[Corresponding author: ]{cedric.lorce@polytechnique.edu}
\affiliation{CPHT, CNRS, \'Ecole polytechnique, Institut Polytechnique de Paris, 91120 Palaiseau, France}

\begin{abstract}

As a follow up of our work on the electromagnetic four-current, we study for the first time the relativistic polarization and magnetization spatial distributions inside a spin-$\frac{1}{2}$ target within the quantum phase-space approach. While the polarization-magnetization tensor is usually defined in terms of the Gordon decomposition of the electromagnetic four-current, a Breit frame analysis reveals that a physically simpler and more natural picture of the system arises when the polarization-magnetization tensor is instead defined in terms of a Sachs decomposition. Relativistic polarization and magnetization distributions for a moving target are compared with their light-front counterparts. In particular, we show that the genuine light-front magnetization distributions are defined in terms of Fourier transforms of the Sachs magnetic form factor, rather than in terms of the Pauli form factor as suggested earlier in the literature. We finally illustrate our results in the case of a nucleon using the electromagnetic form factors extracted from experimental data.

\end{abstract}
\maketitle
\newpage
\section{Introduction}
\label{Introduction}

Nucleons (i.e.~protons and neutrons) are by far the most abundant bound-state systems in nature and are key for studying quantum chromodynamics (QCD), the fundamental theory of strong interactions. A central goal of modern nuclear physics is to explain how nucleons emerge in QCD from first principles~\cite{Gao:2021sml,Li:2022sqg}. Due to the complicated non-perturbative dynamics of their quark and gluon degrees of freedom, nucleons inherit particularly rich and intricate internal structures.

Electromagnetic form factors (FFs) encode fundamental information on the internal electromagnetic structure of hadrons~\cite{Rosenbluth:1950yq,Mcallister:1956ng,Hofstadter:1956qs,Yennie:1957rmp,Durand:1962zza,Hand:1963zz}. Nucleon electromagnetic FFs in particular have been extensively measured over the past decades with very high precision in various scattering experiments~\cite{Chambers:1956zz,Anklin:1994ae,Kelly:2002if,SAMPLE:2003wwa,Maas:2004dh,Qattan:2004ht,G0:2005chy,HAPPEX:2006oqy,Arrington:2007ux,CLAS:2008idi,A1:2010nsl,Zhan:2011ji,Puckett:2011xg,A1:2013fsc,Griffioen:2015hta,Puckett:2017flj,Ye:2017gyb,SANE:2018cub,Xiong:2019umf,Mihovilovic:2019jiz,Atac:2020hdq,PRad:2020oor,Gramolin:2021gln,Zhou:2021gyh,Hague:2021xcc,Atac:2021wqj,Hashamipour:2021kes,A1:2022wzx,Hashamipour:2022noy}. On the theory side, \emph{ab initio} calculations within the lattice QCD approach have also been significantly improved in the last few years~\cite{Alexandrou:2017ypw,Hasan:2017wwt,Shintani:2018ozy,Alexandrou:2018sjm,Jang:2018djx,Jang:2019jkn,Alexandrou:2020aja,Park:2021ypf,Ishikawa:2021eut,Bar:2021crj,Djukanovic:2021cgp,Djukanovic:2021qxp,Alexandrou:2021jok,Gupta:2021lnv,Ruso:2022qes,Lin:2022nnj,Gupta:2023cvo}. For recent reviews on electromagnetic FFs, see Refs.~\cite{Gao:2003ag,Perdrisat:2006hj,Arrington:2006zm,Denig:2012by,Pacetti:2014jai,Punjabi:2015bba,Constantinou:2020hdm,Gao:2021sml,Wang:2022bxo}. 

Spatial distributions of charge and magnetization can be defined in the Breit frame (BF) in terms of 3D Fourier transforms of these electromagnetic FFs~\cite{Ernst:1960zza,Sachs:1962zzc}, but they cannot be considered as probabilistic densities due to relativistic recoil corrections~\cite{Yennie:1957rmp,Breit:1964ga,Kelly:2002if,Burkardt:2000za,Belitsky:2003nz,Jaffe:2020ebz}. Spatial distributions with probabilistic interpretation can however be defined within the light-front (LF) formalism~\cite{Burkardt:2002hr,Miller:2007uy,Carlson:2007xd,Alexandrou:2008bn,Alexandrou:2009hs,Gorchtein:2009qq,Carlson:2009ovh,Miller:2010nz,Miller:2018ybm}, at the cost of losing one spatial dimension and exhibiting distortions induced by the LF perspective. 

Understanding better the relation between 3D BF and 2D LF distributions has been the focus of many recent works, see e.g. Refs.~\cite{Panteleeva:2021iip,Freese:2021mzg,Kim:2021kum,Kim:2022bia,Epelbaum:2022fjc,Panteleeva:2022khw,Li:2013pys,Carlson:2022eps,Freese:2022fat,Panteleeva:2022uii,Alharazin:2022xvp,Freese:2023jcp}. The quantum phase-space formalism distinguishes itself by the fact that the requirement of a strict probabilistic interpretation is relaxed and replaced by a milder quasiprobabilistic picture~\cite{Wigner:1932eb,Hillery:1983ms,Bialynicki-Birula:1991jwl}. This approach is quite appealing since it allows one to define in a consistent way relativistic spatial distributions inside a target with arbitrary spin and arbitrary average momentum~\cite{Lorce:2017wkb,Lorce:2018zpf,Lorce:2018egm,Lorce:2020onh,Lorce:2021gxs,Lorce:2022jyi,Lorce:2022cle,Chen:2022smg,Hong:2023tkv}. In particular, when the average momentum vanishes one recovers the BF picture, while in the limit of infinite average momentum one recovers essentially the LF picture.

In this work, we use the quantum phase-space formalism to study for the first time the relativistic polarization and magnetization spatial distributions inside a spin-$\frac{1}{2}$ target. The paper is organized as follows. In Sec.~\ref{sec:general discussion on polarization and magnetization}, we first briefly review the description of elastic electron-nucleon scattering in terms of electromagnetic FFs, and then discuss the concept of polarization-magnetization tensor. In Sec.~\ref{sec:Quantum phase-space formalism}, we present in detail the quantum phase-space formalism and compare the phase-space picture with the light-front picture. We start our analysis in Sec.~\ref{sec:Breit frame distributions} with the Breit frame distributions of polarization and magnetization for a spin-$\frac{1}{2}$ target. We argue that the polarization-magnetization tensor suggested by the Sachs decomposition of the electromagnetic four-current is physically more transparent than the one suggested by the Gordon decomposition. We proceed in Sec.~\ref{sec:Elastic frame distributions} with the elastic frame distributions of polarization and magnetization, and study in detail their frame dependence, and derive analytic expressions for electric and magnetic dipole moments. For completeness, we also present in Sec.~\ref{sec:Light-front distributions} the light-front distributions and multipole moments, and compare them with the infinite-momentum limit of their elastic frame counterparts. In particular, we explain why the genuine light-front magnetization distributions are given by the 2D Fourier transforms of the Sachs magnetic form factor, rather than the Pauli form factor as suggested earlier in the literature. Finally, we summarize our findings in Sec.~\ref{sec:Summary}, and provide further discussions about charge radii, relativistic centers and multipole decomposition of polarization and magnetization distributions in three Appendices.

\section{Polarization and magnetization for a spin-$\frac{1}{2}$ target}
\label{sec:general discussion on polarization and magnetization}

\begin{figure}[htb!]
	\centering
	{\includegraphics[angle=0,scale=0.47]{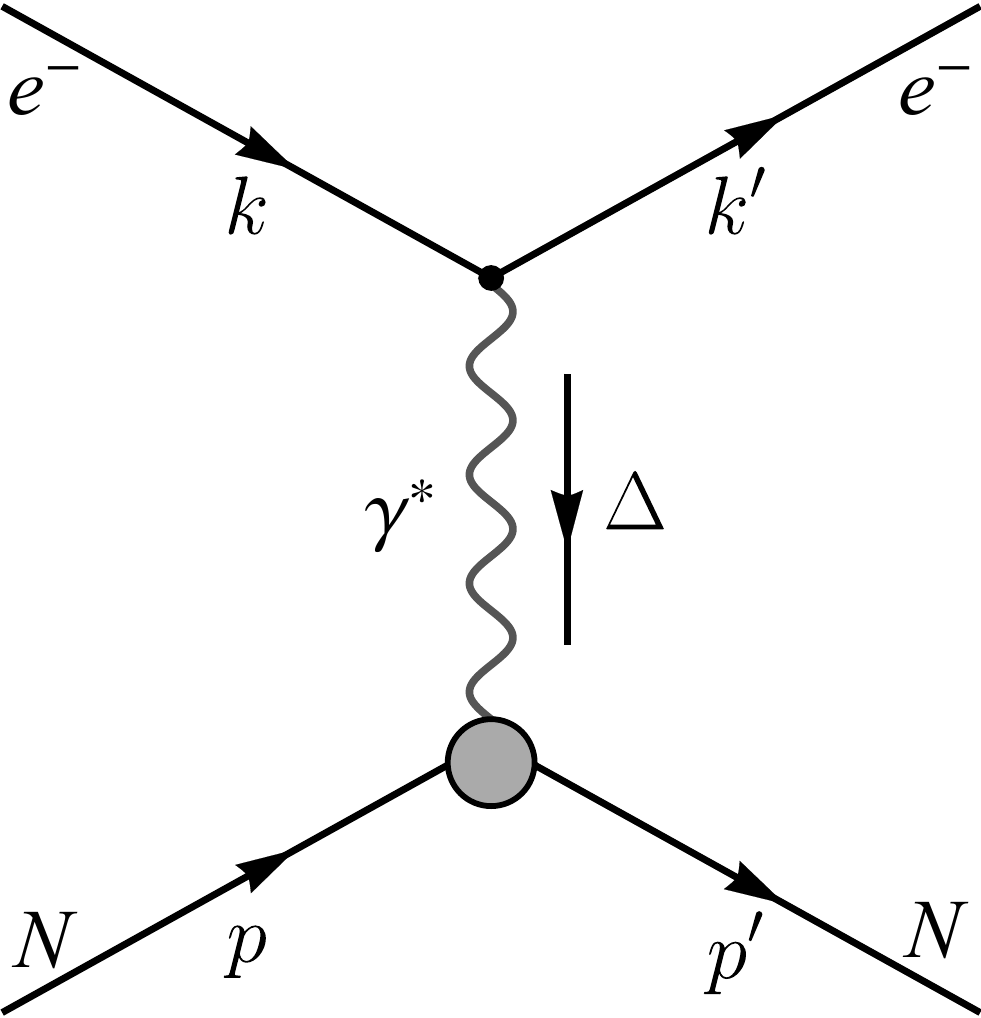}}
	\caption{Feynman diagram of the $t$-channel elastic reaction $e^{-}(k) + N(p) \to e^{-}(k') + N(p')$ in the one-photon-exchange approximation. The four-momentum transfer is $\Delta=k-k'=p'-p$.}
	\label{Fig_FeynmanDiagram}
\end{figure}

Long ago it has been shown that the matrix elements of the electromagnetic four-current operator for a general spin-$\frac{1}{2}$ system can be parametrized as~\cite{Foldy:1952aa,Salzman:1955zz,Yennie:1957rmp}
\begin{equation}\label{genparam}
	\langle p',s'|\hat j^\mu(0)|p,s\rangle=e\,\overline u(p',s')\Gamma^\mu(P,\Delta)u(p,s)
\end{equation}
with $e$ the unit of electric charge (chosen to be that of a proton) and
\begin{equation}\label{DiracPauliparam}
	\Gamma^\mu(P,\Delta)=\gamma^\mu\,F_1(Q^2)+\frac{i\sigma^{\mu\nu}\Delta_\nu}{2M}\,F_2(Q^2),
\end{equation}
where $F_1(Q^2)$ and $F_2(Q^2)$ are Lorentz-invariant functions called Dirac and Pauli form factors (FFs), respectively. For convenience, we introduced the variables $P=\frac{1}{2}(p'+p)$, $\Delta=p'-p$ and $Q^2=-\Delta^2$; see, e.g., the tree-level Feynman diagram in Fig.~\ref{Fig_FeynmanDiagram}. The on-shell conditions $p'^2=p^2=M^2$ imply in particular $P\cdot\Delta=0$ and $P^2+\frac{\Delta^2}{4}=M^2$. There is therefore only one dimensionless Lorentz-invariant variable which we chose as $\tau=Q^2/(4M^2)$. The initial and final canonical polarizations of the system are denoted by $s$ and $s'$, respectively.

In the Breit frame (BF), defined by the condition $\uvec P=\uvec 0$, the amplitudes read~\cite{Yennie:1957rmp,Ernst:1960zza,Sachs:1962zzc}
\begin{equation}\label{BFampl}
	\begin{aligned}
		\langle p'_B,s'_B|\hat j^0(0)|p_B,s_B\rangle&=e\,2M\,\delta_{s'_Bs_B}\,G_E(Q^2),\\
		\langle p'_B,s'_B|\hat{\uvec j}(0)|p_B,s_B\rangle&=e\,(\uvec\sigma_{s'_Bs_B}\times i\uvec\Delta)\,G_M(Q^2),
	\end{aligned}
\end{equation}
where $\uvec\sigma$ are the Pauli matrices and the combinations
\begin{equation}
	\begin{aligned}
		G_E(Q^2)&=F_1(Q^2)-\tau F_2(Q^2),\\
		G_M(Q^2)&=F_1(Q^2)+F_2(Q^2),
	\end{aligned}
\end{equation}
are known as the electric and magnetic Sachs FFs. The spin structure of the amplitudes in the BF turns out to be the same as in the non-relativistic theory. In any other frame, the spin structure becomes more complicated as a result of Wigner rotations~\cite{Lorce:2020onh,Kim:2021kum,Chen:2022smg}. A somewhat related observation is that the differential cross section in the first Born approximation (i.e.~one-photon exchange) can be expressed as~\cite{Yennie:1957rmp,Hand:1963zz}
\begin{equation}\label{crossGEGM}
\frac{\ud\sigma}{\ud\Omega}=\left(\frac{\ud\sigma}{\ud\Omega}\right)_\text{\!Mott}f_\text{recoil}\left[G_E^2(Q^2)+\frac{\tau}{\epsilon}\,G^2_M(Q^2)\right]\frac{1}{1+\tau},
\end{equation}
where $\epsilon=(1+2(1+\tau)\tan^2\frac{\theta}{2})^{-1}$ is the virtual photon polarization with $\theta$ the scattered electron angle in the lab frame. The Mott cross section and the recoil factor are given by
\begin{equation}
\left(\frac{\ud\sigma}{\ud\Omega}\right)_\text{\!Mott}=\frac{\alpha^2\cos^2\frac{\theta}{2}}{4E^2\sin^4\frac{\theta}{2}},\qquad
f_\text{recoil}=\frac{E'}{E}=\frac{1}{1+\frac{2E}{M}\sin^2\frac{\theta}{2}},
\end{equation}
where $\alpha = e^2/(4\pi) \approx 1/137$ is the electromagnetic fine structure constant\footnote{The convention we used throughout this paper is $\hbar=c=1$ with $\mu_0=\epsilon_0=1$.} and $E$ ($E'$) is the initial (final) electron energy in the lab frame. For comparison, in terms of Dirac and Pauli FFs the differential cross section reads~\cite{Rosenbluth:1950yq,Yennie:1957rmp}
\begin{equation}
\frac{\ud\sigma}{\ud\Omega}=\left(\frac{\ud\sigma}{\ud\Omega}\right)_\text{\!Mott}f_\text{recoil}\left\{F_1^2(Q^2)+\tau\left(F_2^2(Q^2)+2\left[F_1(Q^2)+F_2(Q^2)\right]^2\tan^2\tfrac{\theta}{2}\right)\right\}.
\end{equation}
The absence of interference terms in Eq.~\eqref{crossGEGM} makes the separate extraction of $G_E$ and $G_M$ easier, and suggests also that they could be considered as the ``physical'' electromagnetic FFs. A parametrization of Eq.~\eqref{genparam} directly in terms of Sachs FFs reads~\cite{BARNES1962166,Lorce:2020onh}
\begin{equation}\label{sachsparam}
	\Gamma^\mu(P,\Delta)=\frac{MP^\mu}{P^2}\,G_E(Q^2)+\frac{i\epsilon^{\mu\alpha\beta\lambda}\Delta_\alpha P_\beta\gamma_\lambda\gamma_5}{2P^2}\,G_M(Q^2)
\end{equation}
with $\epsilon_{0123}=+1$. It is equivalent to the parametrization~\eqref{DiracPauliparam} on-shell, i.e.~once sandwiched between Dirac spinors. A similar expression for the Dirac theory, i.e.~with $G_E(Q^2)=G_M(Q^2)=Z$, has been considered in position space in Ref.~\cite{Bitar:1967dqa}. The structure of Eq.~\eqref{sachsparam} is particularly interesting since it is reminiscent of a classical current in a polarizable medium, giving further support to the interpretation of the Sachs FFs as the ``physical'' electromagnetic FFs\footnote{Strictly speaking, both Eqs.~\eqref{sachsparam} and~\eqref{crossGEGM} suggest that the actual physical FFs are given by $\bar G_{E,M}(Q^2)\equiv\frac{M}{\sqrt{P^2}}\,G_{E,M}(Q^2)=\frac{1}{\sqrt{1+\tau}}\,G_{E,M}(Q^2)$.}. Similar observations apply to spin-$1$ systems~\cite{Lorce:2022jyi,Yennie:1957rmp,Gourdin:1963} and a generalization of Eq.~\eqref{sachsparam} to higher-spin systems has even been proposed in Ref.~\cite{Theis:1966rja}.

\subsection{Convection and polarization currents}
\label{sec:convpol}

In classical electromagnetism, it is customary to decompose the electromagnetic four-current in position space into ``convection'' and ``polarization'' currents~\cite{vanderlinde2006classical,Gonano2015DefinitionFP}
\begin{equation}\label{totalcurrent}
	J^\mu(x)=J^\mu_c(x)+J^\mu_P(x),\qquad J^\mu_P(x)=\partial_\alpha P^{\alpha\mu}(x).
\end{equation}
The basic idea is that in a polarizable medium some of the charges are somewhat free to move and constitute the convective part of the current, also known as the ``free'' current. The rest of the charges is confined in compact regions, e.g.~around atomic nuclei. Applying an external electromagnetic field to the medium can induce (electric) polarization $\uvec{\mathcal P}$ and magnetization $\uvec M$, generating a new contribution to the total current often called the ``bound'' current. From a relativistic perspective, polarization and magnetization are the two sides of a same coin, the polarization-magnetization tensor
\begin{equation}\label{Pcomp}
	P^{\mu\nu}=\begin{pmatrix}0&\mathcal P_x&\mathcal P_y&\mathcal P_z\\
		-\mathcal P_x&0&-M_z&M_y\\
		-\mathcal P_y&M_z&0&-M_x\\
		-\mathcal P_z&-M_y&M_x&0\end{pmatrix},
\end{equation}
just like the electric and magnetic fields are the two sides of the Faraday tensor $F^{\mu\nu}$. This means that under a Lorentz boost, polarization and magnetization will mix with each other.

Writing Eq.~\eqref{totalcurrent} more explicitly, one obtains
\begin{equation}\label{fourcurrentdetailed}
	\begin{aligned}
		J^0 &=\rho_c -\uvec\nabla\cdot\uvec{\mathcal P} ,\\
		\uvec J &=\rho_c \uvec v +\uvec\nabla\times\uvec M +\partial_0\uvec{\mathcal P}.
	\end{aligned}
\end{equation}
Assuming as usual that surface terms vanish at spatial infinity, we see that the induced polarization does not change the total charge of the system but simply modifies its spatial distribution. Relativistically, this arises from the fact that the divergence of the polarization four-current vanishes identically $\partial_\mu J^\mu_P(x)=\partial_\mu\partial_\alpha P^{\alpha\mu}(x)=0$ owing to the antisymmetry of the polarization-magnetization tensor. In other words, the polarization four-current has the form of what is known in the literature as a \emph{superpotential}.

At the classical level, angular momentum appears only in orbital form. Magnetization arises therefore from loops of charge current, while polarization arises from the separation of electric charges due to the external electric field. At the quantum level, a new form of angular momentum known as {\it spin} enters the game. As a result, a spinning charged particle at rest will also present a permanent magnetic dipole moment (MDM). We should therefore distinguish external and internal contributions to the polarization-magnetization tensor due to, respectively, the external electromagnetic fields and the spin degrees of freedom. At the level of one-photon exchange, we are only sensitive to the spin contribution. The external contribution requires at least two photons and is described at linear order in the electromagnetic field in terms of the medium polarizabilities $P^{\mu\nu}_\text{ext}=\alpha^{\mu\nu\alpha\beta}F_{\alpha\beta}$~\cite{Pieplow:2013}. In this work, we will focus on the internal (or spin) polarization-magnetization tensor.

For a particle at rest, a permanent electric dipole moment (EDM) along the angular momentum breaks time-reversal ($\textsf{T}$) and hence the combined charge-conjugation and parity ($\textsf{CP}$) symmetries. In the Standard Model, these symmetries are known to be broken by the weak interactions and the $\theta$-term in QCD~\cite{Chupp:2017rkp}, but the breaking is so small that one can consider to an excellent approximation that these symmetries remain exact when studying the internal structure of hadrons. It follows that the polarization for a point-like particle at rest must vanish. In the non-relativistic limit, Eq.~\eqref{fourcurrentdetailed} reduces then to
\begin{equation}
	\begin{aligned}
		J^0 &\approx\rho_c,\\
		\uvec J&\approx\rho_c \uvec v+\uvec\nabla\times\uvec M,
	\end{aligned}
\end{equation}
where the term $\uvec\nabla\times\uvec M$ is known as the spin current~\cite{Schwartz:1955,Yennie:1957rmp,Hestenes:1971ra,Wilkes:2020} since $\uvec M\propto\uvec S$ with $\uvec S$ the spin vector. For systems moving with relativistic velocities, one should also include the contributions from $\uvec{\mathcal P}$. The latter do not however contain any new intrinsic information since they simply result from the Lorentz boost of the rest-frame magnetization. 

\subsection{Polarization-magnetization tensor}

Let us now come back to the electromagnetic four-current for a spin-$\frac{1}{2}$ target. In momentum space, the four-divergence turns into a contraction with the four-momentum transfer
\begin{equation}
	\langle p',s'|\partial_\mu \hat O^\mu(x)|p,s\rangle=i\Delta_\mu\langle p',s'|\hat O^\mu(x)|p,s\rangle,
\end{equation}
using the translation invariance property. It is then clear that the parametrization~\eqref{sachsparam} exhibits the same structure as the total current~\eqref{totalcurrent} in classical electromagnetism~\cite{Lorce:2020onh,Li:2013pys}. Accordingly, we identify the convection current with the $G_E$ term and the polarization current with the $G_M$ term. In other words, we write $\Gamma^\mu(P,\Delta)=\Gamma^\mu_c(P,\Delta)+\Gamma^\mu_P(P,\Delta)$ with
\begin{equation}
    \begin{aligned}\label{spinhalf-Vexfun-ConvPol}
      \Gamma^\mu_c(P,\Delta)&= \frac{MP^\mu}{P^2}\,G_E(Q^2),\\
      \Gamma^\mu_P(P,\Delta)&= \frac{i\epsilon^{\mu\alpha\beta\lambda}\Delta_\alpha P_\beta\gamma_\lambda\gamma_5}{2P^2}\,G_M(Q^2).
    \end{aligned}
\end{equation}
This suggests in particular that the polarization-magnetization tensor for a spin-$\frac{1}{2}$ target is given in momentum space by
\begin{equation}\label{polmagT}
	\widetilde P^{\mu\nu}=-\frac{e}{2M}\,\frac{M\,\epsilon^{\mu\nu\beta\lambda} P_\beta }{P^2}\,\overline u(p',s')\gamma_\lambda\gamma_5u(p,s)\,G_M(Q^2).
\end{equation}
Since it involves the axial-vector Dirac bilinear, we will refer to it as the $A$-type polarization-magnetization tensor.

We point out that the identification of a polarization-magnetization tensor from the electromagnetic four-current alone is in fact ambiguous. One reason is that only the divergence of $P^{\mu\nu}$ contributes to $J^\mu$ in Eq.~\eqref{totalcurrent}. As a result, one can alternatively consider the tensor
\begin{equation}\label{Pamb}
    P^{\mu\nu}_{\mathcal A}(x) =P^{\mu\nu}(x) + \epsilon^{\mu\nu\alpha\beta}\partial_\alpha \mathcal A_\beta(x)
\end{equation}
with $\mathcal A^\beta$ an arbitrary axial four-vector field assumed to vanish sufficiently fast at infinity. Our choice in Eq.~\eqref{polmagT} is motivated by its simplicity and by the fact that the relativistic spin appears explicitly in the form of the Dirac axial-vector four-current. An additional ambiguity comes from the equation of motion. Indeed, since Eq.~\eqref{DiracPauliparam} is meant to be sandwiched between free Dirac spinors, we can use the Gordon identity~\cite{Gordon:1928} 
\begin{equation}
    \overline u(p',s')\gamma^\mu u(p,s)=\overline u(p',s')\left[\frac{P^\mu}{M}+\frac{i\sigma^{\mu\nu}\Delta_\nu}{2M}\right] u(p,s)
\end{equation}
and write $\Gamma^\mu(P,\Delta)=\Gamma'^\mu_c(P,\Delta)+\Gamma'^\mu_P(P,\Delta)$ with
\begin{equation}\label{Ttypedec}
    \begin{aligned}
    \Gamma'^\mu_c(P,\Delta)&= \frac{P^\mu}{M}\,F_1(Q^2),\\
    \Gamma'^\mu_P(P,\Delta)&= \frac{i\sigma^{\mu\nu}\Delta_\nu}{2M}\,G_M(Q^2),
    \end{aligned}
\end{equation}
suggesting another \emph{a priori} acceptable definition for the polarization-magnetization tensor
\begin{equation}\label{polmagTbis}
	\widetilde P'^{\mu\nu}=-\frac{e}{2M}\,\overline u(p',s')\sigma^{\mu\nu}u(p,s)\,G_M(Q^2).
\end{equation}
Since it involves the tensor Dirac bilinear, we will refer to it as the $T$-type polarization-magnetization tensor. The decomposition of a current into convection and polarization parts is therefore not unique, and can be understood as a consequence of the on-shell identity
\begin{equation}
   \overline u(p',s')i\sigma^{\mu\nu}\Delta_\nu u(p,s)=\overline u(p',s')\left[\frac{\Delta^2}{2P^2}\,P^\mu+\frac{Mi\epsilon^{\mu\alpha\beta\lambda}\Delta_\alpha P_\beta\gamma_\lambda\gamma_5}{P^2}\right]u(p,s),
\end{equation}
which can easily be derived from the relations given in Refs.~\cite{Lorce:2017isp,Cotogno:2019vjb}. As a result of Gordon's work~\cite{Gordon:1928}, the $T$-type definition~\eqref{polmagTbis} is often the only one considered in the literature, but we will show later that the $A$-type definition~\eqref{polmagT} turns out in fact to be more natural.

As a last remark, we note that in field theory it is customary to describe the full electromagnetic interaction of particles through the single interaction term
\begin{equation}\label{shortform}
    S_\text{int} =\int\ud^4x\,J^\mu(x)A_\mu(x),
\end{equation}
which can be rewritten as follows
\begin{equation}\label{longform}
     S_\text{int} =\int\ud^4x\,J^\mu_c(x)A_\mu(x)-\frac{1}{2}\int\ud^4x\,P^{\mu\nu}(x) F_{\mu\nu}(x)
\end{equation}
using integration by parts. It is then easy to see that the ambiguity mentioned in Eq.~\eqref{Pamb} exists because of the homogeneous Maxwell equation $\epsilon^{\mu\nu\alpha\beta}\partial_\nu F_{\alpha\beta}=0$, which expresses the absence of magnetic charges. Even though the form~\eqref{longform} makes the physics more transparent, it is in practice easier to consider that all the electromagnetic properties can be described in terms of a single electromagnetic four-current $J^\mu$, rather than by a combination of $J^\mu_c$ and $P^{\mu\nu}$. Opinions differ in the literature about whether $J^\mu$ or $J^\mu_c$ should be regarded as the fundamental electromagnetic four-current, just like they differ about whether the (symmetric) Belinfante or the (asymmetric) kinetic energy-momentum tensor should be considered as the fundamental energy-momentum tensor~\cite{Leader:2013jra}. In particular, if one assumes that all forms of magnetism arise from the sole circulation of charges, the polarization-magnetization tensor
\begin{equation}
    P^{\mu\nu}_J(x)\equiv -\frac{1}{2}\left[x^\mu J^\nu(x)-x^\nu J^\mu(x) \right]
\end{equation}
would then seem to be a natural choice for a system sitting at the origin, fixing therefore the form of the polarization current to $J^{\mu}_P=\partial_\alpha P^{\alpha\mu}_J=\frac{1}{2}[J^\mu-\partial_\alpha(x^\alpha J^\mu)]$ and hence the convection current to $J^{\mu}_c=\frac{1}{2}[J^\mu+\partial_\alpha(x^\alpha J^\mu)]$. We will not discuss in detail this option in the present work.

\section{Quantum phase-space formalism}
\label{sec:Quantum phase-space formalism}

Electromagnetic FFs describe the internal charge and magnetization content of a system. While they are objects defined in momentum space and extracted from experimental data involving particles with well-defined momenta, their physical interpretation actually resides in position space. It is therefore important to understand how the concept of spatial distribution arises in quantum field theory.

Let us consider a generic local operator $\hat O(x)$. Its expectation value in a physical state can be written as
\begin{equation}\label{momrepP}
    \langle\Psi|\hat O(x)|\Psi\rangle=\sum_{s',s}\int\frac{\ud^3p'}{(2\pi)^3}\,\frac{\ud^3p}{(2\pi)^3}\,\widetilde\Psi^*(\uvec p',s')\widetilde\Psi(\uvec p,s)\,\frac{\langle p',s'|\hat O(x)|p,s\rangle}{2\sqrt{p'^0p^0}},
\end{equation}
with the four-momentum eigenstates normalized as $\langle p',s'|p,s\rangle=2p^0(2\pi)^3\delta^{(3)}(\uvec p'-\uvec p)\delta_{s's}$ and the momentum-space wave packet $\widetilde\Psi(\uvec p,s)\equiv\langle p,s|\Psi\rangle/\sqrt{2p^0}$ normalized as
\begin{equation}
    \sum_s\int\frac{\ud^3p}{(2\pi)^3}\,|\widetilde\Psi(\uvec p,s)|^2=1.
\end{equation}
The four-momenta being on-shell, the energy components are given by $p^0=\sqrt{\uvec p^2+M^2}$ and $p'^0=\sqrt{\uvec p'^2+M^2}$. 

In a relativistic theory, the Newton-Wigner position operator~\cite{Pryce:1948pf,Newton:1949cq,Foldy:1949wa,Pavsic:2017orp} is the only 3D position operator satisfying usual commutation relations with linear and angular momentum operators, and having mutually commuting components. Although this operator does not transform as part of a Lorentz four-vector, it allows one to localize a relativistic system at a fixed time. The eigenstates of this operator at $t=0$ are related to momentum eigenstates via Fourier transform
\begin{equation}
    |\uvec r,s\rangle=\int\frac{\ud^3p}{(2\pi)^3}\,e^{-i\uvec p\cdot\uvec r}\,\frac{|p,s\rangle}{\sqrt{2p^0}}
\end{equation}
and are normalized as $\langle \uvec r',s'|\uvec r,s\rangle=\delta^{(3)}(\uvec r'-\uvec r)\delta_{s's}$. The position-space wave packet at $t=0$ is then given by
\begin{equation}
    \Psi(\uvec r,s)\equiv\langle \uvec r,s|\Psi\rangle =\int\frac{\ud^3p}{(2\pi)^3}\,e^{i\uvec p\cdot\uvec r}\,\widetilde\Psi(\uvec p,s)
\end{equation}
and satisfies the normalization condition
\begin{equation}
    \sum_s\int\ud^3r\,|\Psi(\uvec r,s)|^2=1.
\end{equation}
In position space, the expectation value~\eqref{momrepP} takes then the familiar form
\begin{equation}\label{momrepR}
    \langle\Psi|\hat O(x)|\Psi\rangle=\sum_{s',s}\int\ud^3r'\,\ud^3r\,\Psi^*(\uvec r',s')\Psi(\uvec r,s)\,\langle \uvec r',s'|\hat O(x)|\uvec r,s\rangle.
\end{equation}
This construction is very similar to the non-relativistic one and reduces to the latter when $p^0\approx p'^0\approx M$.

For a probabilistic interpretation, we need to be able to express the expectation value $\langle\Psi|\hat O|\Psi\rangle$ in a diagonal form\footnote{In spin space, one uses a spin density matrix representation where two canonical polarizations are converted into an unpolarized contribution $\delta_{s's}$ and polarized contributions involving the spin matrices $\uvec S_{s's}$.}. In position space this can be achieved in the case of Galilean symmetry since the latter implies invariance of inertia under a change of frame, and hence a decoupling in momentum space of $\uvec P$- and $\uvec\Delta$-dependences in the matrix elements $\langle p',s'|\hat O(x)|p,s\rangle/(2\sqrt{p'^0p^0})$. One can then write in general
\begin{equation}
\begin{aligned}
    &\int\frac{\ud^3p'}{(2\pi)^3}\,\frac{\ud^3p}{(2\pi)^3}\,\widetilde\Psi^*(\uvec p',s')\widetilde\Psi(\uvec p,s)\, f(\uvec P)\,g(\uvec\Delta)\\
    &=\int\frac{\ud^3P}{(2\pi)^3}\,\frac{\ud^3\Delta}{(2\pi)^3}\,\ud^3r'\,\ud^3r\,\Psi^*(\uvec r',s')\Psi(\uvec r,s)\,e^{-i\uvec P\cdot\uvec z}\,f(\uvec P)\,e^{i\uvec\Delta\cdot\uvec R}\,g(\uvec\Delta)\\
    &= \int\ud^3R\left[ \Psi^*(\uvec R,s')f\!\left(\tfrac{1}{i}\overset{\leftrightarrow}{\uvec\nabla}\right)\!\Psi(\uvec R,s)\right]\int \frac{\ud^3\Delta}{(2\pi)^3}\,e^{i\uvec\Delta\cdot\uvec R}\,g(\uvec\Delta),
    \end{aligned}
\end{equation}
where $f$ and $g$ are two functions, and $A \tensor{\uvec\nabla}  B \equiv \tfrac{1}{2}\left[ A(\uvec\nabla B) - B(\uvec\nabla A)\right]$. Note that the average momentum $\uvec P$ is conjugate to the position shift $\uvec z=\uvec r-\uvec r'$, whereas the momentum transfer $\uvec\Delta$ is conjugate to the average position $\uvec R = (\uvec r+\uvec r')/2$. The ability to perform the $\uvec P$-integration independently of the value of $\uvec\Delta$ corresponds therefore to the ability to provide a density interpretation in position space. 

In a relativistic theory, inertia is a frame-dependent concept and $\uvec P$ is usually entangled with $\uvec\Delta$. It is therefore usually not possible to provide a relativistic density interpretation in 3D position space. The only way out is to switch to the light-front (LF) formalism~\cite{Brodsky:1997de} (or consider the infinite-momentum frame), where a Galilean subgroup of the Lorentz group is singled out by choosing a particular LF direction~\cite{Susskind:1967rg,Kogut:1969xa}, allowing for a density interpretation in impact-parameter space (i.e.~the 2D position space orthogonal to the LF direction)~\cite{Soper:1976jc,Burkardt:2000za,Burkardt:2002hr,Miller:2010nz}. Similar densities were proposed earlier by Fleming~\cite{Fleming:1974af} using a rescaling of the wave packets. An extension of this method has recently been used to define new 3D densities~\cite{Epelbaum:2022fjc,Panteleeva:2022khw,Carlson:2022eps,Alharazin:2022xvp}, but concerns about their physical meaning have triggered some discussions~\cite{Freese:2022fat,Panteleeva:2022uii}.

Despite their nice probabilistic interpretation, LF densities in impact-parameter space have however some shortcomings. First, the probabilistic interpretation is limited by the Galilean subgroup. Considering for example the electromagnetic four-current operator $\hat j^\mu$, a probabilistic interpretation can be attributed to the LF charge density $\hat j^+=(\hat j^0+\hat j^3)/\sqrt{2}$ but not to the longitudinal LF current $\hat j^-=(\hat j^0-\hat j^3)/\sqrt{2}$, see e.g.~Ref.~\cite{Chen:2022smg}. Second, in the non-relativistic regime it is in general not clear how to relate the LF densities to the standard non-relativistic 3D densities, even when the system is in average at rest. Third, LF densities appear to be distorted for transversely polarized targets~\cite{Burkardt:2002hr,Carlson:2007xd,Carlson:2009ovh,Gorchtein:2009qq,Alexandrou:2008bn,Alexandrou:2009hs,Lorce:2009bs}, a phenomenon which can be understood to some extent as an artifact coming from looking at $\hat j^+$ instead of $\hat j^0$. Last but not least, even for unpolarized targets the structure of LF densities can sometimes be difficult to conciliate with an intuitive picture of the system. A typical example is the appearance of an unexpected negative core in the LF charge distribution of a neutron~\cite{Miller:2007uy}. These additional LF distortions have recently been understood as artifacts caused by the Melosh-Wigner spin rotation\footnote{Melosh-Wigner rotations are also at the origin of some relations between transverse-momentum dependent parton distributions and orbital angular momentum observed in various models of the nucleon~\cite{Lorce:2011zta,Lorce:2011kn}.}~\cite{Lorce:2020onh,Lorce:2022jyi,Chen:2022smg}.

Because of Lorentz symmetry, the notion of relativistic spatial distribution necessarily depends on the target average momentum $\uvec P$, hindering therefore in general a probabilistic interpretation in position space. We are therefore naturally led to switch our perspective to a phase-space picture, which is \emph{quasi}probabilistic at the quantum level owing to Heisenberg's uncertainty relations. Following the quantum phase-space formalism~\cite{Wigner:1932eb,Hillery:1983ms,Bialynicki-Birula:1991jwl}, one rewrites Eq.~\eqref{momrepP} as
\begin{equation}\label{PSrep}
    \langle\Psi|\hat O(x)|\Psi\rangle=\sum_{s',s}\int\frac{\ud^3P}{(2\pi)^3}\,\ud^3R\,\rho^{s's}_\Psi(\uvec R,\uvec P)\,\langle\hat O\rangle^{s's}_{\uvec R,\uvec P}(x),
\end{equation}
where
\begin{equation}
    \begin{aligned}
        \rho^{s's}_\Psi(\uvec R,\uvec P)&\equiv\int\ud^3z\,e^{-i\uvec P\cdot\uvec z}\,\Psi^*(\uvec R - \tfrac{\uvec z}{2},s')\Psi(\uvec R + \tfrac{\uvec z}{2},s)\\
        &=\int\frac{\ud^3q}{(2\pi)^3}\,e^{-i\uvec q\cdot\uvec R}\,\widetilde\Psi^*(\uvec P+\tfrac{\uvec q}{2},s')\widetilde\Psi(\uvec P-\tfrac{\uvec q}{2},s)
    \end{aligned}
\end{equation}
is the Wigner distribution interpreted as the quantum weight (positive or negative) for finding the system at average position $\uvec R$ with average momentum $\uvec P$. This construction does not rely particularly on Galilean or Lorentz symmetries, and hence makes the connection with the non-relativistic theory straightforward. Probabilistic densities are recovered upon integration over average position or momentum variables
\begin{equation}
    \begin{aligned}
        \int\ud^3R\,\rho_\Psi^{s's}(\uvec R,\uvec P)&=\widetilde\Psi^*(\uvec P,s')\widetilde\Psi(\uvec P,s),\\
        \int\frac{\ud^3P}{(2\pi)^3}\,\rho_\Psi^{s's}(\uvec R,\uvec P)&=\Psi^*(\uvec R,s')\Psi(\uvec R,s).
    \end{aligned}
\end{equation}
A compelling feature of the quantum phase-space formalism is that wave-packet details are cleanly factorized in Eq.~\eqref{PSrep}. We can then interpret the phase-space amplitude
\begin{equation}
    \langle\hat O\rangle^{s's}_{\uvec R,\uvec P}(x)=\int\frac{\ud^3\Delta}{(2\pi)^3}\,e^{i\uvec\Delta\cdot\uvec R}\,\frac{\langle P+\tfrac{\Delta}{2},s'|\hat O(x)|P-\tfrac{\Delta}{2},s\rangle}{2\sqrt{p'^0p^0}}
\end{equation}
as the internal distribution associated with a state localized in the Wigner sense around average position $\uvec R$ and average momentum $\uvec P$~\cite{Lorce:2018zpf,Lorce:2018egm,Lorce:2021gxs}. Whenever the $\uvec P$-dependence of $\langle\hat O\rangle^{s's}_{\uvec R,\uvec P}(x)$ is simple (typically when Galilean symmetry is at play), we can extract it and use
\begin{equation}
    \int\frac{\ud^3P}{(2\pi)^3}\,\rho_\Psi^{s's}(\uvec R,\uvec P)f(\uvec P)=\Psi^*(\uvec R,s')f\!\left(\tfrac{1}{i}\overset{\leftrightarrow}{\uvec\nabla}\right)\!\Psi(\uvec R,s)
\end{equation}
to obtain genuine internal densities (i.e.~internal distributions with a probabilistic interpretation), see e.g.~Ref.~\cite{Freese:2022fat} for a recent detailed discussion. 

By relaxing the requirement of probabilistic interpretation, the quantum phase-space formalism overcomes the shortcomings associated with the LF densities, shows that the latter are closely related to the instant-form distributions defined in the infinite-momentum frame (IMF), and explains the various LF distortions as a result of relativistic kinematical effects associated with spin~\cite{Lorce:2020onh,Lorce:2022jyi,Chen:2022smg}.

\section{Breit frame distributions}
\label{sec:Breit frame distributions}

From a phase-space perspective, the BF can be regarded as the average rest frame of the system. Since the energy transfer constrained by $\Delta^0=\uvec P\cdot\uvec\Delta/P^0$ vanishes when $\uvec P=\uvec 0$, internal distributions in the BF do not depend on $x^0$. BF distributions are therefore defined as
\begin{equation}\label{BFdistr}
    O_B(\uvec r)\equiv\langle\hat O\rangle^{s'_Bs_B}_{\uvec 0,\uvec 0}(\uvec r)=\int\frac{\ud^3\Delta}{(2\pi)^3}\,e^{-i\uvec\Delta\cdot\uvec r}\,\frac{\langle p'_B,s'_B|\hat O(0)|p_B,s_B\rangle}{2P^0_B},
\end{equation}
where $\uvec r=\uvec x-\uvec R$ is the distance relative to the center of the system, $\uvec p'_B=-\uvec p_B=\uvec\Delta/2$ and $P^0_B=p'^0_B=p^0_B=M\sqrt{1+\tau}$.

Applying the general definition~\eqref{BFdistr} to the electromagnetic four-current operator, one obtains using the BF amplitudes in Eq.~\eqref{BFampl}~\cite{Yennie:1957rmp,Friar1975,Lorce:2020onh}
\begin{equation}\label{BFdensities}
	\begin{aligned}
		J^0_B(\uvec r)&=e\int\frac{\ud^3\Delta}{(2\pi)^3}\,e^{-i\uvec\Delta\cdot\uvec r}\,\frac{M}{P^0_B}\,G_E(\uvec\Delta^2),\\
		\uvec J_B(\uvec r)&=e\,\frac{\uvec\nabla\times\uvec\sigma}{2M}\int\frac{\ud^3\Delta}{(2\pi)^3}\,e^{-i\uvec\Delta\cdot\uvec r}\,\frac{M}{P^0_B}\,G_M(\uvec\Delta^2),
	\end{aligned}
\end{equation}
where explicit spin indices have been omitted for better legibility. These relativistic distributions differ from the conventional ones introduced by Sachs~\cite{Ernst:1960zza,Sachs:1962zzc}, where the factor $M/P^0_B=1/\sqrt{1+\tau}$ has been removed by hand. A detailed discussion of these BF distributions for a nucleon target can be found in Refs.~\cite{Lorce:2020onh,Chen:2022smg}.

\subsection{$A$-type polarization-magnetization tensor}
\label{sec:$A$-type polarization-magnetization tensor}

We can now apply the same formalism to the polarization-magnetization tensor $P^{\mu\nu}$. Evaluating Eq.~\eqref{polmagT} in the BF leads to
\begin{equation}
\begin{aligned}\label{spinhalf3DBFPvMv}
     \widetilde{\mathcal P}^i_B &= \widetilde P^{0i}_B=0,
     \\
    \widetilde M^i_B  &=-\frac{1}{2}\,\epsilon^{ijk}\widetilde P^{jk}_B=e\left[\sigma^i-\frac{\Delta^i(\uvec\Delta\cdot\uvec\sigma)}{4P^0_B(P^0_B+M)}\right]G_M(Q^2).
\end{aligned}
\end{equation}
The corresponding relativistic 3D distributions are then given by
\begin{equation}\label{simplepicture}
	\begin{aligned}
		\uvec{\mathcal P}_B(\uvec r) &= \uvec 0,\\
		\uvec M_B(\uvec r) &= \frac{e}{2M}\int\frac{\ud^3\Delta}{(2\pi)^3}\,e^{-i\uvec\Delta\cdot\uvec r}\left[\uvec\sigma-\frac{\uvec\Delta(\uvec\Delta\cdot\uvec\sigma)}{4P^0_B(P^0_B+M)}\right]\frac{M}{P^0_B}\,G_M(\uvec\Delta^2).
	\end{aligned}
\end{equation}
We see that $\rho_P\equiv -\uvec\nabla\cdot\uvec{\mathcal P}$, the polarization contribution to the charge distribution, vanishes in the BF simply because the BF polarization distribution itself vanishes. The BF magnetization distribution has two terms. Taking the curl eliminates the second term and we find
\begin{equation}
    \uvec J_B(\uvec r)=\uvec\nabla\times\uvec M_B(\uvec r),
\end{equation}
as expected for a system in its average rest frame. 

In magnetostatics, it is customary to define an {\it effective} magnetic charge distribution
\begin{equation}
	\rho_M\equiv-\uvec\nabla\cdot\uvec M,
\end{equation}
by analogy with the polarization charge distribution $\rho_P$. Using the results in Eq.~\eqref{simplepicture}, we find that the BF effective magnetic charge distribution is given by
\begin{equation}\label{effectiveMagCharge}
    \rho_{M,B}(\uvec r)=\frac{e}{2M}\int\frac{\ud^3\Delta}{(2\pi)^3}\,e^{-i\uvec\Delta\cdot\uvec r}\,(i\uvec\Delta\cdot\uvec\sigma)\left(\frac{M}{P^0_B}\right)^2G_M(\uvec\Delta^2).
\end{equation}
Contrary to the BF charge distribution $J^0_B(\uvec r)$, the BF effective magnetic charge distribution is spin-dependent and is not spherically symmetric. The target polarization provides a preferred spatial direction which reduces spherical symmetry to axial symmetry.

\begin{figure}[tb!]
	\centering
	{\includegraphics[angle=0,scale=0.38]{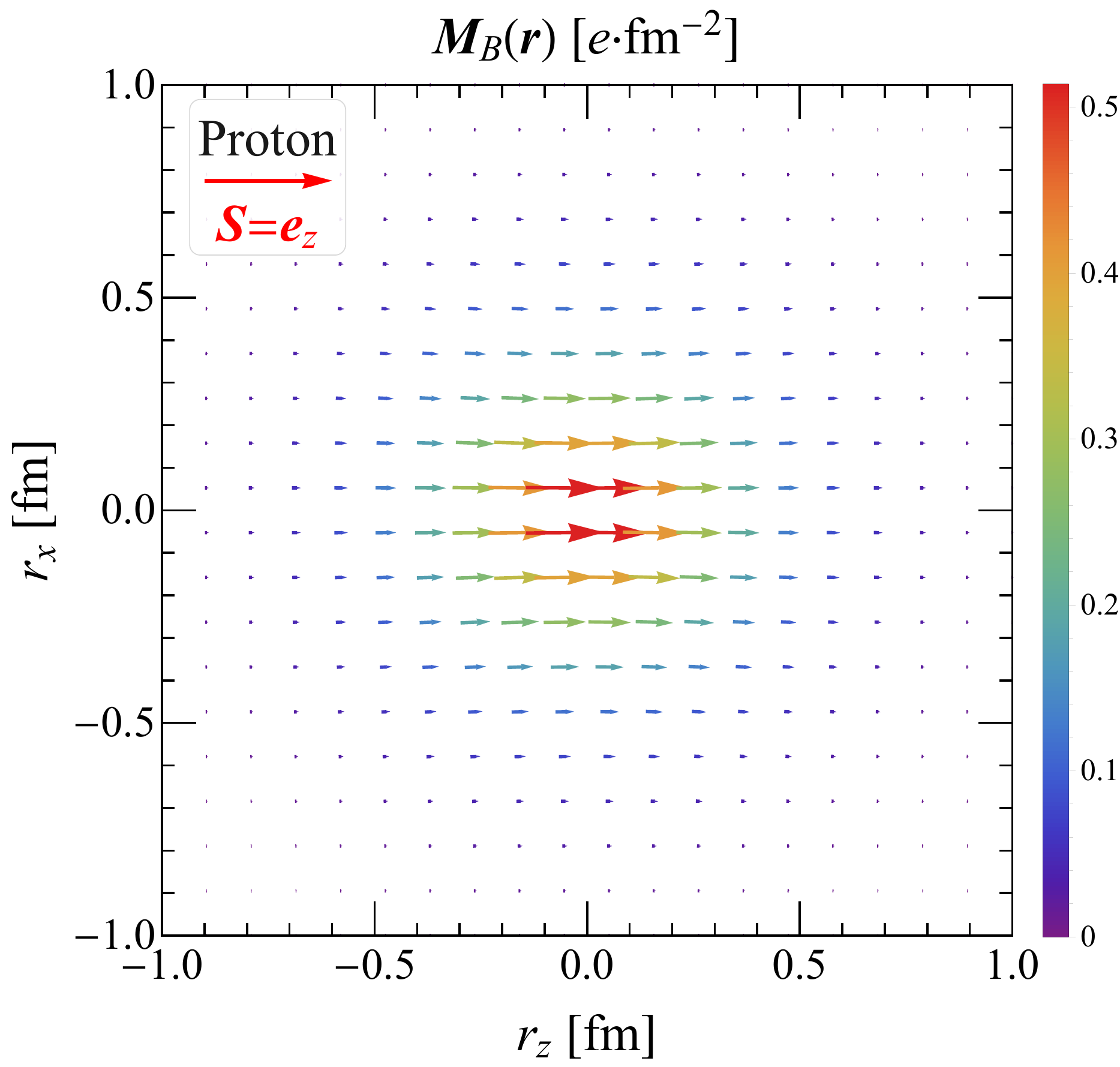}\ \ }
	{\includegraphics[angle=0,scale=0.38]{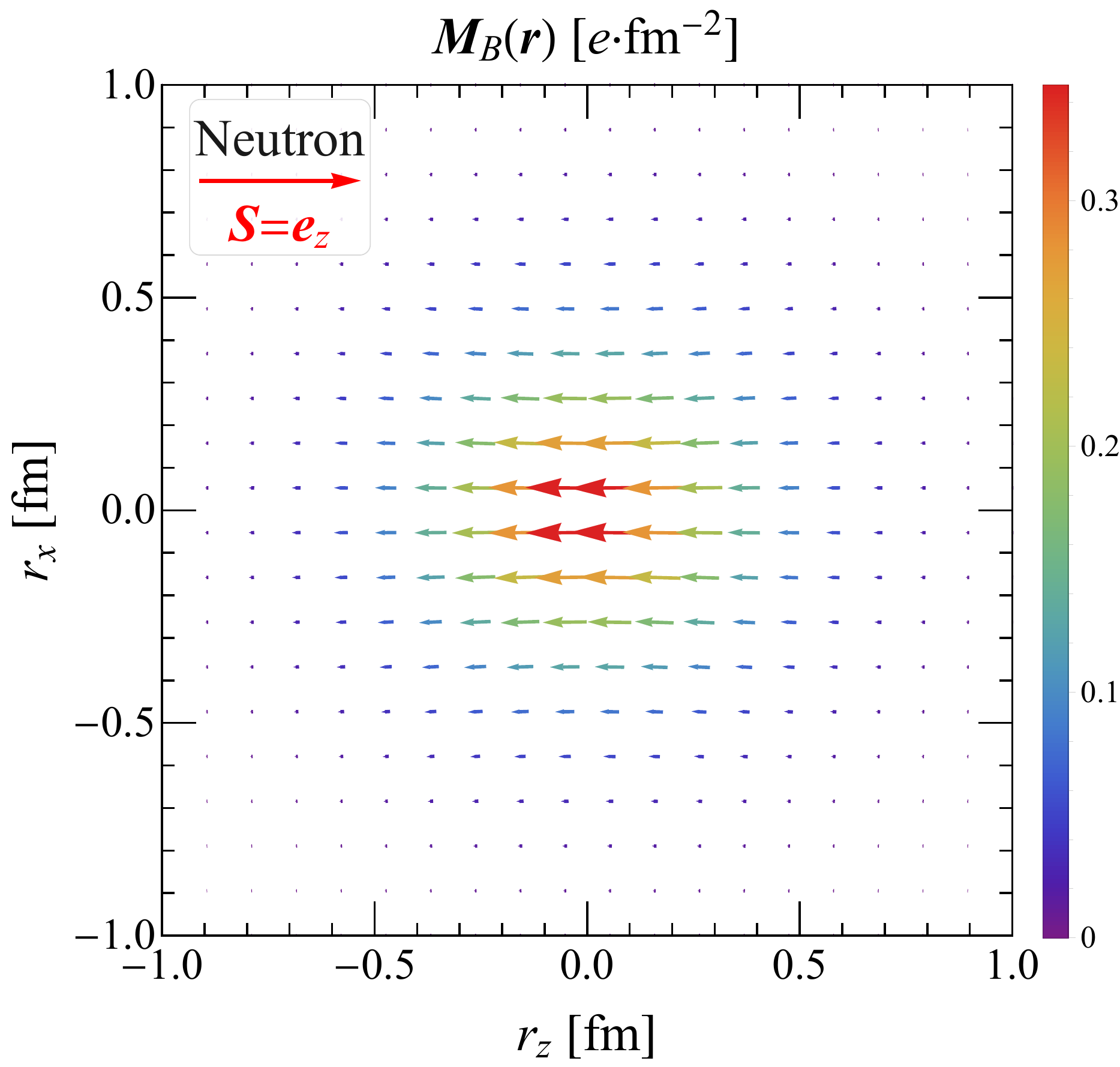}}
	{\includegraphics[angle=0,scale=0.38]{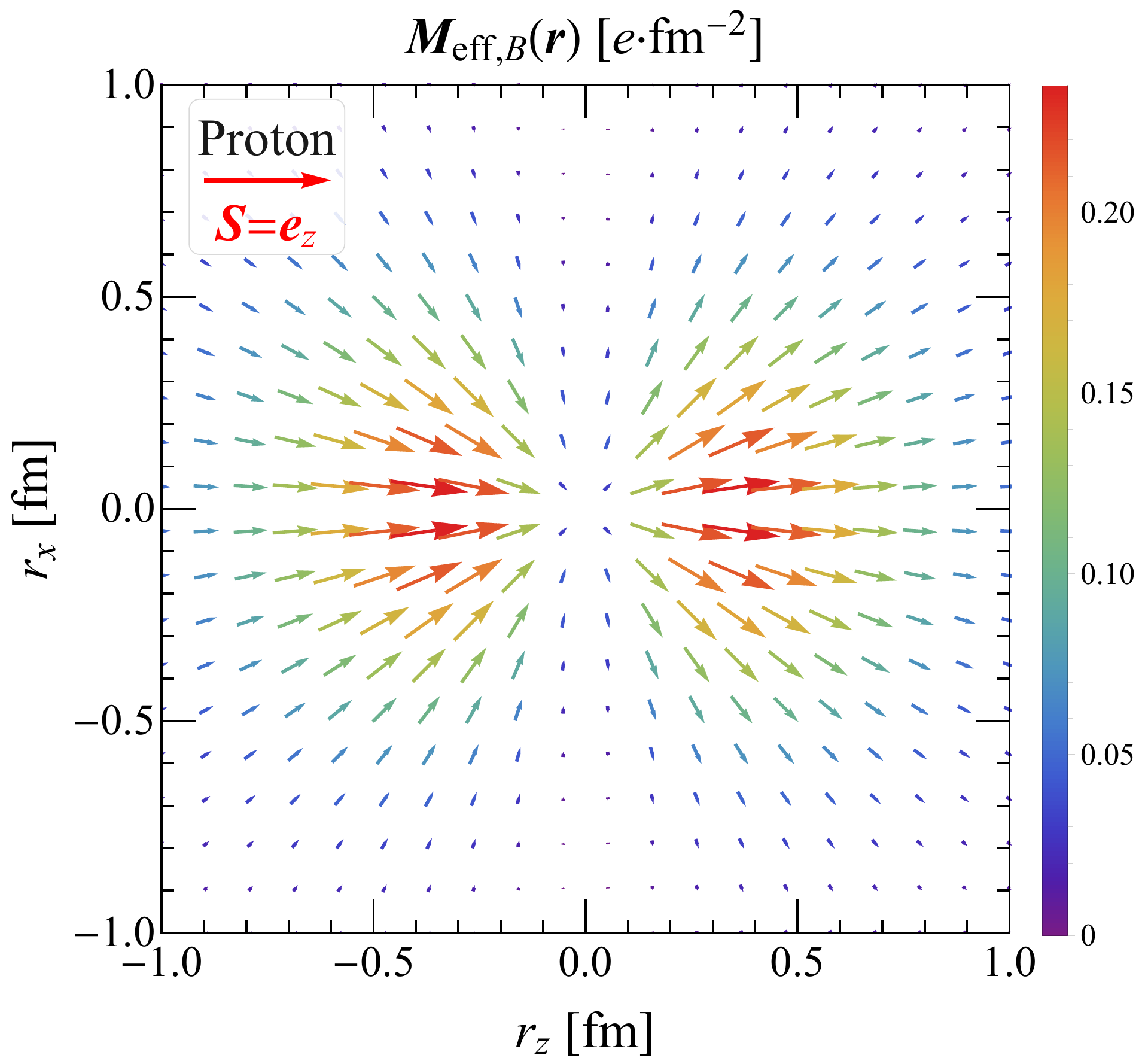}}
	{\ \includegraphics[angle=0,scale=0.38]{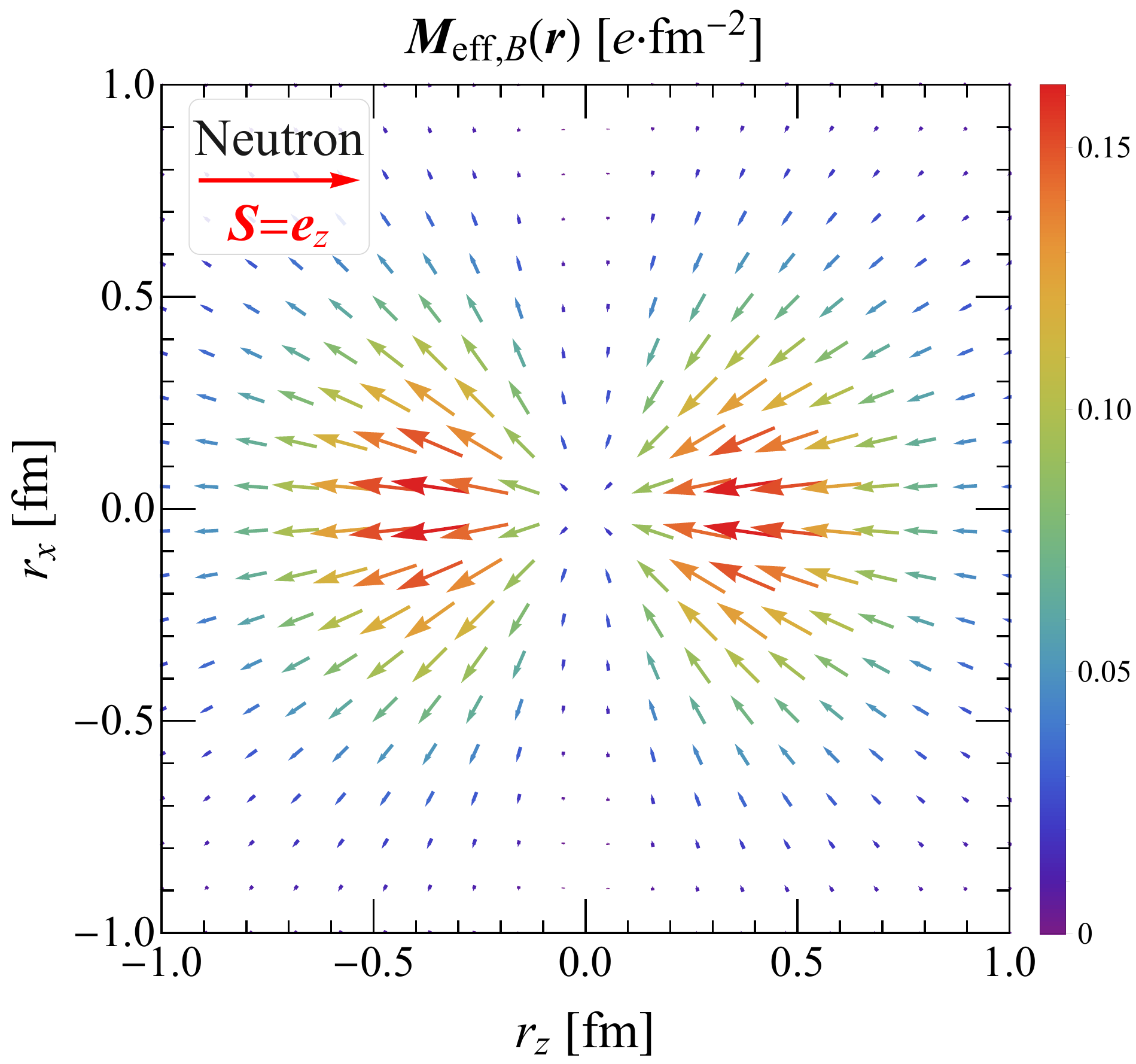}}
	{\includegraphics[angle=0,scale=0.38]{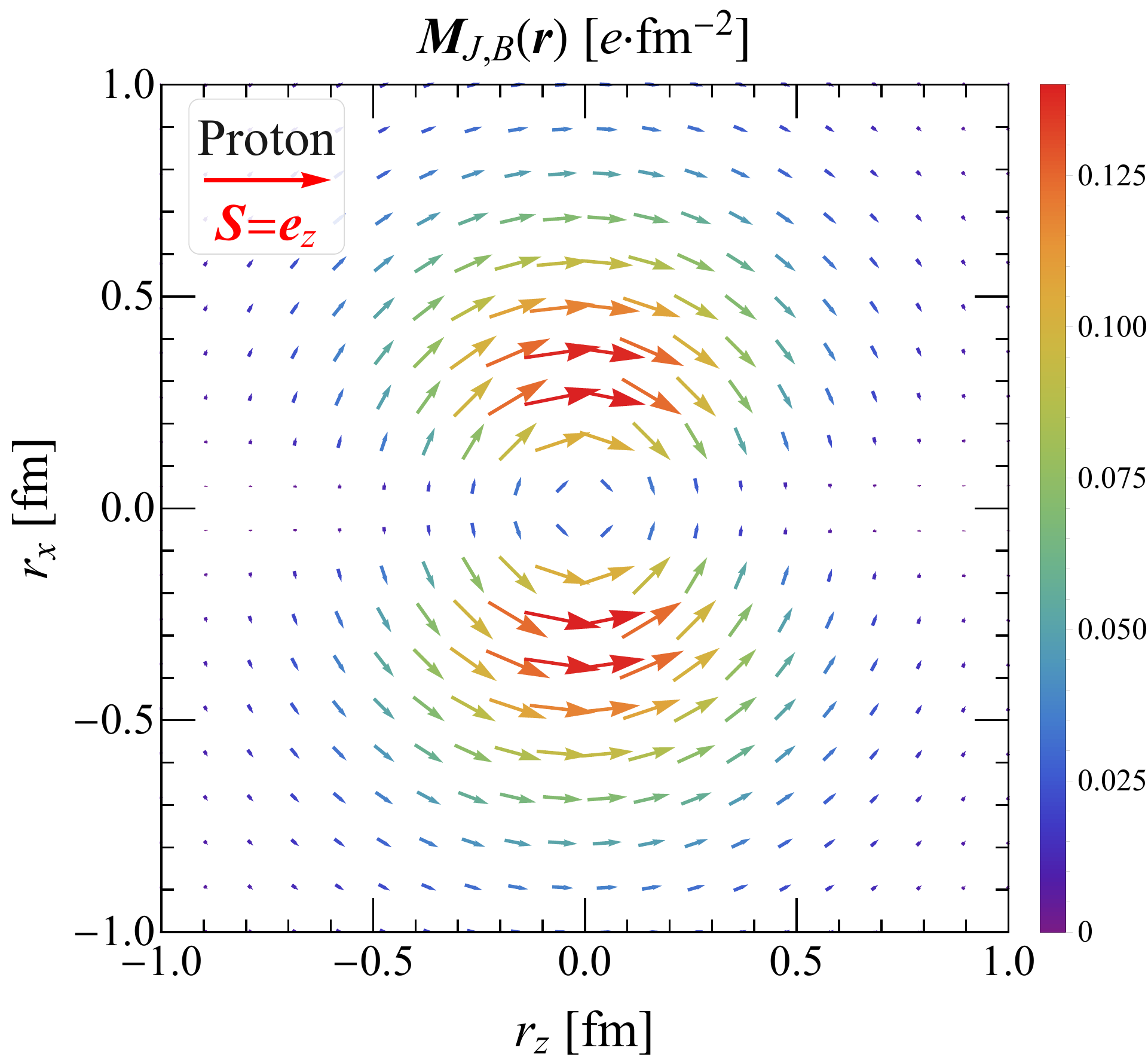}}
	{\includegraphics[angle=0,scale=0.38]{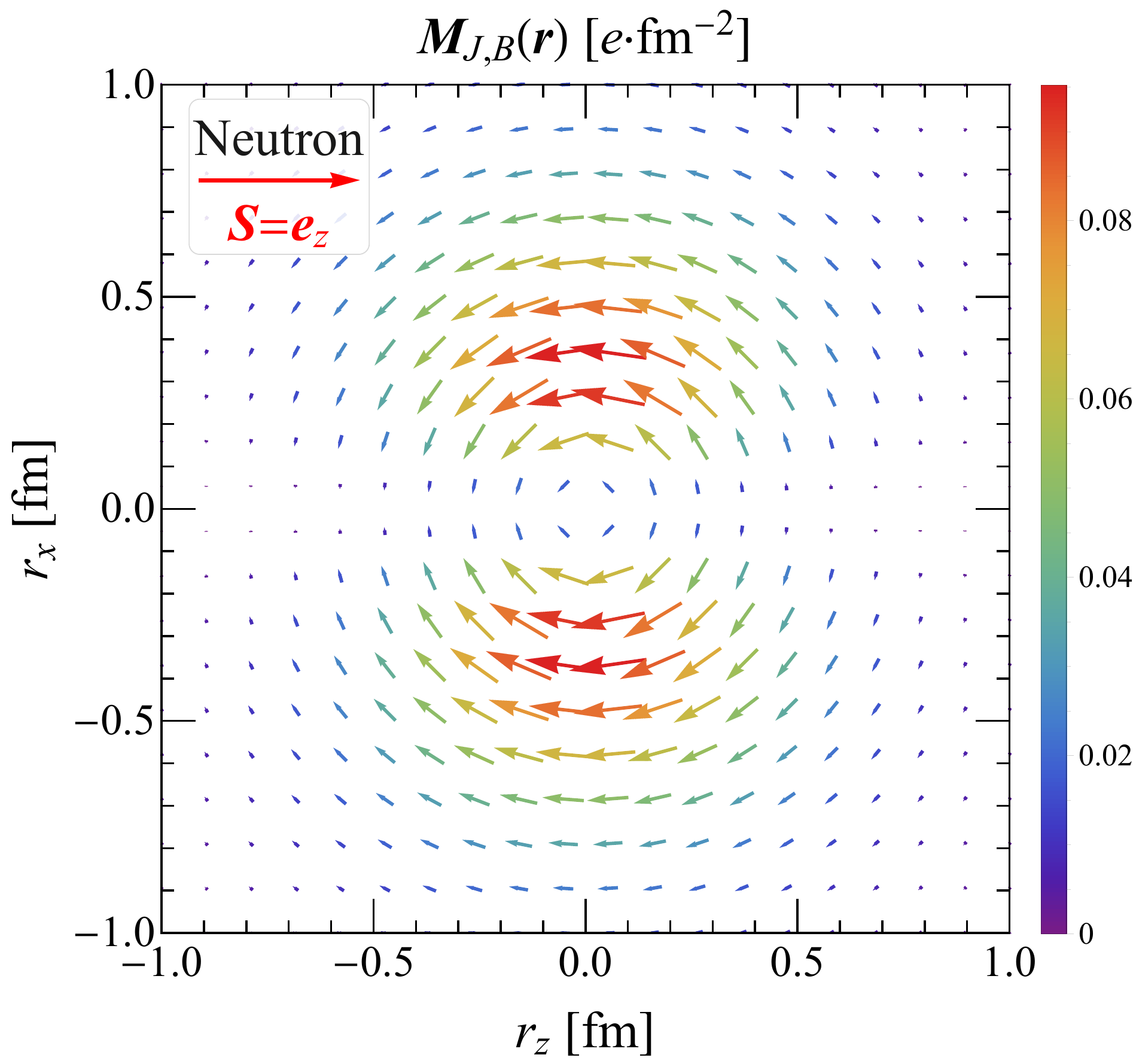}}
	\caption{Comparisons between three kinds of magnetization distributions in the Breit frame $\uvec M_B(\uvec r)$, $\uvec M_{\text{eff},B}(\uvec r)\equiv \uvec r \rho_{M,B}(\uvec r)$ and $\uvec M_{J,B}(\uvec r)\equiv \frac{1}{2}\,\uvec r \times \uvec J_{B}(\uvec r)$ inside a proton (left panels) or a neutron (right panels) polarized along the $z$-direction. The vector plots give the direction and magnitude of the magnetization distributions, evaluated in the $r_y=0$ plane using the parametrization for the nucleon electromagnetic form factors given in Ref.~\cite{Bradford:2006yz}.}
	\label{Fig_Mv3DBF2D}
\end{figure}

Discrete spacetime symmetries impose that the total EDM must vanish in the average rest frame. We indeed find
\begin{equation}\label{spinhalf-BFEDM}
	\uvec d_B=\int\ud^3r\,\uvec r\,J^0_B(\uvec r)=\int\ud^3r\,\uvec r\,\rho_{P,B}(\uvec r)=\int\ud^3r\,\uvec{\mathcal P}_B(\uvec r)=\uvec 0.
\end{equation}
In contrast, the total MDM in the rest frame is not required to vanish and can be expressed in at least three different but equivalent ways,
\begin{equation}\label{spinhalf-BFMDM}
    \uvec\mu_{B}=\int\ud^3r\,\uvec M_{B}(\uvec r) =\int\ud^3r\,\uvec r\,\rho_{M,B}(\uvec r)=\int\ud^3r\,\frac{\uvec r\times\uvec J_B(\uvec r)}{2}=\uvec \sigma\,G_M(0)\,\frac{e}{2M},
\end{equation}
provided that the surface terms vanish at infinity. Note however that at the level of spatial distributions the three integrands, namely $\uvec M_B(\uvec r)$, $\uvec M_{\text{eff},B}(\uvec r)\equiv \uvec r \rho_{M,B}(\uvec r)$ and $\uvec M_{J,B}(\uvec r) \equiv \frac{1}{2}\, \uvec r \times \uvec J_{B}(\uvec r)$, look quite different, see Fig.~\ref{Fig_Mv3DBF2D}. Since these BF spatial distributions are axially symmetric about the polarization axis, it is sufficient to show a section containing the latter. Strictly speaking, $\uvec M_{J,B}(\uvec r)$ should be interpreted as the contribution to the MDM at $\uvec r=\uvec 0$ due to the current element at position $\uvec r$. Similarly, $\uvec M_{\text{eff},B}(\uvec r)$ corresponds to the contribution to the MDM at $\uvec r=\uvec 0$ due to the effective magnetic charge element at position $\uvec r$. Only $\uvec M_B(\uvec r)$ can be thought of as the genuine spatial distribution of magnetization.

\subsection{$T$-type polarization-magnetization tensor}
\label{sec:T-type polarization-magnetization tensor}

For comparison, we consider here the $T$-type definition $P'^{\mu\nu}$ for the polarization-magnetization tensor. Evaluating Eq.~\eqref{polmagTbis} in the BF gives
\begin{equation}
\begin{aligned}\label{spinhalf3DBFPvMvbis}
     \widetilde{\mathcal P}'^i_B &= \widetilde P'^{0i}_B=e\,\frac{i\Delta^i}{2M}\,G_M(Q^2),\\
    \widetilde M'^i_B  &=-\frac{1}{2}\,\epsilon^{ijk}\widetilde P'^{jk}_B=e\left[\sigma^i+\frac{\Delta^i(\uvec\Delta\cdot\uvec\sigma)}{4M(P^0_B+M)}\right]G_M(Q^2),
\end{aligned}
\end{equation}
and so the corresponding relativistic 3D distributions read
\begin{equation}\label{complexpicture}
	\begin{aligned}
		\uvec{\mathcal P}'_B(\uvec r) &=\frac{e}{2M} \int\frac{\ud^3\Delta}{(2\pi)^3}\,e^{-i\uvec\Delta\cdot\uvec r}\,\frac{i\uvec\Delta}{2M}\,\frac{M}{P^0_B}\,G_M(\uvec\Delta^2),\\
		\uvec M'_B(\uvec r) &= \frac{e}{2M}\int\frac{\ud^3\Delta}{(2\pi)^3}\,e^{-i\uvec\Delta\cdot\uvec r}\left[\uvec\sigma+\frac{\uvec\Delta(\uvec\Delta\cdot\uvec\sigma)}{4M(P^0_B+M)}\right]\frac{M}{P^0_B}\,G_M(\uvec\Delta^2).
	\end{aligned}
\end{equation}
This time we have a non-vanishing polarization distribution, but since it is time-independent we still get a pure spin current according to Eq.~\eqref{fourcurrentdetailed}
\begin{equation}
    \uvec J_B(\uvec r)=\uvec\nabla\times\uvec M'_B(\uvec r).
\end{equation}
This can also be seen directly from the expressions for the $A$-type and $T$-type magnetization distributions since they differ only in the part that does not contribute to the curl.

Remembering that $\uvec p'_B=-\uvec p_B=\uvec\Delta/2$ and $p'^0_B=p^0_B=P^0_B$, we recognize in Eq.~\eqref{spinhalf3DBFPvMvbis} the characteristic structure of spin defined relative to the center of energy $\left(\uvec\sigma+\frac{\uvec p_B(\uvec p_B\cdot\uvec\sigma)}{M(p^0_B+M)}\right)$, while we find in Eq.~\eqref{spinhalf3DBFPvMv} the characteristic structure of spin defined relative to the center of mass $\left(\uvec\sigma-\frac{\uvec p_B(\uvec p_B\cdot\uvec\sigma)}{p^0_B(p^0_B+M)}\right)$, see Ref.~\cite{Lorce:2021gxs}\footnote{Pushing the logic further suggests that the combination $P''^{\mu\nu}=(\sqrt{P^2}P^{\mu\nu}+M P'^{\mu\nu})/(\sqrt{P^2}+M)$ could be interpreted as the polarization-magnetization tensor defined relative to the center of spin.}. Therefore, part of the ambiguity in the definition of the polarization-magnetization tensor comes from the choice made for the center of the system, which in turn defines the internal angular momentum or spin. Even though the system in the BF is in average at rest, the initial and final momenta are non-zero. Contrary to the center of mass, the position of the center of energy inside a spinning system depends on its momentum~\cite{Lorce:2018zpf,Lorce:2021gxs}, see Appendix~\ref{App-Relativistic centers of the nucleon}. For $\uvec\Delta\neq \uvec 0$ the center of energy is in general shifted relative to the center of mass, but the shift in the initial state is exactly opposite to that in the final state, so that the \emph{average} position of the center of energy coincides in the BF with that of the center of mass. However, the initial and final shifts affect the appearance of the spatial distributions and imply that $\uvec M'_B(\uvec r)\neq \uvec M_B(\uvec r)$. 

\begin{figure}[tb]
	\centering
	{\ \includegraphics[angle=0,scale=0.38]{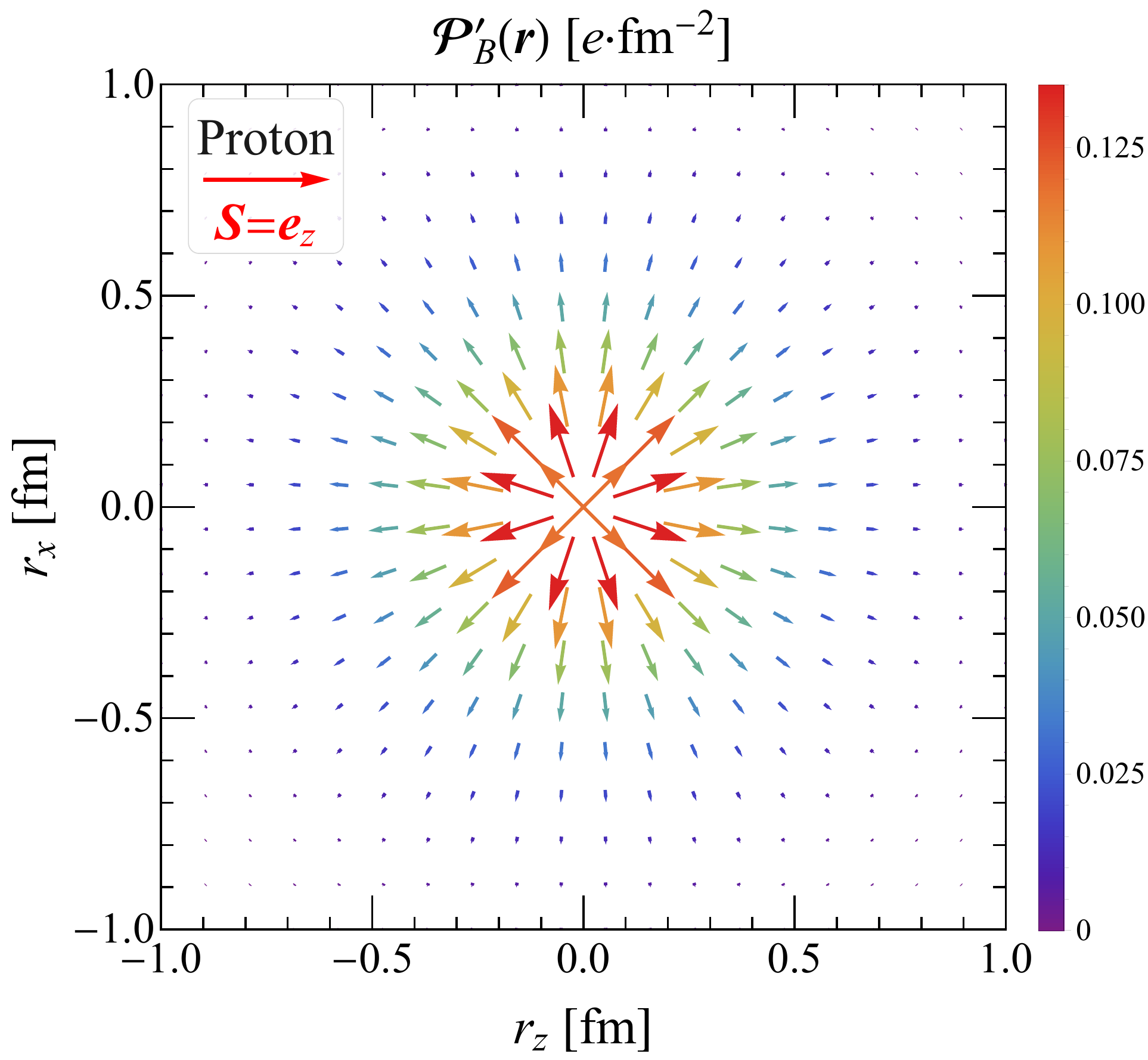}}
	{\includegraphics[angle=0,scale=0.38]{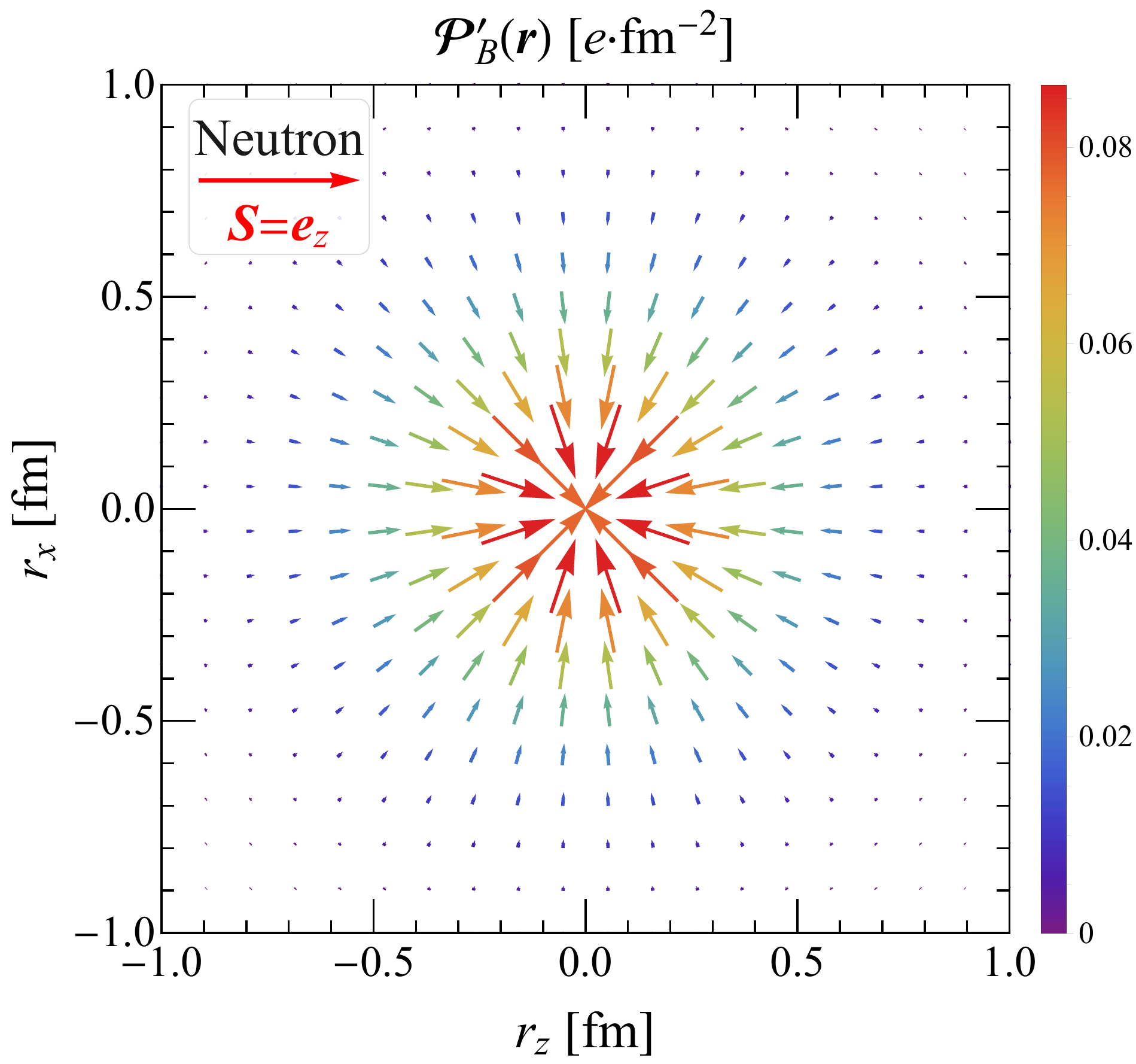}}
	{\includegraphics[angle=0,scale=0.38]{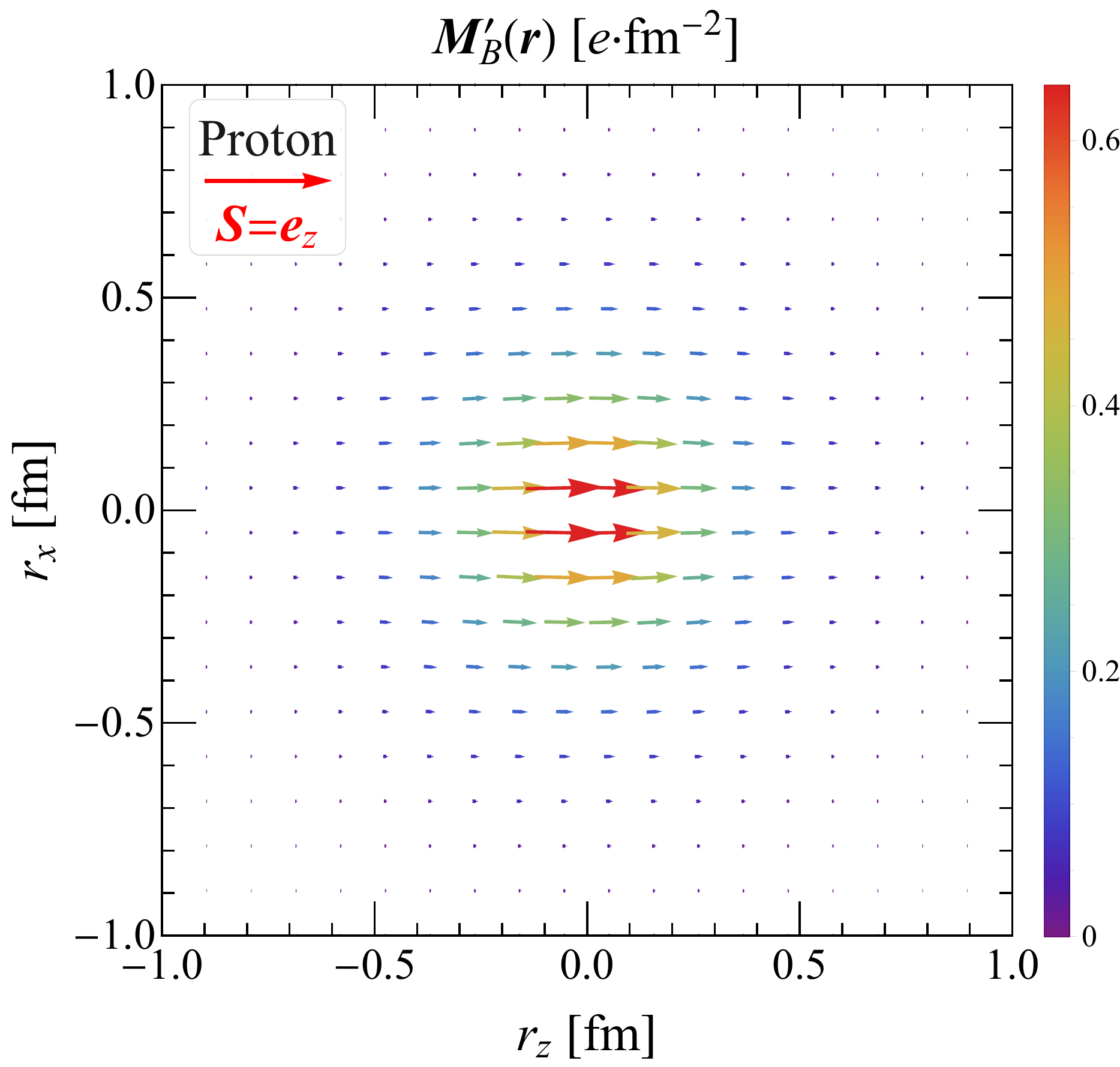}\ \ }
	{\includegraphics[angle=0,scale=0.38]{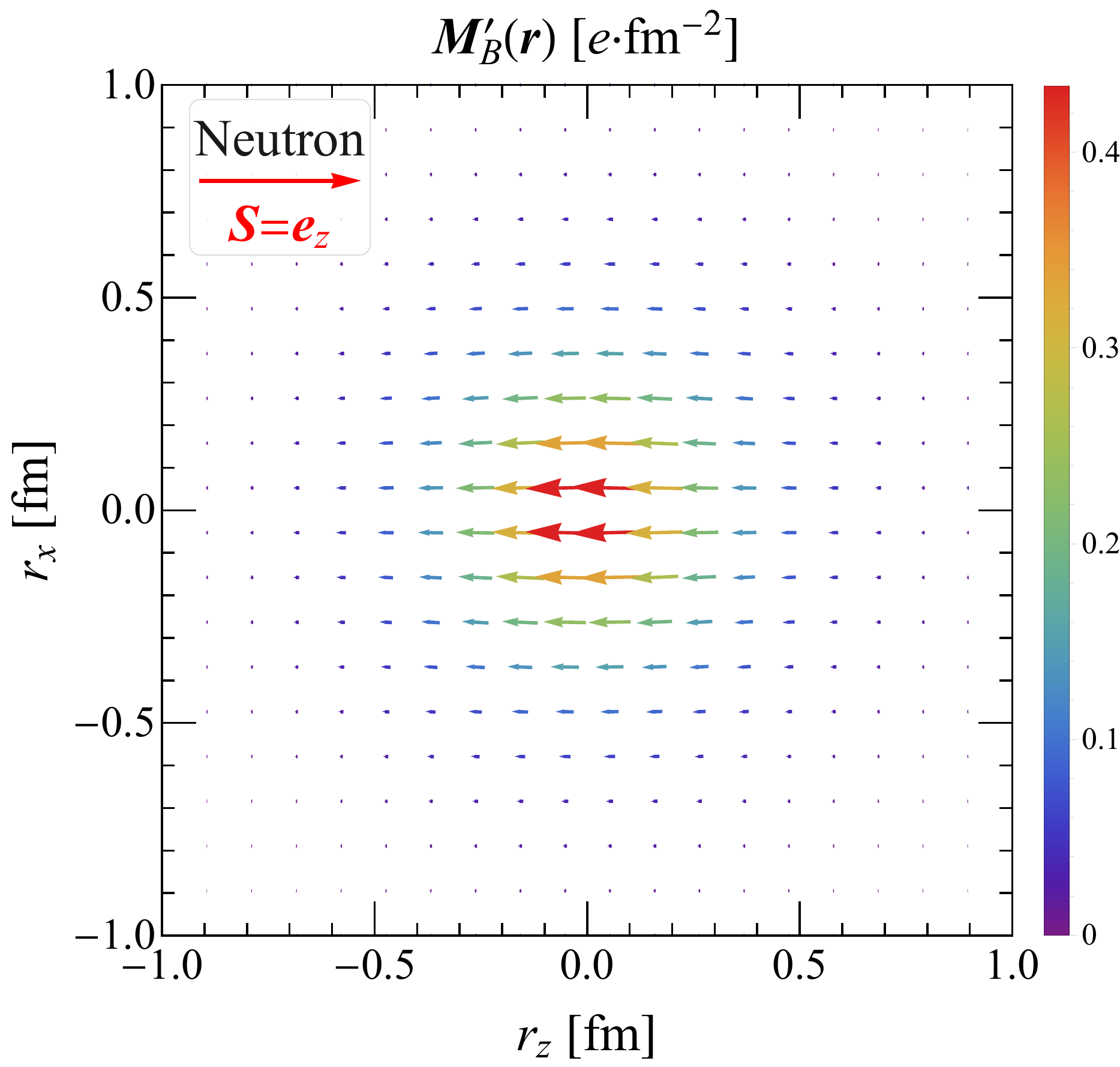}}
	\caption{Breit frame $T$-type polarization and magnetization distributions $\uvec{\mathcal{P}}_B'(\uvec r)$ and $\uvec M_B'(\uvec r)$, see Eq.~\eqref{complexpicture}, inside a proton (left panels) or a neutron (right panels) polarized along the $z$-direction. The vector plots give the direction and magnitude of the distributions, evaluated in the $r_y=0$ plane using the parametrization for the nucleon electromagnetic form factors given in Ref.~\cite{Bradford:2006yz}.}
	\label{Fig_GordonPvMv3DBF2D}
\end{figure}

In Fig.~\ref{Fig_GordonPvMv3DBF2D}, we show the spatial distributions of the $T$-type polarization and magnetization in the BF. Except at the origin, the $T$-type polarization distribution does not vanish and has the structure of a spherical hedgehog. The $T$-type magnetization distribution is indeed similar to but different from the $A$-type magnetization distribution shown in Fig.~\ref{Fig_Mv3DBF2D}.

\begin{figure}[tb]
	\centering
	{\includegraphics[angle=0,scale=0.45]{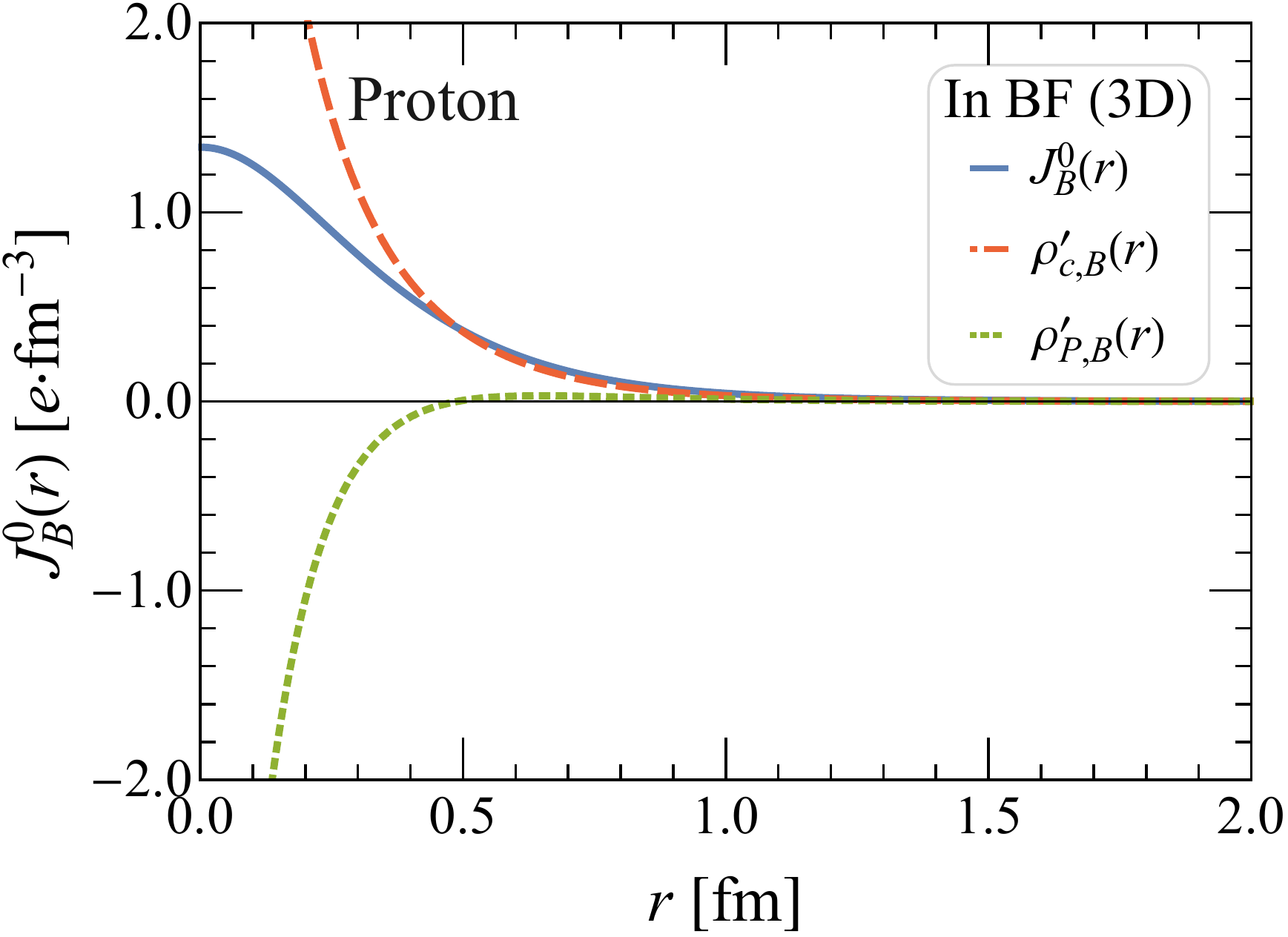}}
	{\includegraphics[angle=0,scale=0.45]{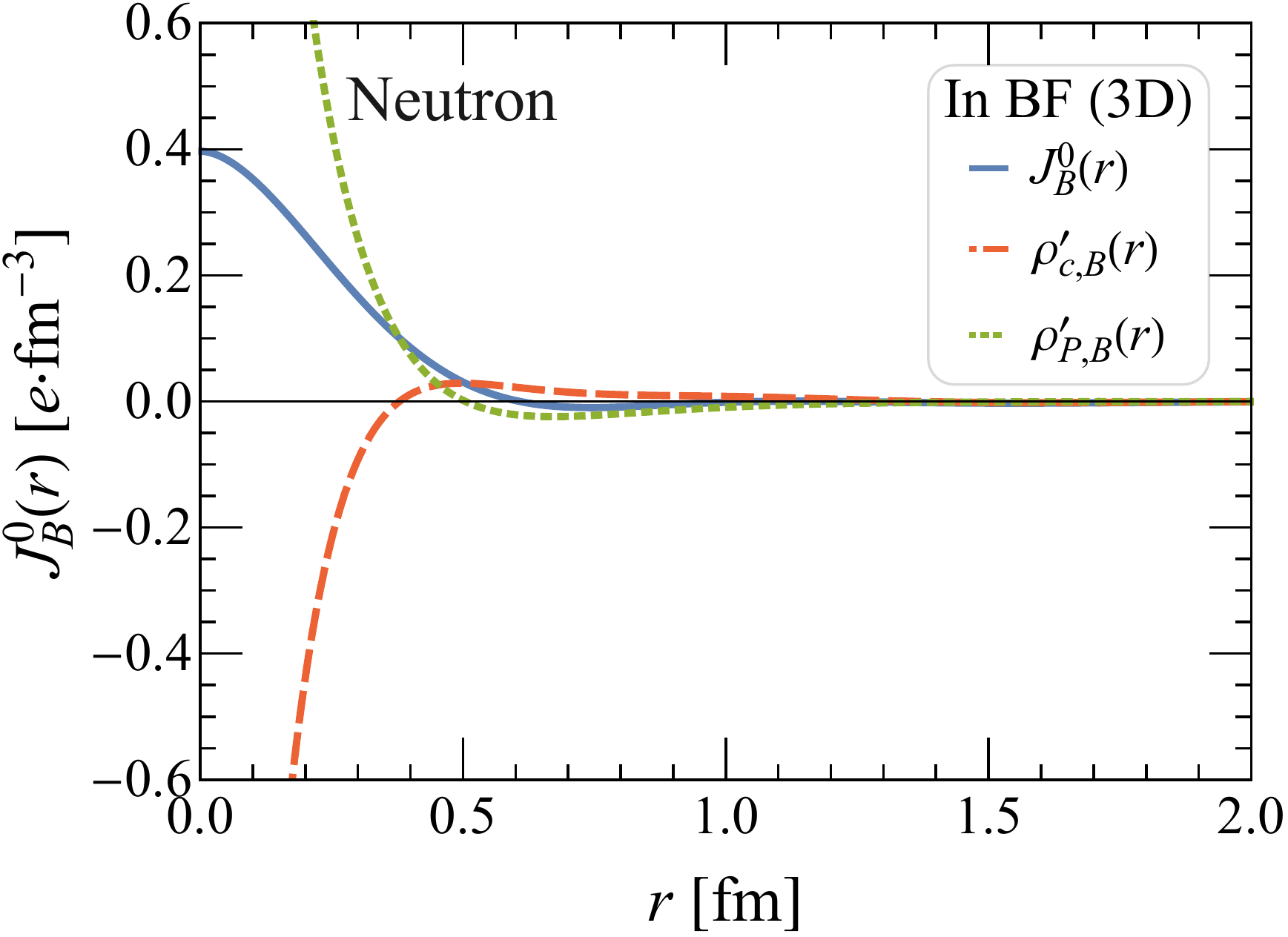}}
	{\includegraphics[angle=0,scale=0.45]{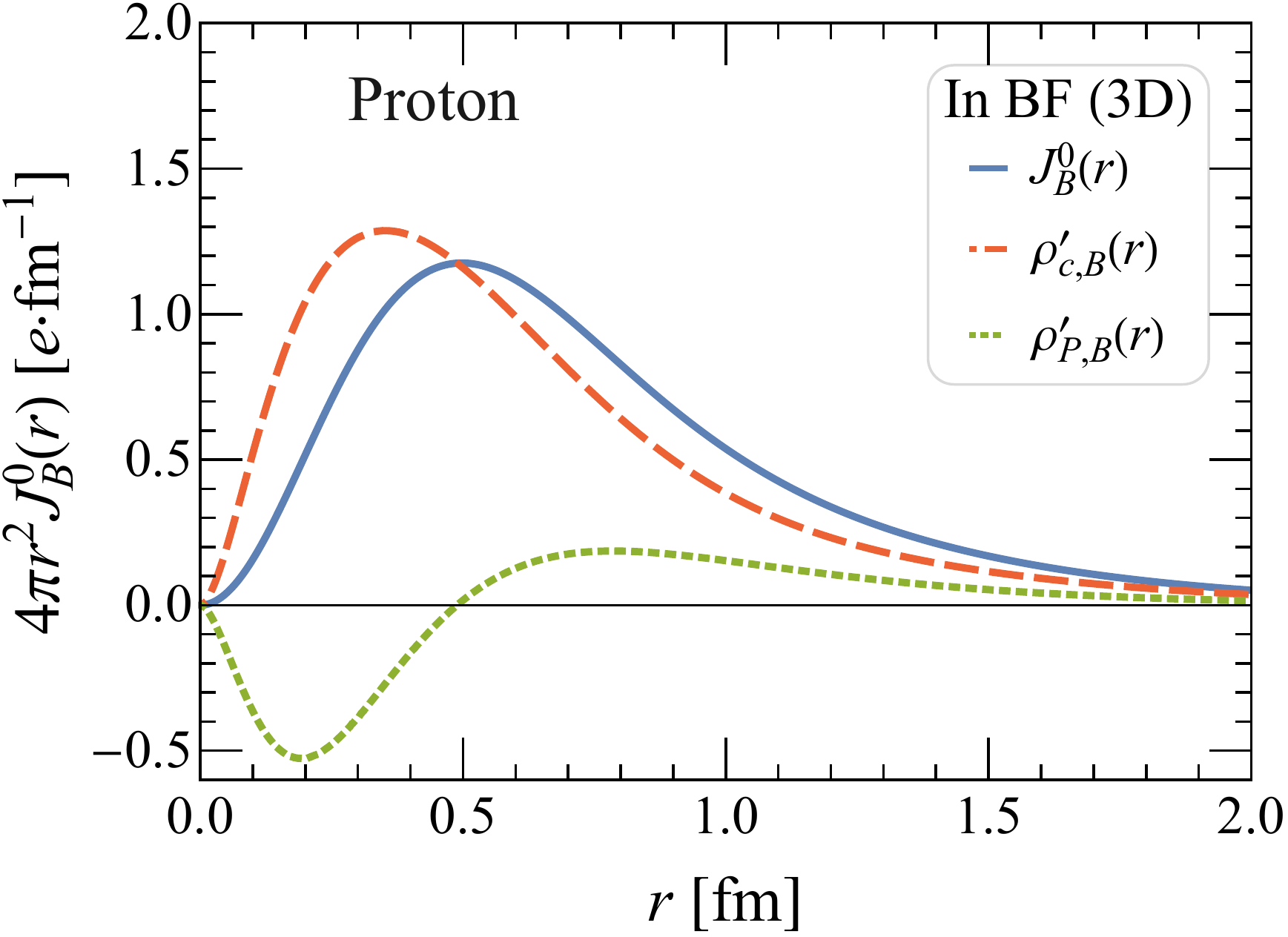}}
	{\includegraphics[angle=0,scale=0.45]{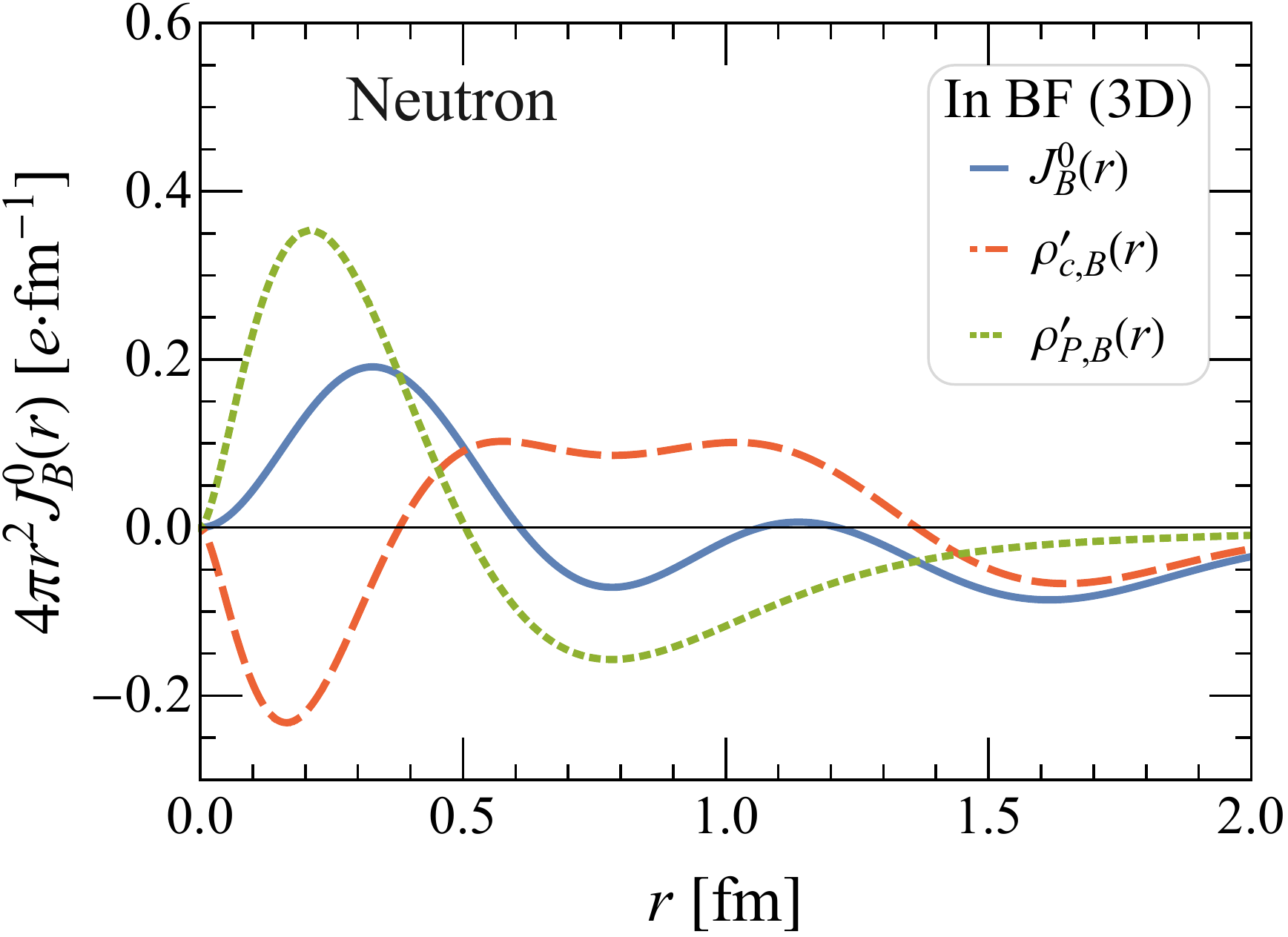}}
	\caption{Decomposition of the Breit frame charge distribution into $T$-type convection and polarization contributions, see Eqs.~\eqref{spinhalf-BFJ0cTtype} and \eqref{spinhalf-BFJ0PTtype}, inside a proton (left panels) or a neutron (right panels), based on the parametrization for the nucleon electromagnetic form factors given in Ref.~\cite{Bradford:2006yz}.}
	\label{Fig_Nucleon3DBFJ0Ttype}
\end{figure}

In the picture based on the $T$-type decomposition~\eqref{Ttypedec} of the electromagnetic four-current, the total charge distribution
\begin{equation}\label{spinhalf-BFJ0Ttype}
    J^0_B(\uvec r)=\rho'_{c,B}(\uvec r)+\rho'_{P,B}(\uvec r)
\end{equation} 
consists in a convection charge distribution driven by the Dirac FF
\begin{equation}\label{spinhalf-BFJ0cTtype}
    \rho'_{c,B}(\uvec r)=e\int\frac{\ud^3\Delta}{(2\pi)^3}\,e^{-i\uvec\Delta\cdot\uvec r}\,\frac{P^0_B}{M}\,F_1(\uvec\Delta^2)
\end{equation}
and a non-vanishing polarization charge distribution given by
\begin{equation}\label{spinhalf-BFJ0PTtype}
    \rho'_{P,B}(\uvec r)=-\uvec\nabla\cdot\uvec{\mathcal P}'_B=-e\int\frac{\ud^3\Delta}{(2\pi)^3}\,e^{-i\uvec\Delta\cdot\uvec r}\,\tau\,\frac{M}{P^0_B}\,G_M(\uvec\Delta^2).
\end{equation}
Both of these contributions are spherically symmetric and are represented in Fig.~\ref{Fig_Nucleon3DBFJ0Ttype}. While the proton charge distribution is dominated by the convection contribution, the neutron charge distribution appears to be globally dominated by the polarization contribution for $r\lesssim 1.4$ fm and by the convection contribution for $r\gtrsim 1.4$ fm.
We observe in particular a large cancellation between the convection and polarization charge distributions close to the center of the nucleon, suggesting that the $T$-type decomposition is not really natural (at least in the BF). 

We also find that the $T$-type effective magnetic charge distribution
\begin{equation}\label{spinhalf-effectiveMagTtype}
    \rho'_{M,B}(\uvec r)=-\uvec\nabla\cdot\uvec M'_B=\frac{e}{2M}\int\frac{\ud^3\Delta}{(2\pi)^3}\,e^{-i\uvec\Delta\cdot\uvec r}\,(i\uvec\Delta\cdot\uvec\sigma)\,G_M(\uvec\Delta^2)
\end{equation}
differs from the $A$-type one~\eqref{effectiveMagCharge} by a relativistic kinematical factor $(M/P^0_B)^2$, which reflects the difference in the Lorentz boost properties between spin defined relative to the center of mass and spin defined relative to the center of energy~\cite{Lorce:2021gxs}.

Finally, the $T$-type BF EDM and MDM
\begin{equation}
\begin{aligned}\label{complex-EDMMDM}
	\uvec d_B'&=\int\ud^3r\,\uvec r\,\rho'_{ P,B}(\uvec r)=\int\ud^3r\,\uvec{\mathcal P}_B'(\uvec r)=\uvec 0,\\
	\uvec\mu_{B}'&=\int\ud^3r\,\uvec M_{B}'(\uvec r) =\int\ud^3r\,\uvec r\,\rho_{M,B}'(\uvec r)=\uvec \sigma\,G_M(0)\,\frac{e}{2M},
\end{aligned}
\end{equation}
are the same as the $A$-type ones, see Eqs.~\eqref{spinhalf-BFEDM} and \eqref{spinhalf-BFMDM}. The reason is that integrating over whole position space amounts to setting $\uvec\Delta=\uvec 0$ in momentum space. As one can see from the on-shell identity~\cite{Lorce:2017isp,Cotogno:2019vjb}
\begin{equation}
   \overline u(p',s')\sigma^{\mu\nu} u(p,s)=\overline u(p',s')\left[\frac{i\Delta^{[\mu}\gamma^{\nu]}}{2M}+\frac{\epsilon^{\mu\nu\beta\lambda} P_\beta\gamma_\lambda\gamma_5}{M}\right]u(p,s),
\end{equation}
where we used the shorthand notation $a^{[\mu}b^{\nu]}\equiv a^{\mu} b^{\nu}-a^{\nu}b^{\mu}$, the difference between $\widetilde P^{\mu\nu}$ and $\widetilde P'^{\mu\nu}$ vanishes in the forward limit $\uvec\Delta\to\uvec 0$, and so the $A$-type and $T$-type polarization-magnetization tensors agree on the integrated quantities but disagree on how these quantities are distributed over space. The results in Eq.~\eqref{complex-EDMMDM} should also be expected from the fact that the EDM and MDM can be expressed directly in terms of the electromagnetic four-current, and hence should not depend on how the latter is decomposed into convection and polarization contributions.

In conclusion, even if defining the polarization-magnetization tensor in terms of the tensor Dirac bilinear seems a priori natural, the associated picture turns out to be more complicated than the one based on the axial-vector Dirac bilinear. For this reason, we consider that the $A$-type polarization-magnetization tensor gives a more physical picture than the $T$-type one.

\section{Elastic frame distributions}
\label{sec:Elastic frame distributions}

BF distributions provide our best proxy for picturing a system at rest around the origin. If we are however interested in the internal structure of a moving system, we can use the so-called elastic frame (EF) distributions introduced in Ref.~\cite{Lorce:2017wkb}. They are defined as
\begin{equation}\label{EFdef}
    O_\text{EF}(\uvec b_\perp;P_z)\equiv\int\ud r_z\,\langle\hat O\rangle^{s's}_{\uvec 0,\uvec P}(r)=\int\frac{\ud^2\Delta_\perp}{(2\pi)^2}\,e^{-i\uvec\Delta_\perp\cdot\uvec b_\perp}\,\frac{\langle p',s'|\hat O(0)|p,s\rangle}{2P^0}\bigg|_{\Delta_z=0},
\end{equation}
where the $z$-axis has been chosen for convenience along $\uvec P=(\uvec 0_\perp,P_z)$, and $\uvec r=(\uvec b_\perp,r_z)$ is the distance relative to the center of the system, which has been set at the origin $\uvec R=\uvec 0$. Integrating over the longitudinal coordinate amounts to setting the momentum transfer in the longitudinal direction to zero, which in turn implies a vanishing energy transfer $\Delta^0=\uvec P\cdot\uvec\Delta/P^0=0$ and hence a time-independent distribution.

At $P_z=0$, the EF distributions coincide with the BF distributions projected onto the transverse plane
\begin{equation}
    O_\text{EF}(\uvec b_\perp;0)=\int\ud r_z\,O_B(\uvec r).
\end{equation}
In the limit $P_z\to\infty$, we obtain the IMF distributions
\begin{equation}
    O_\text{IMF}(\uvec b_\perp)\equiv\lim_{P_z\to\infty}O_\text{EF}(\uvec b_\perp;P_z)
\end{equation}
which coincide most of the time with the distributions defined within the LF formalism, up to some trivial factors~\cite{Lorce:2018egm,Lorce:2020onh,Kim:2021kum,Lorce:2022jyi,Chen:2022smg,Kim:2022bia,Hong:2023tkv}. EF distributions provide therefore a nice and clear interpolation between BF and LF distributions.

To understand how the distributions change with $P_z$, we need to know how matrix elements for different sets of initial and final momenta are related to each other. For the electromagnetic four-current operator, Poincar\'e symmetry implies that~\cite{Jacob:1959at,Durand:1962zza}
\begin{equation}	\label{EFLorentzTrans-Spinj}
		\langle p',s'|\hat j^\mu(0)|p,s\rangle = \sum_{s_{B}',s_{B}} D^{\dag(j)}_{s's_{B}'}(p_{B}',\Lambda)D^{(j)}_{s_{B}s}(p_{B},\Lambda) \,\Lambda^{\mu}_{\phantom{\mu}\nu} \,\langle p_{B}',s_{B}'|\hat j^{\nu}(0)|p_{B},s_{B}\rangle,
\end{equation}
where $p^{(\prime)\mu}=\Lambda^{\mu}_{\phantom{\mu}\nu}p^{(\prime)\nu}_B$ and $D^{(j)}$ is the Wigner rotation matrix for spin-$j$ targets. For the polarization-magnetization tensor, we can write in a similar way
\begin{equation}\label{tensorLT}
		(\widetilde P^{\mu\nu})_{s's}= \sum_{s_{B}',s_{B}} D^{\dag (j)}_{s's_{B}'}(p_{B}',\Lambda)D^{(j)}_{s_{B}s}(p_{B},\Lambda) \,\Lambda^{\mu}_{\phantom{\mu}\alpha}\,\Lambda^{\nu}_{\phantom{\nu}\beta} \,(\widetilde P^{\alpha\beta}_B)_{s'_Bs_B}.
\end{equation}
In the case of a spin-$\frac{1}{2}$ system in the EF, the Wigner rotation matrix takes the form
\begin{equation}\label{Wignerhalf}
    D^{(1/2)}_{s_{B}s}(p_{B},\Lambda)=D^{\dag(1/2)}_{s's'_{B}}(p'_{B},\Lambda)=\begin{pmatrix}\cos\frac{\theta}{2}&-e^{-i\phi_\Delta}\sin\frac{\theta}{2}\\ e^{i\phi_\Delta}\sin\frac{\theta}{2}&\cos\frac{\theta}{2}\end{pmatrix},
\end{equation}
with $\uvec\Delta=(Q\cos\phi_\Delta,Q\sin\phi_\Delta,0)$, and the Wigner rotation angle $\theta$ satisfies~\cite{Chen:2022smg}
\begin{equation}\label{WignerAngleSinCos}
    \cos\theta=\frac{P^0+M(1+\tau)}{(P^0+M)\sqrt{1+\tau}},\qquad \sin\theta=-\frac{\sqrt{\tau}P_z}{(P^0+M)\sqrt{1+\tau}},
\end{equation}
where the EF energy is given by $P^0=p'^0=p^0=\sqrt{P^2_z+M^2(1+\tau)}$. When $P_z\neq 0$, the Wigner rotation depends on the momentum transfer $\uvec\Delta$, and hence distorts the spatial distributions after the Fourier transform~\cite{Chen:2022smg}.

\subsection{Elastic frame polarization and magnetization}
\label{sec:Elastic-frame polarization and magnetization}

Since the BF analysis in the previous section revealed that the $A$-type definition of the polarization-magnetization tensor from the physics perspective was more natural than the $T$-type one, we will consider only the former in the following. Evaluating Eq.~\eqref{polmagT} in the generic EF leads to
\begin{equation}
    \begin{aligned}\label{polmagEF}
        \widetilde M_{z,\text{EF}}&=e\,\sigma_z\,G_M(Q^2),\\
        \widetilde{\uvec M}_{\perp,\text{EF}}&=e\,\gamma\left[\frac{(\uvec e_z\times i\uvec\Delta)_\perp}{|\uvec \Delta_\perp|}\left(\frac{(\uvec\sigma\times i\uvec\Delta)_z}{|\uvec \Delta_\perp|}\,\cos\theta-\sin\theta\right)+\frac{\uvec\Delta_\perp(\uvec\Delta_\perp\cdot\uvec\sigma_\perp)}{\uvec \Delta^2_\perp\sqrt{1+\tau}}\right]G_M(Q^2),\\
         \widetilde{\uvec{\mathcal P}}_\text{EF}&=\uvec\beta\times \widetilde{\uvec M}_\text{EF},
    \end{aligned}
\end{equation}
where $\gamma=P^0/\sqrt{P^2}$ and $\uvec\beta=\uvec P/P^0$. We see that the Wigner rotation mixes $(\uvec\sigma_{s's}\times i\uvec\Delta)_z$ and $\delta_{s's}$, but leaves $(\sigma_z)_{s's}$ and $(\uvec\Delta_\perp\cdot\uvec\sigma_{s's})$ unchanged\footnote{This would have been less clear if we had written the transverse magnetization amplitudes as
\begin{equation*}
    \widetilde{\uvec M}_{\perp,\text{EF}} = e\,\gamma \left[ \uvec \sigma_\perp  \cos\theta - \frac{(\uvec e_z \times i\uvec\Delta)_\perp}{|\uvec \Delta_\perp|}\,  \sin\theta- \frac{\uvec\Delta_\perp (\uvec\Delta_\perp \cdot \uvec\sigma_\perp)}{4M\sqrt{1+\tau}(P^0+M)} \right]G_M(Q^2).
\end{equation*}}, as can be checked using Eq.~\eqref{Wignerhalf}.
Beside the Wigner rotation, we recognize the familiar structure of the Lorentz transformation of a rest-frame MDM (or of a pure magnetic field). Moreover, the expression for the polarization amplitudes is reminiscent of the classical expression for an induced EDM $\uvec d=\uvec v\times\uvec\mu$. Comparing with the BF amplitudes~\eqref{spinhalf3DBFPvMv} in the limit $\Delta_z\to 0$, we see that Eq.~\eqref{polmagEF} is fully consistent with the general expectation~\eqref{tensorLT}.

Following the general definition~\eqref{EFdef}, the EF polarization and magnetization distributions are given by
\begin{equation}
    \begin{aligned}\label{spinhalfPvMv}
        \uvec{\mathcal P}_\text{EF}(\uvec b_\perp;P_z)&=\int\frac{\ud^2\Delta_\perp}{(2\pi)^2}\,e^{-i\uvec\Delta_\perp\cdot\uvec b_\perp}\,\frac{1}{2P^0}\,\widetilde {\uvec {\mathcal P}}_\text{EF}(\uvec\Delta_\perp;P_z),\\
        \uvec M_\text{EF}(\uvec b_\perp;P_z)&=\int\frac{\ud^2\Delta_\perp}{(2\pi)^2}\,e^{-i\uvec\Delta_\perp\cdot\uvec b_\perp}\,\frac{1}{2P^0}\,\widetilde {\uvec M}_\text{EF}(\uvec\Delta_\perp;P_z),
    \end{aligned}
\end{equation}
which coincide at $P_z=0$ with the projections of the BF polarization and magnetization distributions (\ref{simplepicture}) onto the transverse plane, respectively. The longitudinal components assume a particularly simple form
\begin{equation}
	\begin{aligned}\label{spinhalfMvL}
		\mathcal P_{z,\text{EF}}(\uvec b_\perp;P_z) 
		&=0,\\
		M_{z,\text{EF}}(\uvec b_\perp;P_z) 
		&= \frac{e}{2M}\,\sigma_z \int\frac{\ud^2\Delta_\perp}{(2\pi)^2}\,e^{-i\uvec\Delta_\perp\cdot\uvec b_\perp}\,\frac{M}{P^0}\, G_M(\uvec\Delta_\perp^2),
	\end{aligned}
\end{equation}
because they do not mix with other components under a Lorentz boost. Since the polarization distribution vanishes in the BF~\eqref{simplepicture}, so does $\mathcal P_{z,\text{EF}}$.

\begin{figure}[tb!]
	\centering
	{\includegraphics[angle=0,scale=0.38]{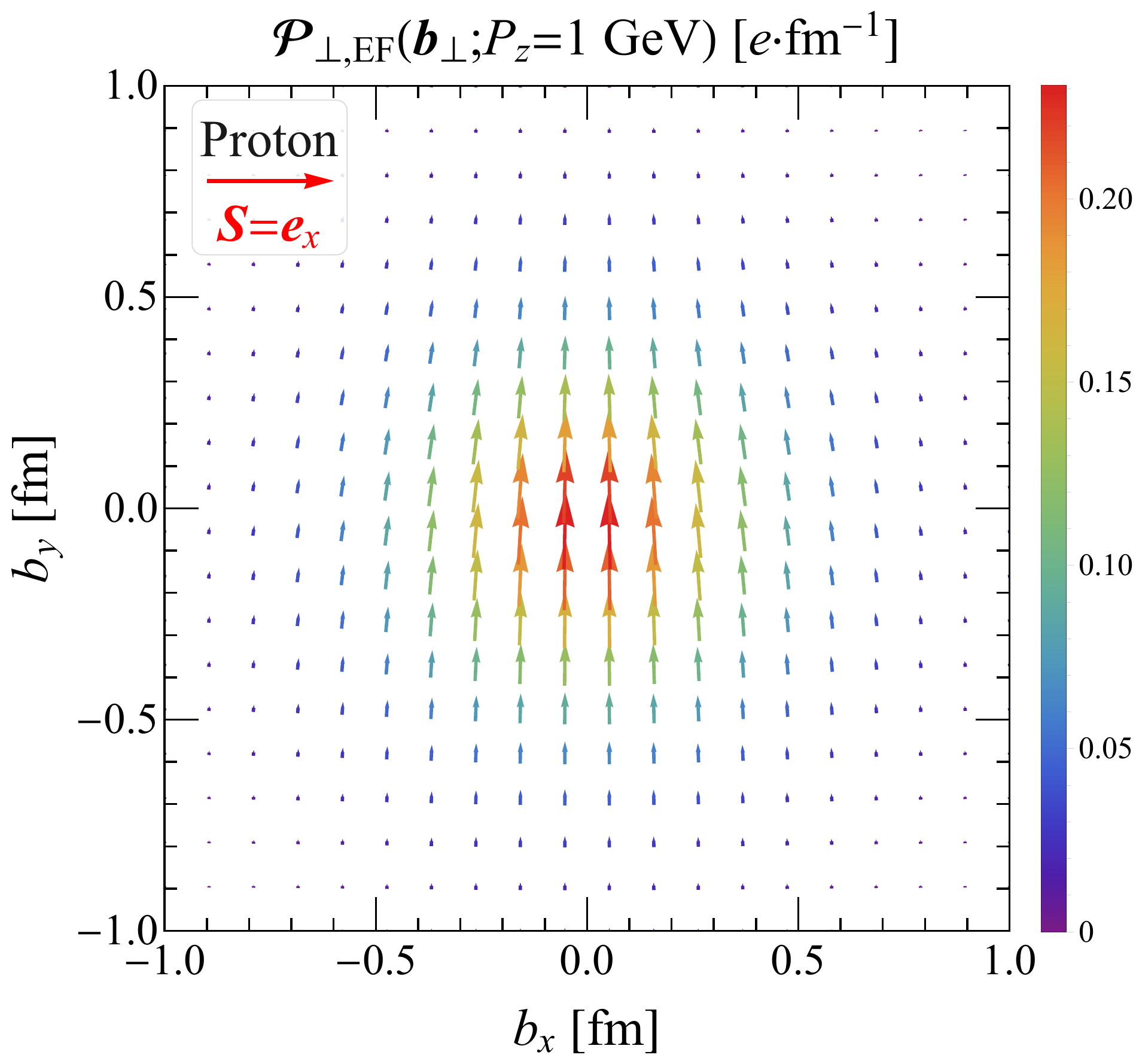}}
	{\includegraphics[angle=0,scale=0.38]{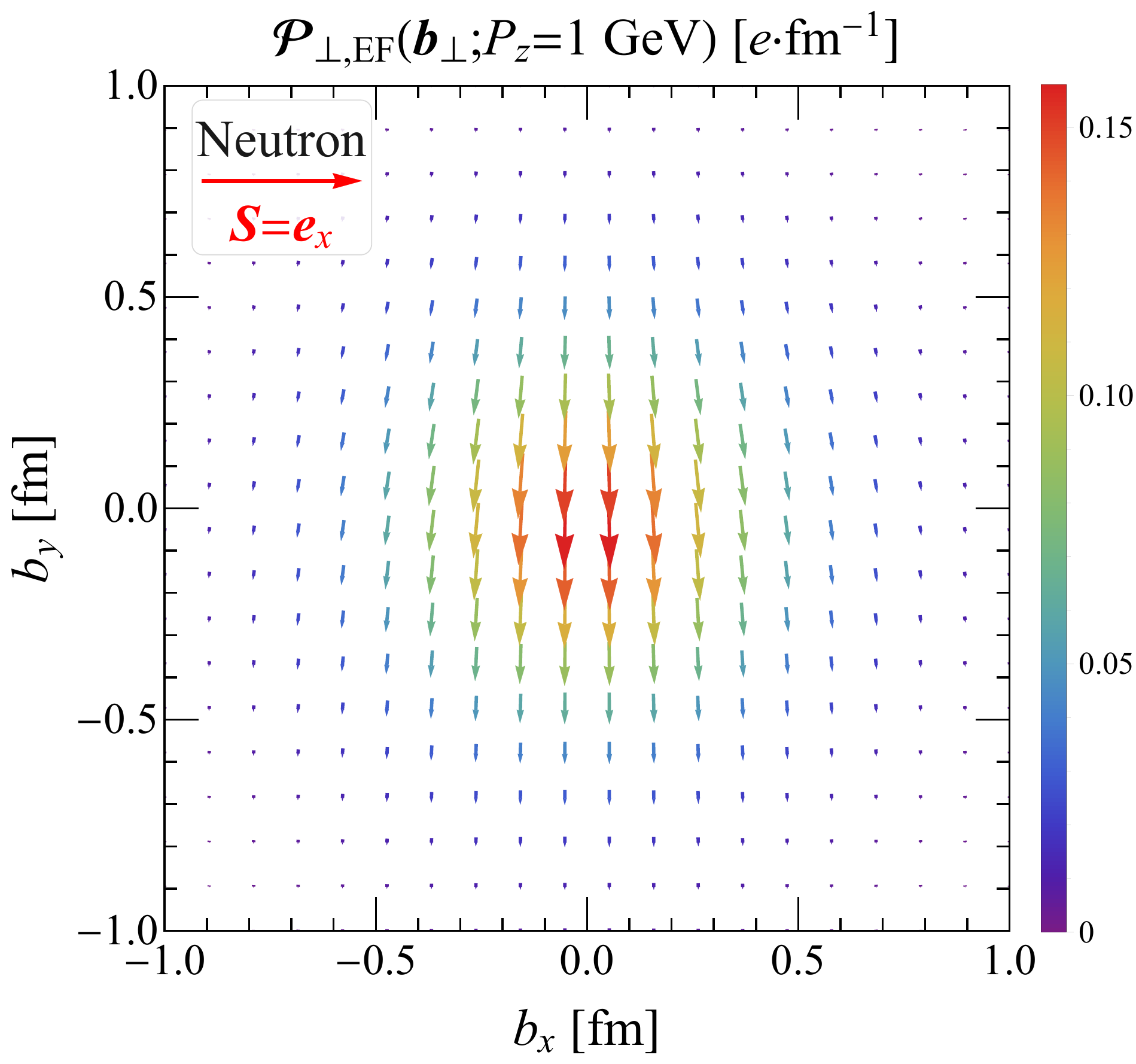}}
	{\includegraphics[angle=0,scale=0.38]{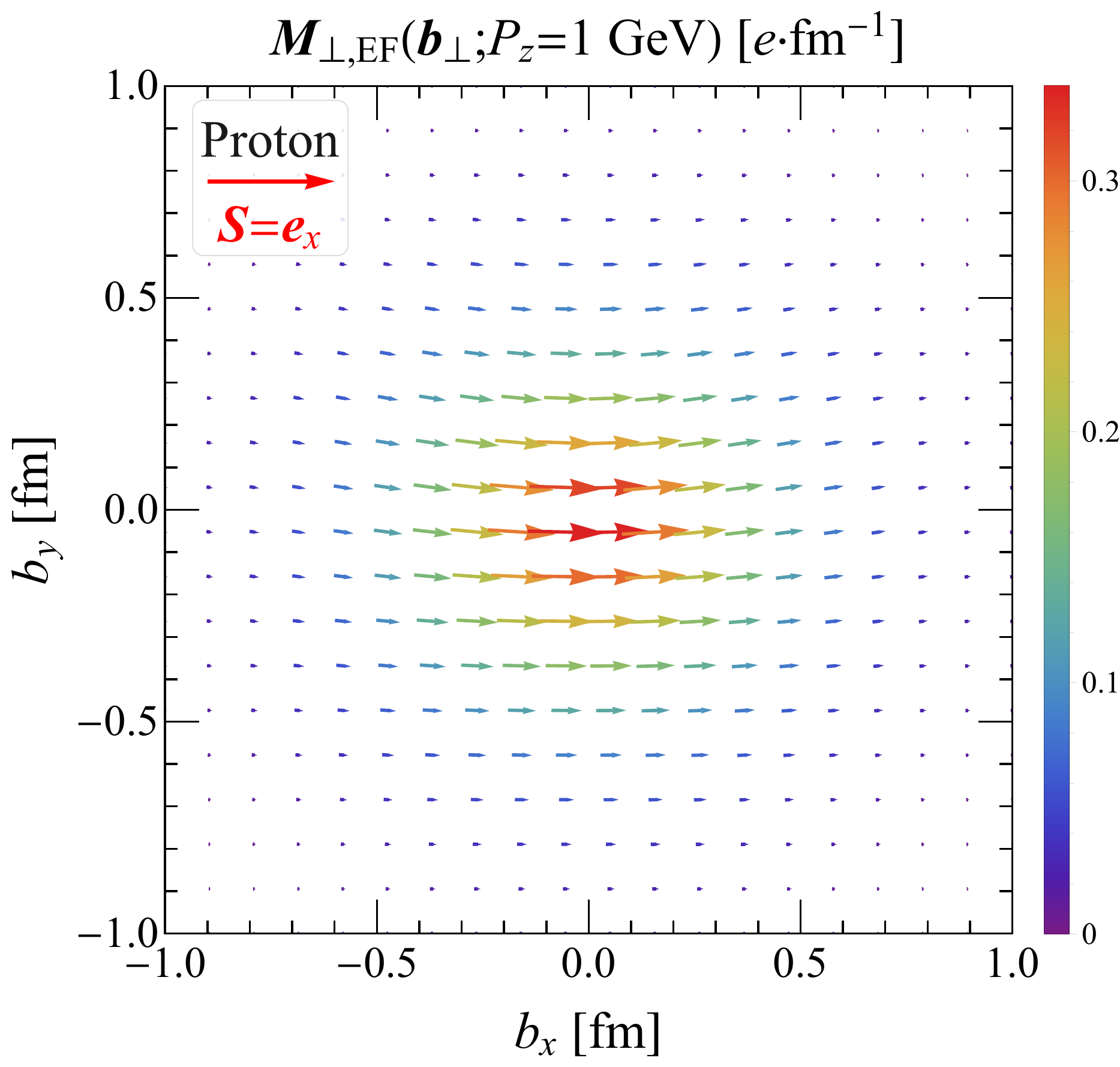}}
	{\includegraphics[angle=0,scale=0.38]{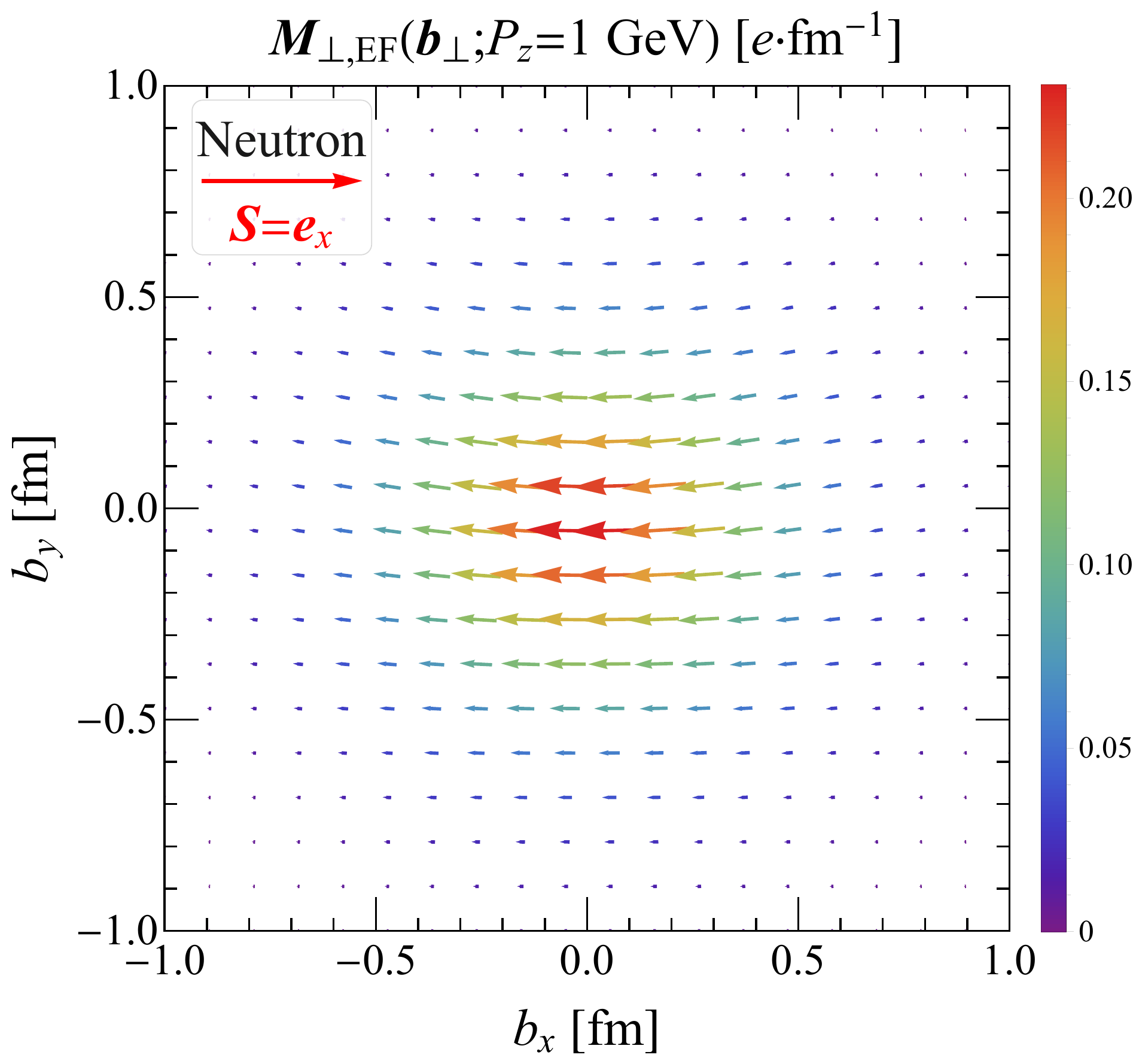}}
	\caption{Elastic frame transverse polarization and magnetization distributions $\uvec{\mathcal P}_{\perp,\text{EF}}(\uvec b_\perp;P_z)$ and $\uvec M_{\perp,\text{EF}}(\uvec b_\perp;P_z)$ in the transverse plane, see Eq.~\eqref{spinhalfPvMv}, inside a proton (left panels) or a neutron (right panels) polarized along the $x$-direction and with momentum $P_z=1~\text{GeV}$. Based on the parametrization for the nucleon electromagnetic form factors given in Ref.~\cite{Bradford:2006yz}. }
	\label{Fig_2DEFMvT}
\end{figure}

\begin{figure}[tb!]
	\centering
	{\includegraphics[angle=0,scale=0.4500]{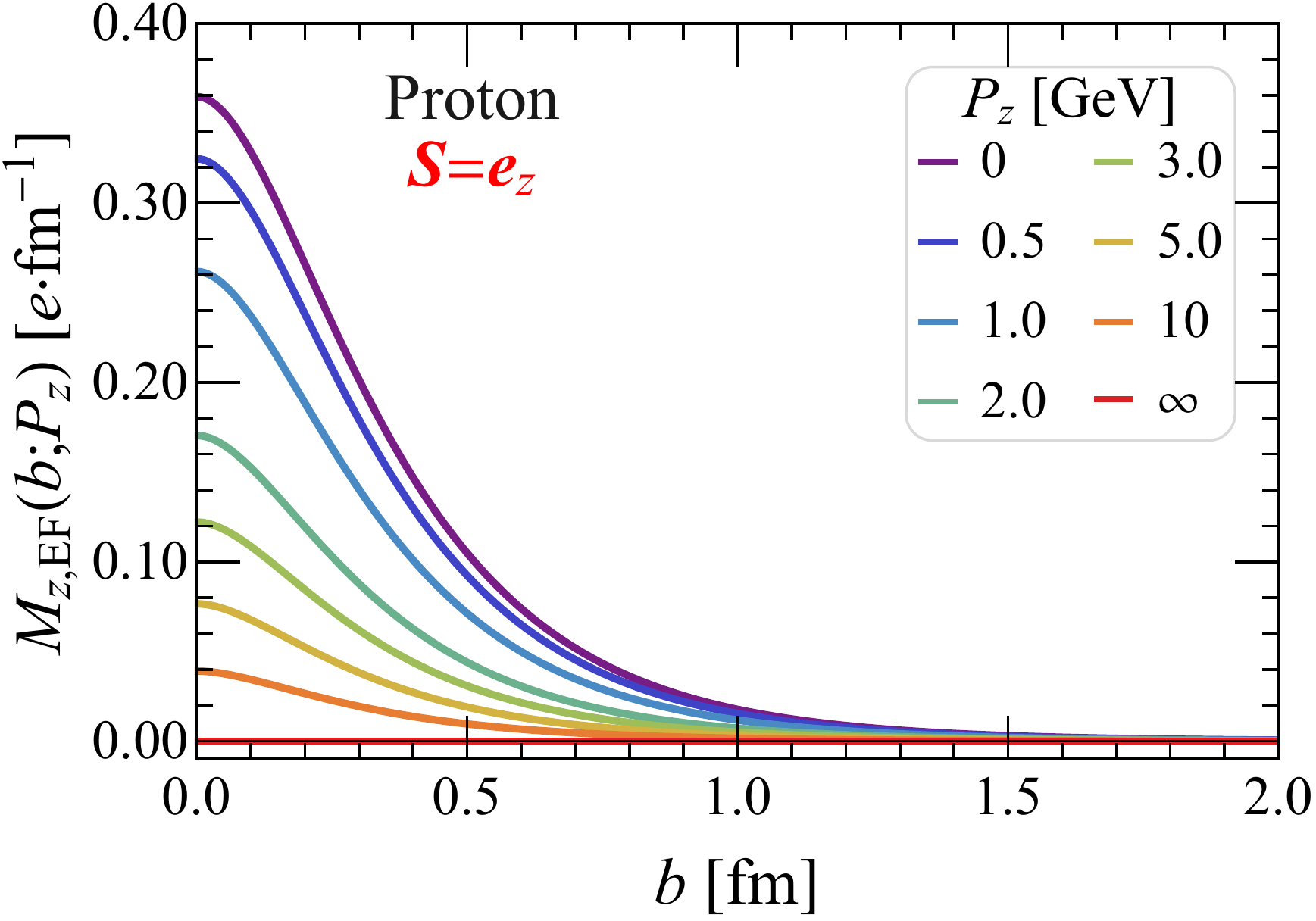}}
	{\includegraphics[angle=0,scale=0.4634]{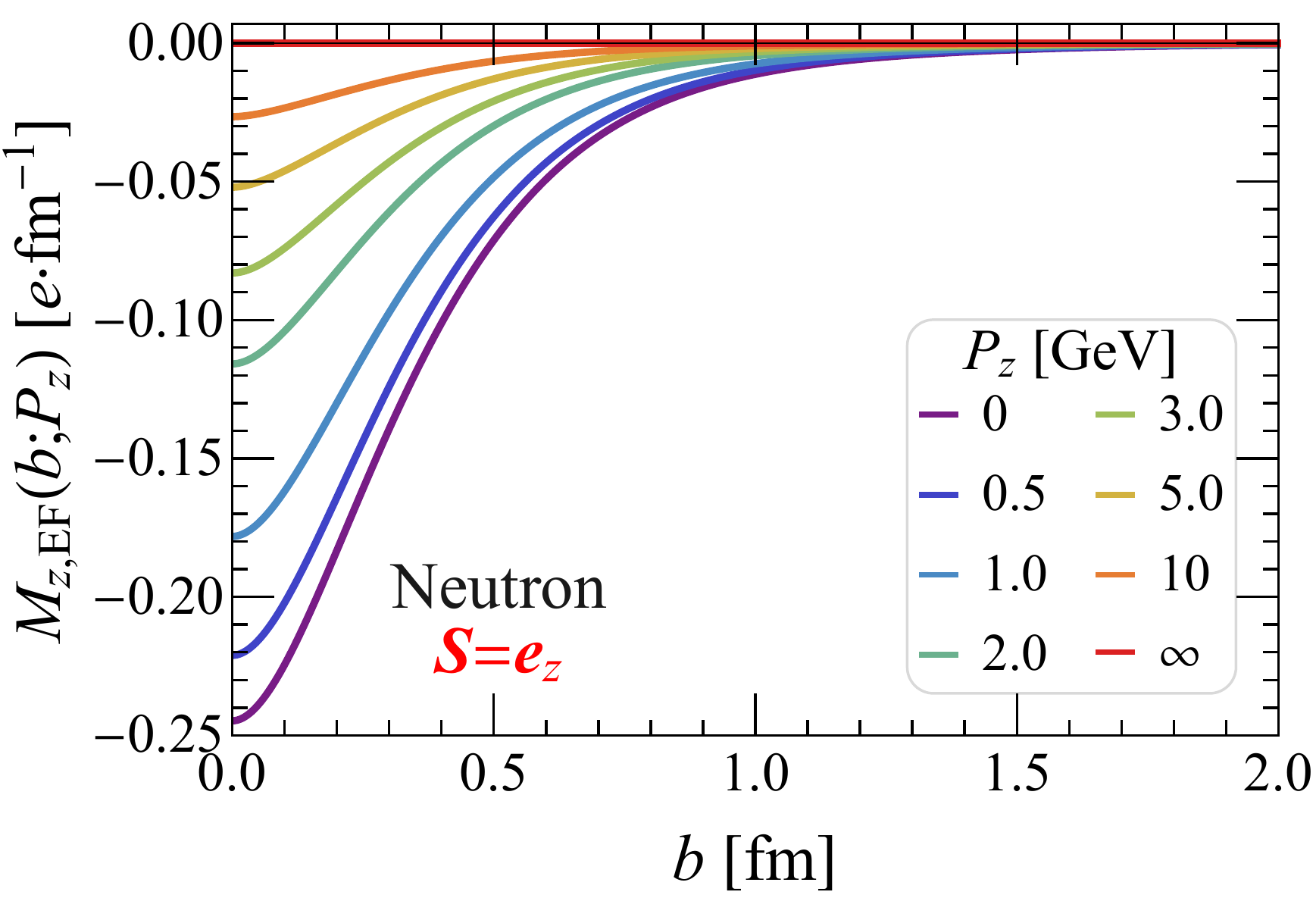}}
	\caption{Elastic frame longitudinal magnetization distribution $M_{z,\text{EF}}(\uvec b_\perp;P_z)$, see Eq.~\eqref{spinhalfMvL}, inside a longitudinally polarized proton (left panel) or neutron (right panel) for different values of average momentum $P_z$. Based on the parametrization of nucleon electromagnetic form factors given in Ref.~\cite{Bradford:2006yz}.}
	\label{Fig_2DEFMz}
\end{figure}

In Fig.~\ref{Fig_2DEFMvT}, we show the EF spatial distributions of transverse polarization and magnetization in the transverse plane from Eq.~\eqref{spinhalfPvMv} for a nucleon polarized along the $x$-axis and moving with average momentum $P_z=1$ GeV. Note that the vector fields point toward slightly different directions at different positions in the transverse plane as a result of the Wigner rotation, see Appendix~\ref{App-Multipole decomposition}. In addition, the momentum dependence of the axially symmetric longitudinal magnetization distribution inside the longitudinally polarized nucleon is sketched in Fig.~\ref{Fig_2DEFMz}. As $P_z$ increases, the magnitude of the longitudinal magnetization decreases as a consequence of the relativistic factor $M/P^0$ in Eq.~\eqref{spinhalfMvL}.

A comparison of these results with the EF distributions of the electromagnetic four-current studied in Ref.~\cite{Chen:2022smg}
\begin{equation}\label{spinhalfEFJ0Jv}
    \begin{aligned}
        J^0_\text{EF}(\uvec b_\perp;P_z)&=e\int\frac{\ud^2\Delta_\perp}{(2\pi)^2}\,e^{-i\uvec\Delta_\perp\cdot\uvec b_\perp}\left[\cos\theta+\frac{(\uvec\sigma\times i\uvec\Delta)_z}{|\uvec \Delta_\perp|}\,\sin\theta\right]\frac{G_E(\uvec\Delta_\perp^2)}{\sqrt{1+\tau}}\\
        &+e\int\frac{\ud^2\Delta_\perp}{(2\pi)^2}\,e^{-i\uvec\Delta_\perp\cdot\uvec b_\perp}\,\frac{P_z}{P^0}\left[-\sin\theta+\frac{(\uvec\sigma\times i\uvec\Delta)_z}{|\uvec \Delta_\perp|}\,\cos\theta\right]\frac{\sqrt{\tau}\,G_M(\uvec\Delta_\perp^2)}{\sqrt{1+\tau}},\\
        J_{z,\text{EF}}(\uvec b_\perp;P_z)&=e\int\frac{\ud^2\Delta_\perp}{(2\pi)^2}\,e^{-i\uvec\Delta_\perp\cdot\uvec b_\perp}\,\frac{P_z}{P^0}\left[\cos\theta+\frac{(\uvec\sigma\times i\uvec\Delta)_z}{|\uvec \Delta_\perp|}\,\sin\theta\right]\frac{G_E(\uvec\Delta_\perp^2)}{\sqrt{1+\tau}}\\
        &+e\int\frac{\ud^2\Delta_\perp}{(2\pi)^2}\,e^{-i\uvec\Delta_\perp\cdot\uvec b_\perp}\left[-\sin\theta+\frac{(\uvec\sigma\times i\uvec\Delta)_z}{|\uvec \Delta_\perp|}\,\cos\theta\right]\frac{\sqrt{\tau}\,G_M(\uvec\Delta_\perp^2)}{\sqrt{1+\tau}},\\
        \uvec J_{\perp,\text{EF}}(\uvec b_\perp;P_z)&=e\,\sigma_z\int\frac{\ud^2\Delta_\perp}{(2\pi)^2}\,e^{-i\uvec\Delta_\perp\cdot\uvec b_\perp}\,\frac{(\uvec e_z\times i\uvec\Delta)_\perp}{2P^0}\,G_M(\uvec\Delta_\perp^2),
    \end{aligned}
\end{equation}
indicates that the EF polarization four-current distributions (given by the $G_M$-dependent terms) can be expressed as\footnote{Note that acting with $\nabla_z$ on any 2D EF distribution gives zero.}
\begin{equation}
\begin{aligned}\label{2DEFrhoPJvM}
    \rho_{P,\text{EF}}(\uvec b_\perp;P_z)&=-\uvec\nabla\cdot\uvec{\mathcal P}_\text{EF}(\uvec b_\perp;P_z),\\
    \uvec J_{P,\text{EF}}(\uvec b_\perp;P_z) &= \uvec \nabla \times \uvec M_\text{EF}(\uvec b_\perp;P_z).
    \end{aligned}
\end{equation}

\begin{figure}[t!]
	\centering
	{\includegraphics[angle=0,scale=0.46]{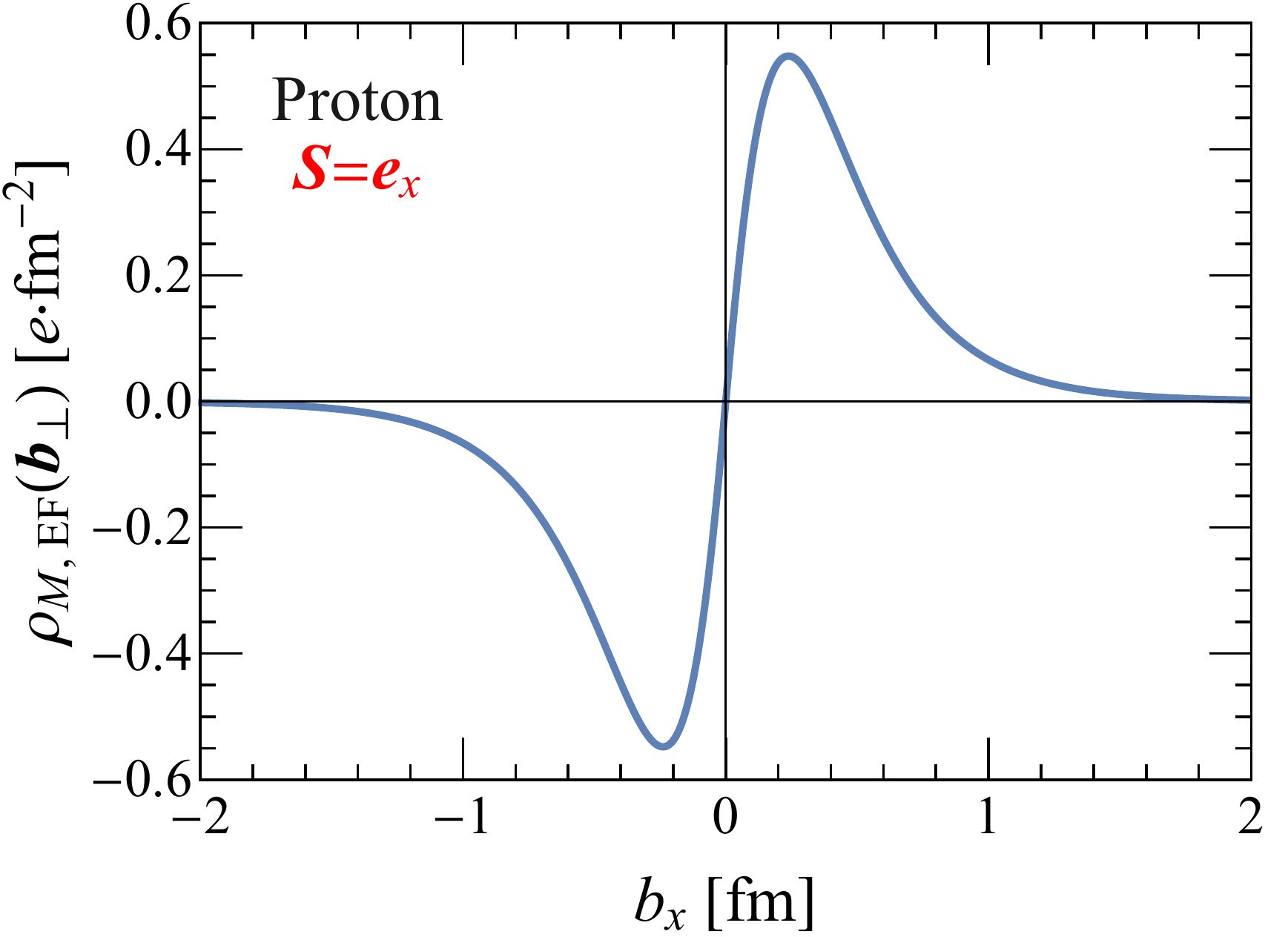}}
	{\includegraphics[angle=0,scale=0.46]{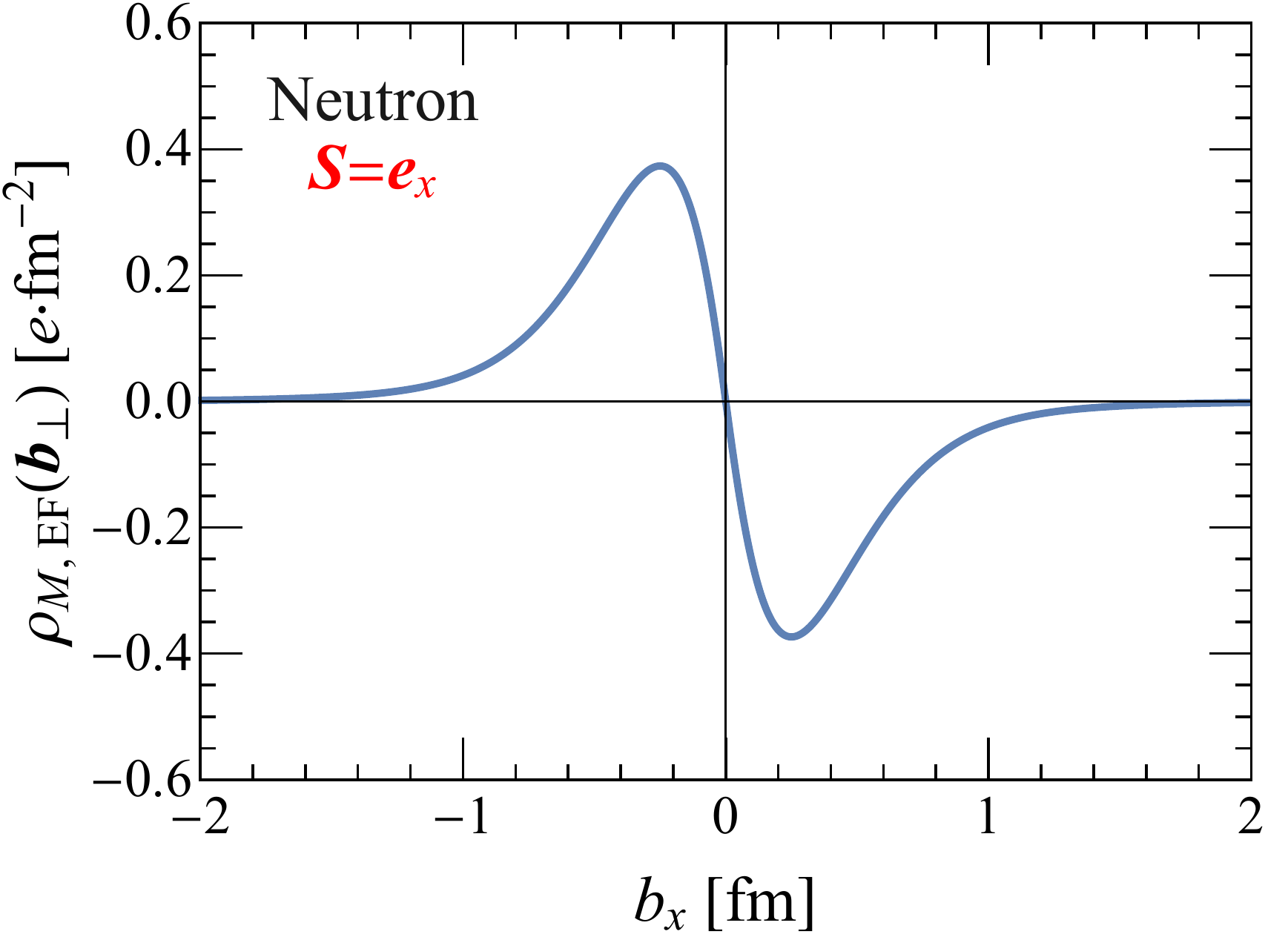}}
	\caption{Elastic frame effective magnetic charge distribution $\rho_{M,\text{EF}}=-\uvec\nabla \cdot \uvec M_{\text{EF}}$, see Eq.~\eqref{2DEFrhoM}, at $b_y=0$ inside a proton (left panel) or a neutron (right panel) polarized along the $x$-direction. Based on the parametrization for the nucleon electromagnetic form factors given in Ref.~\cite{Bradford:2006yz}.}
	\label{Fig_2DEFrhoMag}
\end{figure}
\begin{figure}[t!]
	\centering
	{\includegraphics[angle=0,scale=0.46]{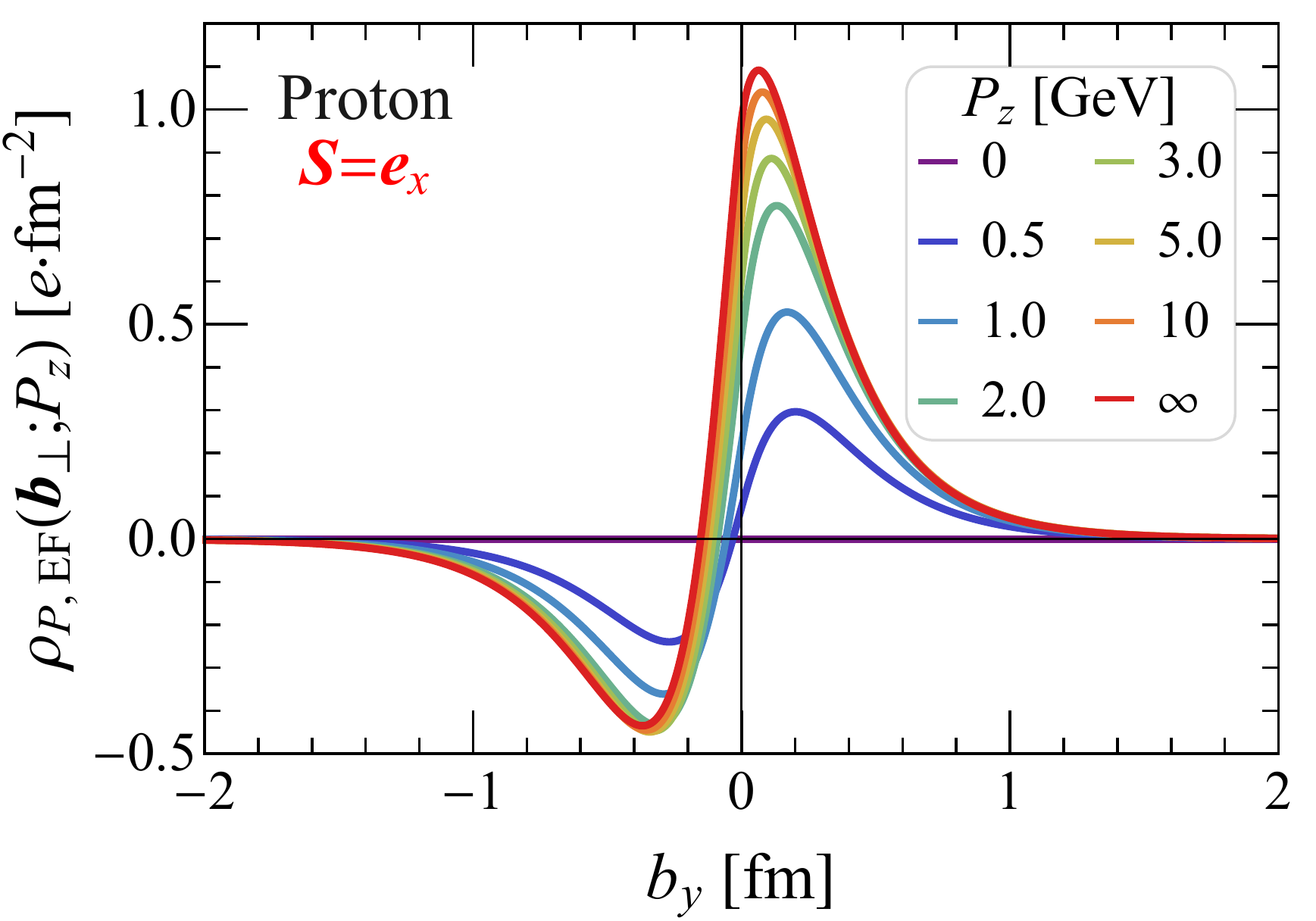}}
	{\includegraphics[angle=0,scale=0.46]{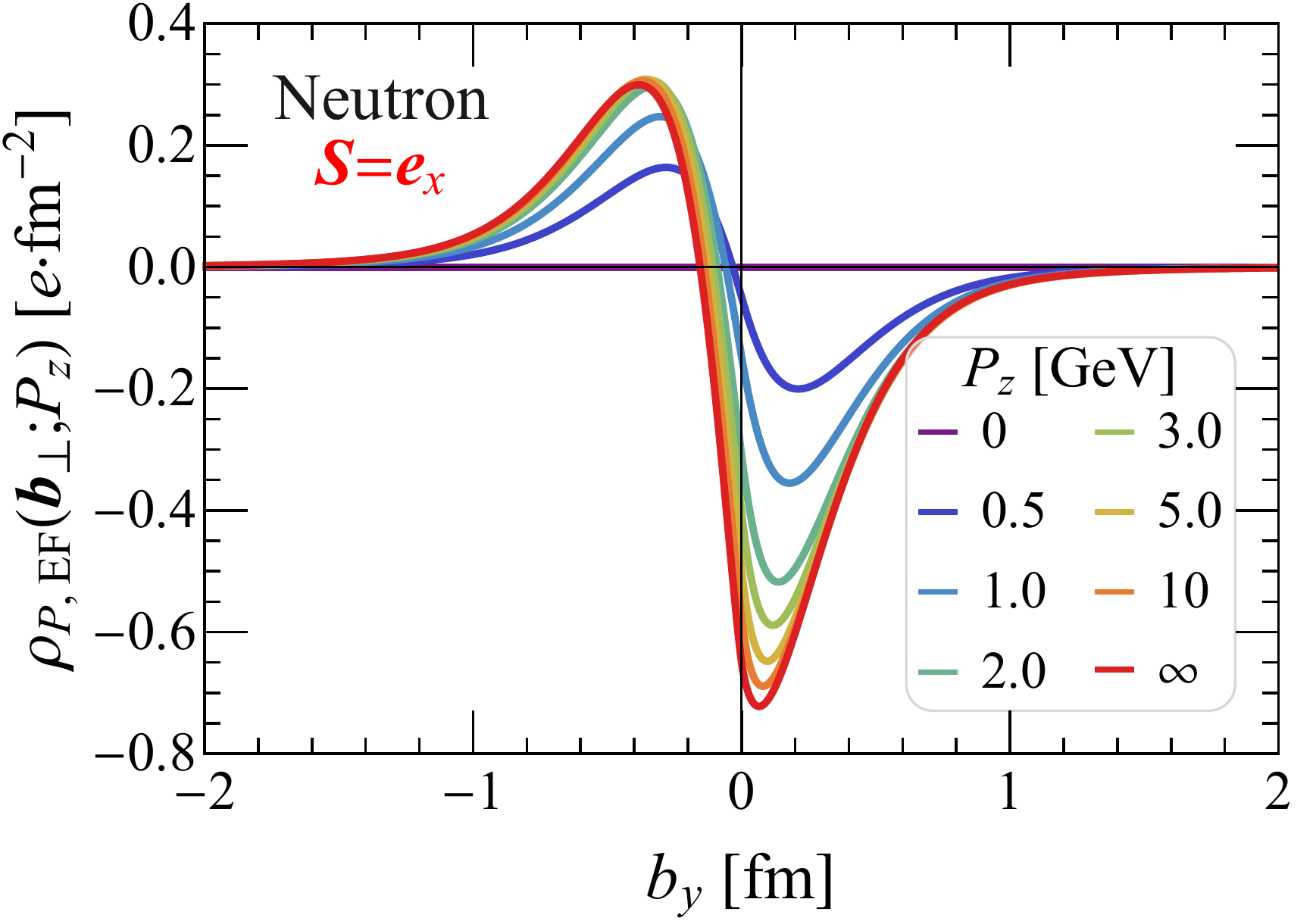}}
	\caption{Elastic frame polarization charge distribution $\rho_{P,\text{EF}}=-\uvec\nabla \cdot \uvec{\mathcal{P}}_{\text{EF}}$, see Eq.~\eqref{2DEFrhoP}, at $b_x=0$ inside a proton (left panel) or a neutron (right panel) polarized along the $x$-direction for different values of the average momentum $P_z$. Based on the parametrization for the nucleon electromagnetic form factors given in Ref.~\cite{Bradford:2006yz}.}
    \label{Fig_2DEFrhoPol}
\end{figure}

By analogy with the 3D case~\eqref{effectiveMagCharge}, we can also define a 2D effective magnetic charge distribution as follows
\begin{equation}\label{2DEFrhoM}
    \rho_{M,\text{EF}}(\uvec b_\perp;P_z)\equiv-\uvec\nabla\cdot\uvec M_\text{EF}(\uvec b_\perp;P_z)=\frac{e}{2M}\int\frac{\ud^2\Delta_\perp}{(2\pi)^2}\,e^{-i\uvec\Delta_\perp\cdot\uvec b_\perp}\,(i\uvec\Delta_\perp\cdot\uvec\sigma_\perp)\,\frac{G_M(\uvec\Delta_\perp^2)}{1+\tau}.
\end{equation}
Interestingly, it does not depend on $P_z$ (remember that $\uvec\Delta_\perp\cdot\uvec\sigma$ is invariant under the Wigner rotation) and coincides with the projection of the BF effective magnetic charge distribution~\eqref{effectiveMagCharge} onto the transverse plane. In Fig.~\ref{Fig_2DEFrhoMag}, we show the $P_z$-independent spatial distribution of the 2D relativistic effective magnetic charge distribution from Eq.~\eqref{2DEFrhoM} inside a transversely polarized nucleon. Likewise, we show in Fig.~\ref{Fig_2DEFrhoPol} the $P_z$-dependent spatial distributions of the 2D relativistic polarization charge distribution from Eq.~\eqref{2DEFrhoPJvM} 
\begin{equation}\label{2DEFrhoP}
    \rho_{P,\text{EF}}(\uvec b_\perp;P_z) = e\int\frac{\ud^2\Delta_\perp}{(2\pi)^2}\,e^{-i\uvec\Delta_\perp\cdot\uvec b_\perp}\,\frac{P_z}{P^0}\left[-\sin\theta+\frac{(\uvec\sigma\times i\uvec\Delta)_z}{|\uvec \Delta_\perp|}\,\cos\theta\right]\frac{\sqrt{\tau}\,G_M(\uvec\Delta_\perp^2)}{\sqrt{1+\tau}}
\end{equation}
inside a transversely polarized nucleon. 

\subsection{Elastic frame electric and magnetic dipole moments}
\label{sec:Elastic frame electric and magnetic dipole moments}

The EF MDM is obtained by integrating the EF magnetization distribution over the transverse plane,
\begin{equation}\label{MDM-EF}
    \uvec \mu_{\text{EF}}(P_z)=\int\ud^2b_\perp\,\uvec M_\text{EF}(\uvec b_\perp; P_z) =\frac{1}{2E_P}\,\widetilde{\uvec M}_\text{EF}(\uvec 0_\perp;P_z),
\end{equation}
where we remind that $E_P=\sqrt{M^2+\uvec P^2}$. By analogy with the 3D BF expressions, we can alternatively define the longitudinal EF MDM as
\begin{equation}\label{MDM-EF-Muz}
    \mu_{z,\text{EF}}(P_z)=\int\ud^2b_\perp\,\frac{\left[\uvec b_\perp\times\uvec J_\text{EF}(\uvec b_\perp;P_z)\right]_z}{2}=\sigma_z\,\frac{M}{E_P}\,G_M(0)\,\frac{e}{2M},
\end{equation}
which agrees with the longitudinal component in Eq.~\eqref{MDM-EF}. A similar expression for the transverse MDM would require a 3D definition of the EF current, which is beyond the scope of the present work. We can however use the 2D effective magnetic charge distribution~\eqref{2DEFrhoM} and alternatively define the transverse EF MDM as
\begin{equation}
    \uvec\mu_{\perp,\text{EF}}(P_z)=\int\ud^2b_\perp\,\uvec b_\perp\,\rho_{M,\text{EF}}(\uvec b_\perp;P_z)=\uvec\sigma_\perp\,G_M(0)\,\frac{e}{2M},
\end{equation}
which agrees with the transverse components in Eq.~\eqref{MDM-EF}. A similar expression for the longitudinal MDM would require a 3D definition of the EF effective magnetic charge distribution, which is also beyond the scope of the present work.

From the familiar Lorentz transformation of the magnetic field, one might naively think that a global Lorentz factor $\gamma_P=E_P/M$ is missing in the expressions for $\uvec\mu_\text{EF}(P_z)$. It is in fact compensated by the Lorentz contraction factor $1/\gamma_P$ associated with the volume element. We expect that similar expressions should hold for spin-$j$ targets, namely
\begin{equation}
	\begin{aligned}
		\mu^{(j)}_{z,\text{EF}}(P_z)&= \Sigma_z\,\frac{M}{E_P}\, G_{M1}(0)\,\frac{e}{2M},\\
		\uvec \mu^{(j)}_{\perp,\text{EF}}(P_z)&= \uvec\Sigma_\perp\,G_{M1}(0)\,\frac{e}{2M},
	\end{aligned}
\end{equation}
where $G_{M1}(Q^2)$ is the BF magnetic dipole FF for a spin-$j$ system~\cite{Lorce:2009bs}, and $\uvec\Sigma_{s's}$ are the generalization of the Pauli matrices to higher spin\footnote{The spin matrices for a spin-$j$ target are generically given by $\uvec S_{s's}=j\,\uvec\Sigma_{s's}$.}.
\newline

\begin{figure}[tb!]
	\centering
	{\includegraphics[angle=0,scale=0.450]{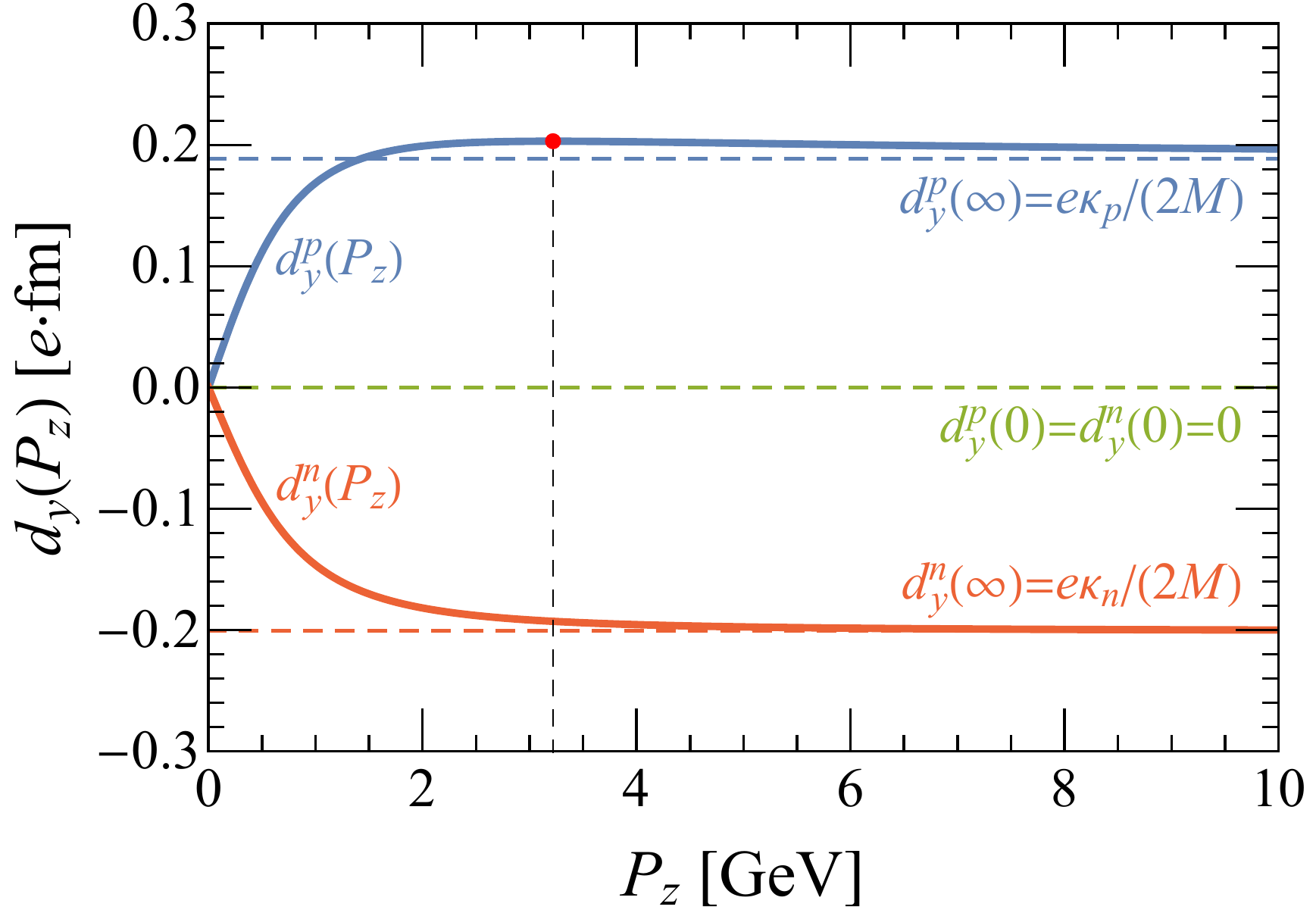}}
	{\includegraphics[angle=0,scale=0.448]{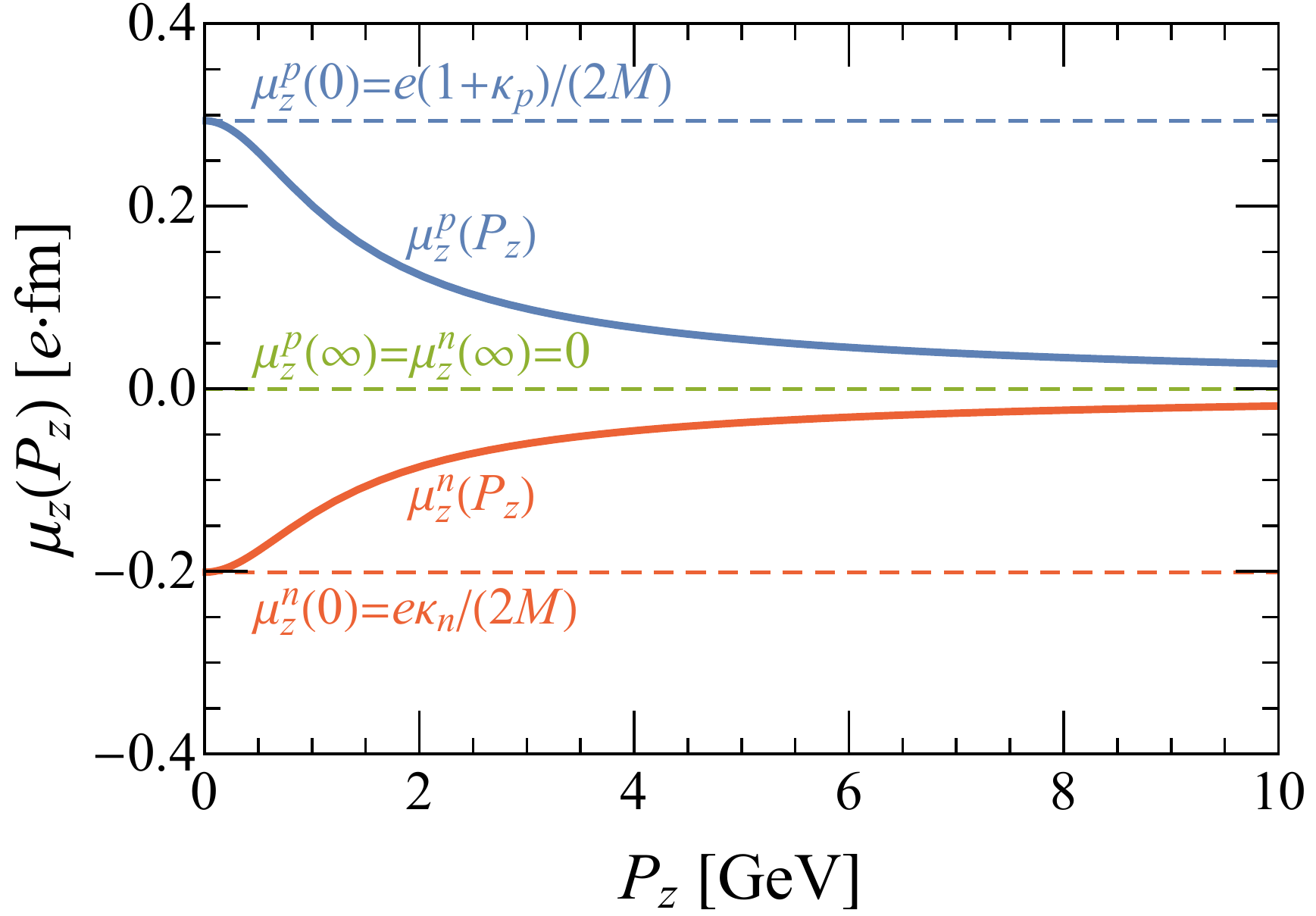}}
	\caption{Transverse electric dipole moment $d_y(P_z)$, see Eq.~\eqref{EDM-EF-dy}, inside a nucleon polarized along the $x$-axis (left panel) and longitudinal magnetic dipole moment $\mu_z(P_z)$, see Eq.~\eqref{MDM-EF-Muz}, inside a longitudinally polarized nucleon (right panel), as functions of the nucleon average momentum $P_z$. $\kappa_{p,n}\equiv G^{p,n}_M(0)-G^{p,n}_E(0)$ stand for the proton and neutron anomalous magnetic dipole moments. At $P_z \approx 3.22~\text{GeV}$, the proton transverse electric dipole moment reaches its maximum value $d^p_{y,\text{max}}\approx 0.203~e\!\cdot \!\text{fm}$.}
	\label{Fig_2DEFEDMMDM}
\end{figure}

Let us now discuss the (transverse) EF EDM. It is defined as
\begin{equation}\label{EFEDM}
    \uvec d_{\perp,\text{EF}}(P_z)=\int\ud^2b_\perp\,\uvec b_\perp\,J^0_\text{EF}(\uvec b_\perp;P_z).
\end{equation}
For a spin-$\frac{1}{2}$ target, we find that it is explicitly given by
\begin{equation}\label{EDM-EF-dy}
    \uvec d_{\perp,\text{EF}}(P_z)=(\uvec e_z\times\uvec\sigma)_\perp\,\frac{P_z}{E_P}\left[G_M(0)-\frac{E_P}{E_P+M}\,G_E(0)\right]\frac{e}{2M}.
\end{equation}
This analytic expression agrees with the numerical results for the nucleon obtained in Ref.~\cite{Kim:2021kum}. The first contribution corresponds to the longitudinal boost of a rest-frame transverse MDM and has the expected form $\frac{\uvec P}{E_P}\times\uvec\mu_\text{EF}(0)$. The second contribution comes from the Wigner rotation and can be understood in terms of a sideways shift of the center of spin\footnote{The center of spin is given by the expectation value of the Newton-Wigner operator~\cite{Newton:1949cq}. The angular momentum referring to this point coincides with spin in an arbitrary frame.}, defining the origin of our coordinate system, with respect to the relativistic center of mass in a moving frame~\cite{Lorce:2018zpf,Lorce:2021gxs}. Its magnitude is precisely the relative distance between these two points, see Appendix~\ref{App-Relativistic centers of the nucleon}, multiplied by the total charge of the system as if the latter were concentrated at the relativistic center of mass. Since the sideways shift is proportional to the spin value, we expect the induced EDM for a spin-$j$ target to read
\begin{equation}\label{spinjEDM}
    \uvec d^{(j)}_{\perp,\text{EF}}(P_z)=(\uvec e_z\times\uvec\Sigma)_\perp\,\frac{P_z}{E_P}\left[G_{M1}(0)-\frac{E_P}{E_P+M}\,2j\,G_{E0}(0)\right]\frac{e}{2M},
\end{equation}
where $G_{E0}(Q^2)$ is the BF electric monopole FF for a spin-$j$ system. This generic expression agrees with the result found for spin-$1$ targets~\cite{Lorce:2022jyi}. In Fig.~\ref{Fig_2DEFEDMMDM}, we show the momentum dependence of the transverse EDM in a transversely polarized nucleon and of the longitudinal MDM in a longitudinally polarized nucleon. The maximum transverse EDM for a proton is reached for $P_z=\frac{M}{2}\sqrt{(\eta+\sqrt{\eta^2+4\eta})^2-4}$ with $\eta= G^p_M(0)/G^p_E(0)=1+\kappa_p \approx 2.793$.

\section{Light-front distributions}
\label{sec:Light-front distributions}

For completeness, we finally study the polarization-magnetization distributions within the LF formalism, where LF components are defined as $x^\mu=[x^+,x^-,\uvec x_\perp]$ with $x^\pm\equiv(x^0\pm x^3)/\sqrt{2}$. As a result, scalar products read $p\cdot x=p^+x^-+p^-x^+-\uvec p_\perp\cdot\uvec x_\perp$ and the constrained momentum component is then given by $p^-=(\uvec p^2_\perp+M^2)/(2p^+)$. 

It is possible to define $x^+$-independent LF distributions~\cite{Burkardt:2002hr,Miller:2010nz,Lorce:2017wkb} by considering the so-called symmetric LF frame specified by the conditions\footnote{One can relax the condition $\uvec P_\perp=\uvec 0_\perp$ provided that LF distributions are restricted to $x^+=0$, as stressed recently in Refs.~\cite{Freese:2021czn,Freese:2022fat}. Note however that LF boosts are kinematical operations and so the description at $\uvec P_\perp\neq\uvec 0_\perp$ can be related in a straightforward way to the description at $\uvec P_\perp=\uvec 0_\perp$, just like in the non-relativistic theory.} $\uvec P_\perp=\uvec 0_\perp$ and $\Delta^+=0$, which ensure that the LF energy transfer $\Delta^-=(\uvec P_\perp\cdot\uvec\Delta_\perp-P^-\Delta^+)/P^+$ vanishes. Similarly to Eq.~\eqref{EFdef}, the LF distributions are defined as
\begin{equation}
    O_\text{LF}(\uvec b_\perp;P^+)\equiv\int\frac{\ud^2\Delta_\perp}{(2\pi)^2}\,e^{-i\uvec\Delta_\perp\cdot\uvec b_\perp}\,\frac{_\text{LF}\langle p',\lambda'|\hat O(0)|p,\lambda\rangle_\text{LF}}{2P^+}\bigg|_{\Delta^+=|\uvec P_\perp|=0},
\end{equation}
where the LF helicity states are related to the canonical spin states via the Melosh rotation $|p,\lambda\rangle_\text{LF}=\sum_s |p,s\rangle\,\mathcal M_{s\lambda}$ with
\begin{equation}\label{Meloshrotation}
    \mathcal M_{s\lambda}=\frac{(\sqrt{2}p^++M)\,\delta_{s\lambda}-i(\uvec p_\perp \times\uvec\sigma_{s\lambda})_z}{\sqrt{2\sqrt{2}p^+(p^0+M)}}
\end{equation}
in the case of a spin-$\frac{1}{2}$ system~\cite{Melosh:1974cu}. As already discussed in Sec.~\ref{sec:Quantum phase-space formalism}, a key feature of the LF formalism is that the symmetry subgroup associated with the transverse LF plane is Galilean. As a result, LF distributions can in some cases be interpreted as probabilistic densities. The pictures provided by these LF densities cannot however be considered as realistic representations of the system at rest, even when $P^-=P^+=M\sqrt{1+\tau}/\sqrt{2}$, because they are distorted by relativistic artefacts caused by the Melosh rotation~\cite{Lorce:2020onh,Lorce:2022jyi,Chen:2022smg}.

\subsection{Light-front polarization and magnetization}
\label{sec:Light-front polarization and magnetization}

We have seen in Eq.~\eqref{Pcomp} that polarization and magnetization correspond to the following components of the antisymmetric polarization-magnetization tensor $P^{\mu\nu}$
\begin{equation}
    \mathcal P^\mu=P^{0\mu},\qquad M^\mu=-\frac{1}{2}\,\epsilon^{\mu\alpha\beta 0}P_{\alpha\beta}.
\end{equation}
Note that despite what the notation suggests, $\mathcal P^\mu$ and $M^\mu$ are not Lorentz four-vectors. In particular, we have by construction $\mathcal P^0=M^0=0$ in any frame. In the LF formalism, it is therefore natural to define LF polarization and magnetization components as follows
\begin{equation}\label{LF-PolMagDef}
        \mathcal P^\mu_\text{LF}=P^{+\mu},\qquad M^\mu_\text{LF}=-\frac{1}{2}\,\epsilon^{\mu\alpha\beta -}P_{\alpha\beta}.
\end{equation}
More explicitly, we have
\begin{equation}\label{spinhalf-LFPol0}
    \mathcal P^+_\text{LF}=0,\qquad \mathcal P^i_{\perp,\text{LF}}=P^{+i}=\frac{\mathcal P^i_\perp-\epsilon^{ij}_\perp M^j_\perp}{\sqrt{2}},\qquad \mathcal P^-_\text{LF}=P^{+-}=-\mathcal P_z,
\end{equation}
and
\begin{equation}\label{spinhalf-LFMag0}
    M^+_\text{LF}=-\frac{1}{2}\,\epsilon^{ij}_\perp P^{ij}=M_z,\qquad M^i_{\perp,\text{LF}}=-\epsilon^{ij}_\perp P^{-j}=\frac{M^i_\perp-\epsilon^{ij}_\perp \mathcal P^j_\perp}{\sqrt{2}},\qquad M^-_\text{LF}=0,
\end{equation}
which is similar to the decomposition of the generalized angular momentum tensor into LF boost and angular momentum operators\footnote{In the literature, the LF angular momentum operators are unfortunately often defined \emph{without} the transverse Levi-Civita symbol, missing therefore the axial-vector nature of angular momentum.}~\cite{Kogut:1969xa,Brodsky:1997de}.

For the LF polarization and magnetization amplitudes, the evaluation of Eq.~\eqref{polmagT} in the symmetric LF frame with LF helicity states gives
\begin{equation}
	\begin{aligned}\label{LFmagpolampl}
		\widetilde M ^+_\text{LF}&=e\,(\sigma_z)_{\lambda'\lambda}\,G_M(Q^2),\\
		\widetilde{\uvec M}_{\perp,\text{LF}}&=e\,\frac{P^-}{M(1+\tau)}\left[(\uvec\sigma_\perp)_{\lambda'\lambda}+\delta_{\lambda'\lambda}\,\frac{(\uvec e_z\times i\uvec\Delta)_\perp}{2M}\right]G_M(Q^2),\\
		\widetilde{\mathcal P}^-_\text{LF}&=0,\\
		\widetilde{\mathcal P}^i_{\perp,\text{LF}}&=-\frac{P^+}{P^-}\,\epsilon^{ij}_\perp \widetilde{M}^j_{\perp,\text{LF}}.
	\end{aligned}
\end{equation}

\begin{figure}[tb!]
	\centering
	{\includegraphics[angle=0,scale=0.38]{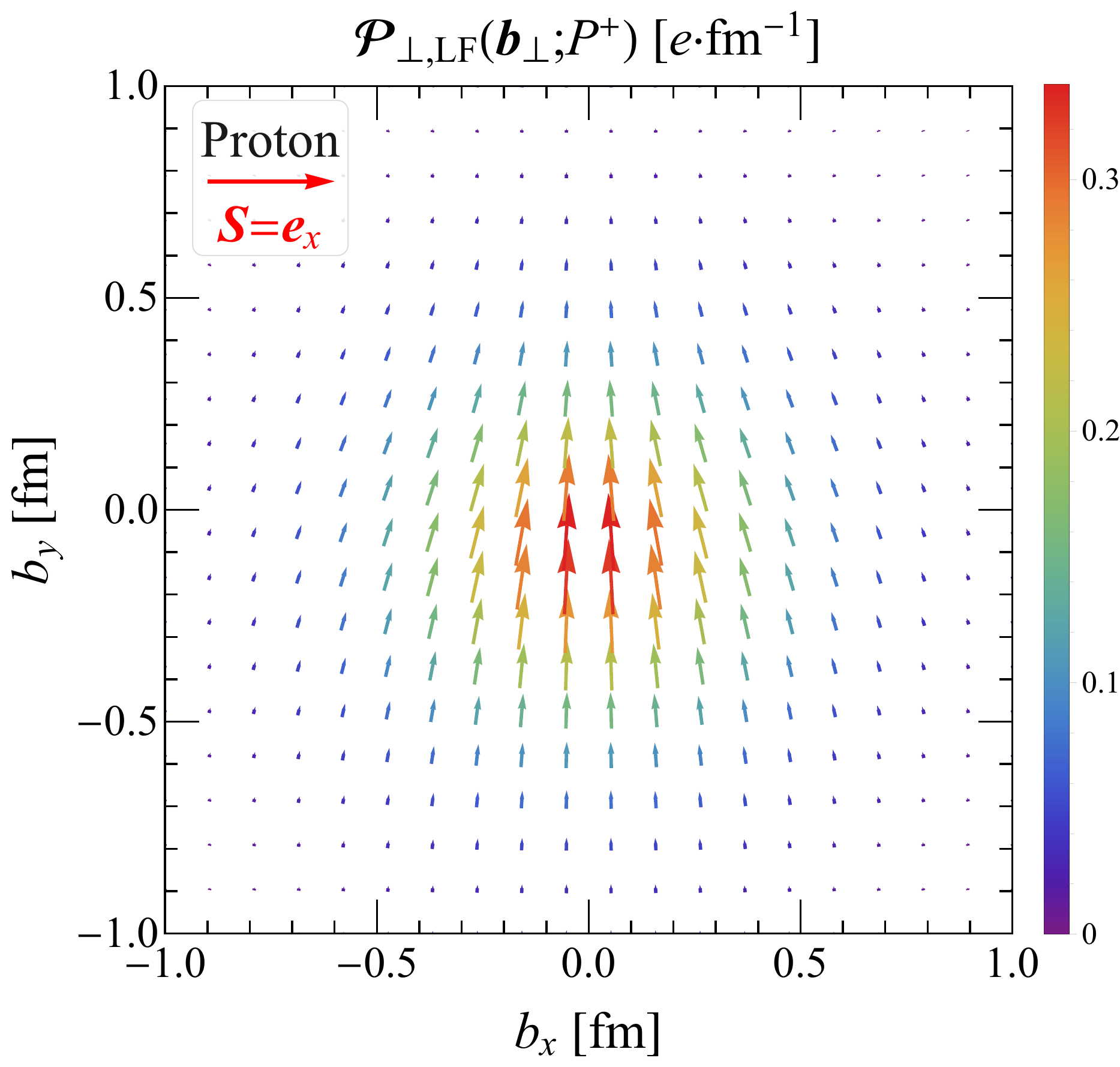}}
	{\ \includegraphics[angle=0,scale=0.38]{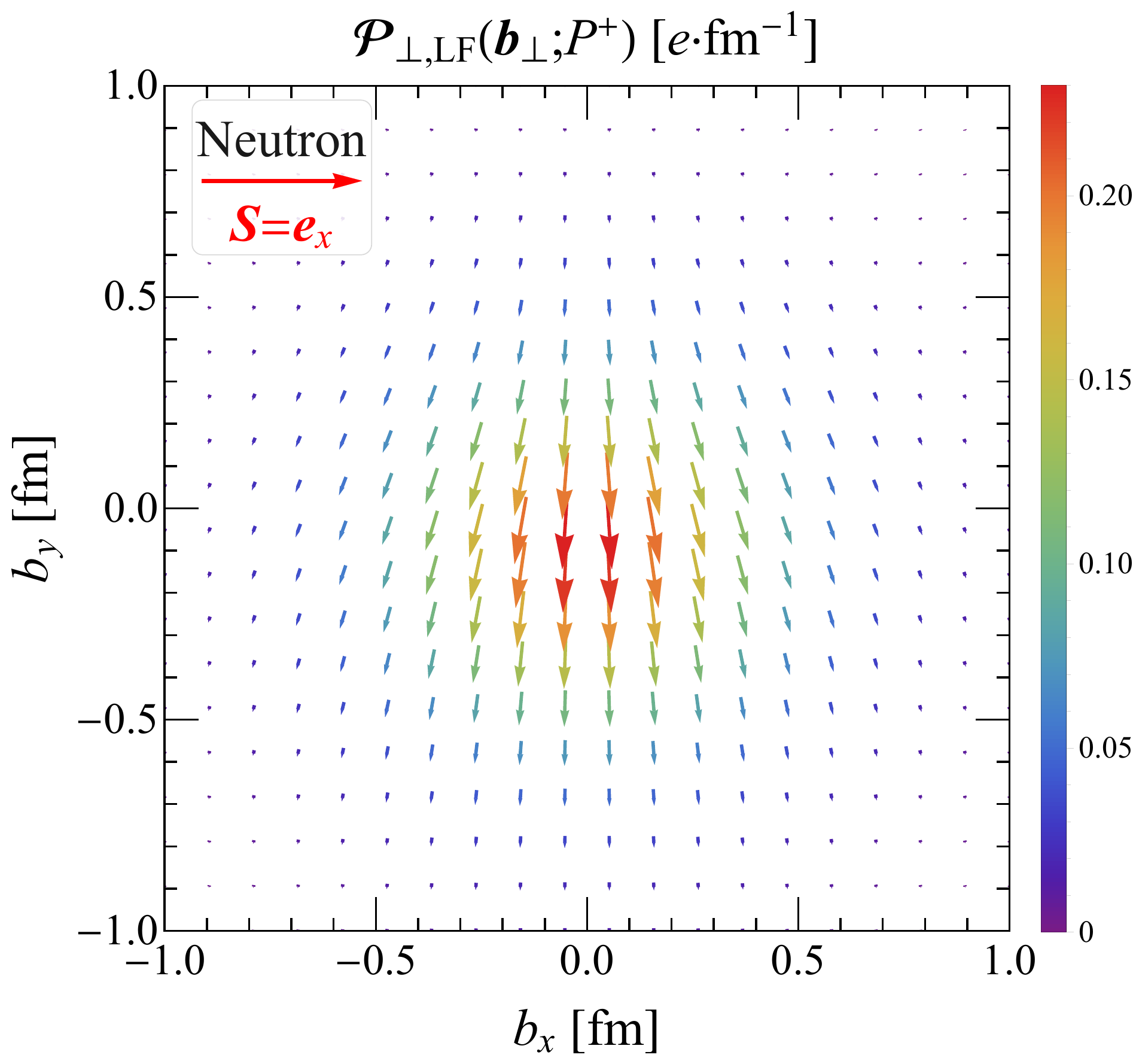}}
	{\includegraphics[angle=0,scale=0.38]{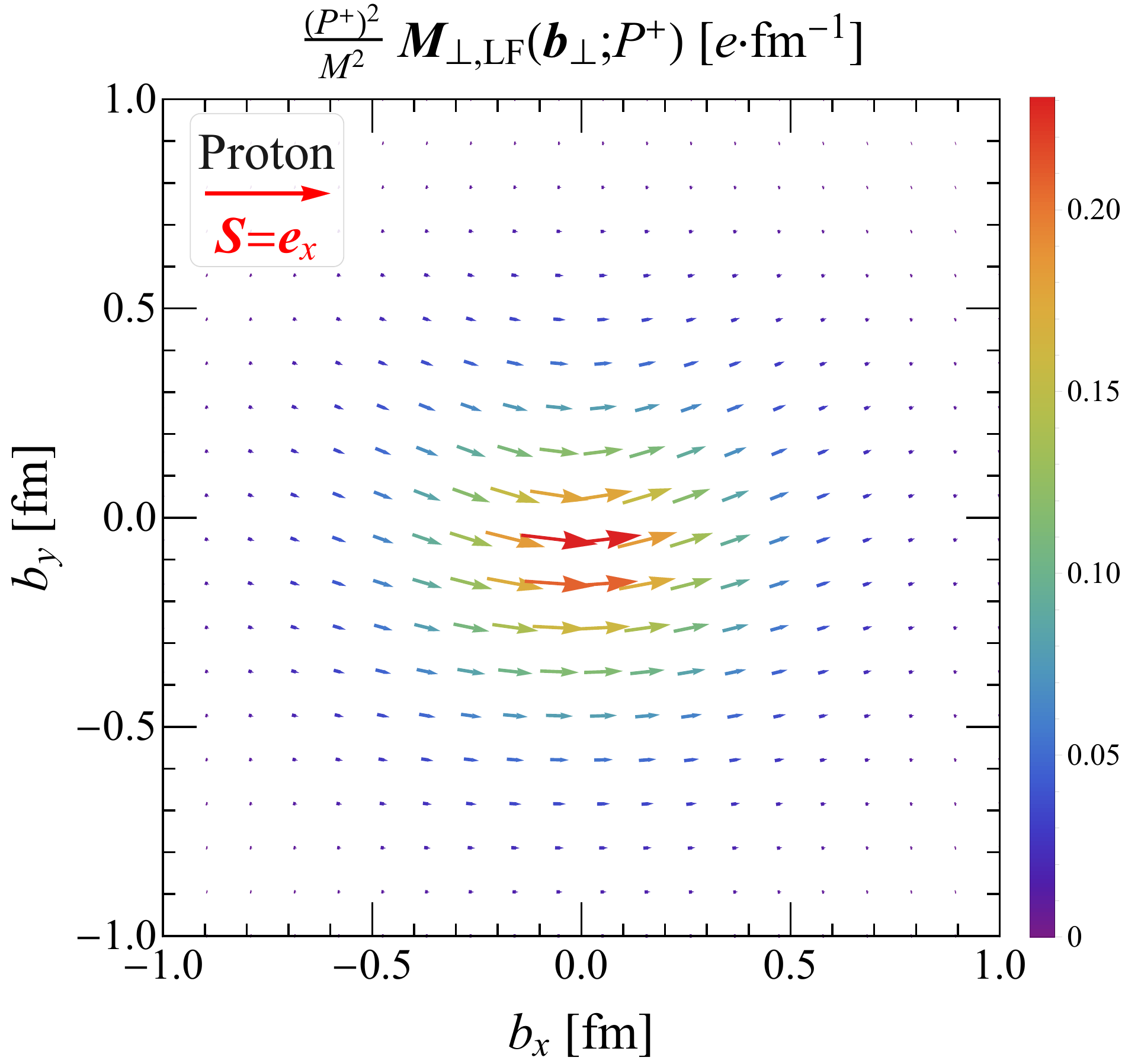}}
	{\includegraphics[angle=0,scale=0.38]{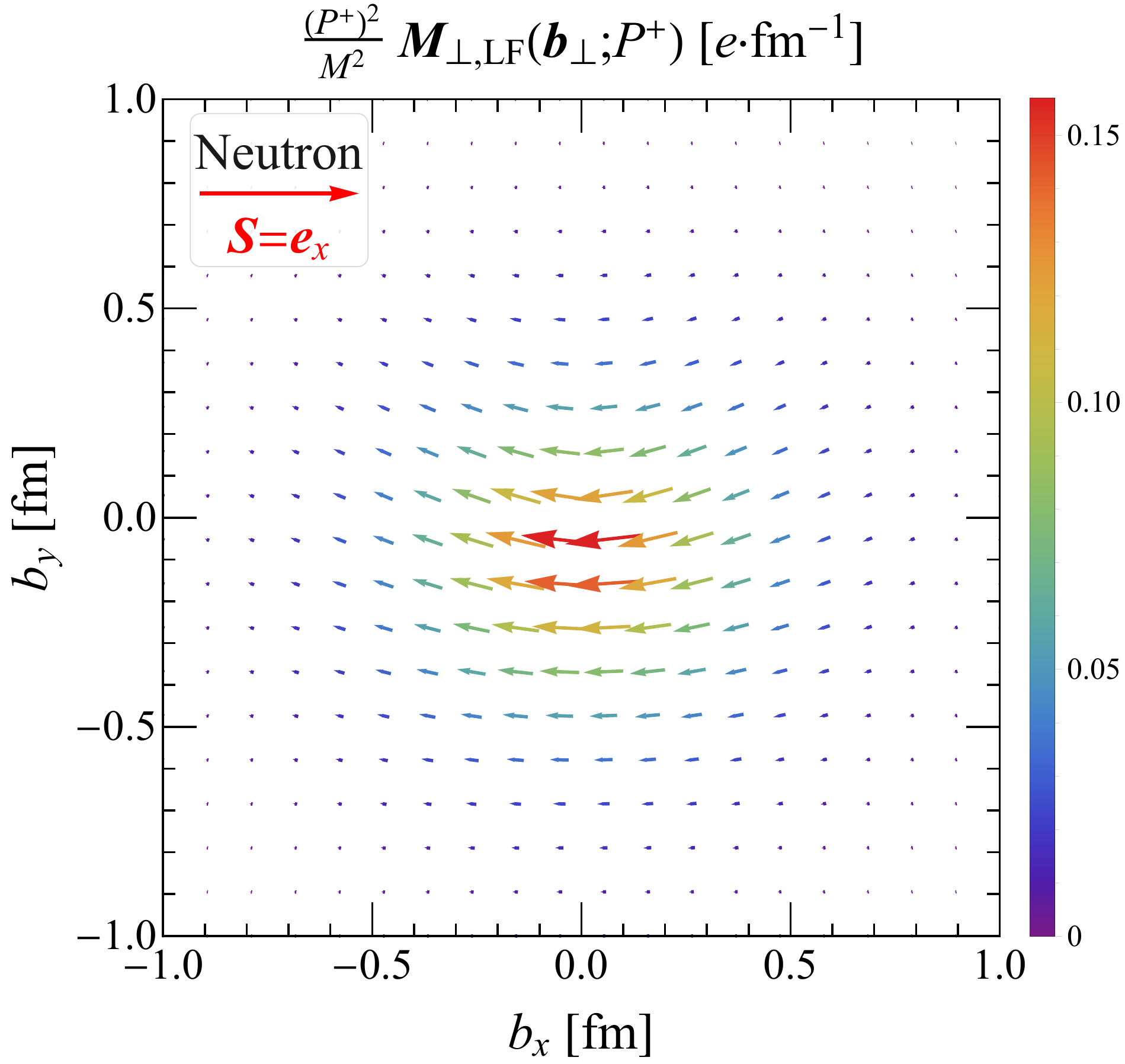}}
	\caption{Light-front transverse polarization and (scaled) magnetization distributions $\uvec{\mathcal P}_{\perp,\text{LF}}(\uvec b_\perp;P^+)$ and $\frac{(P^+)^2}{M^2}\uvec M_{\perp,\text{LF}}(\uvec b_\perp;P^+)$ in the transverse plane, see Eq.~\eqref{spinhalf-LFPolMag}, inside a proton (left panels) or a neutron (right panels) polarized along the $x$-direction. Based on the parametrization for the nucleon electromagnetic form factors given in Ref.~\cite{Bradford:2006yz}.}
	\label{Fig_2DLFMvTPvT}
\end{figure}

Similarly to Eq.~\eqref{spinhalfPvMv}, the LF polarization and magnetization distributions are then obtained by following 2D Fourier transforms
\begin{equation}
    \begin{aligned}\label{spinhalf-LFPolMag}
       \mathcal P^\mu_\text{LF}(\uvec b_\perp;P^+)&=\int\frac{\ud^2\Delta_\perp}{(2\pi)^2}\,e^{-i\uvec\Delta_\perp\cdot\uvec b_\perp}\,\frac{1}{2P^+}\,\widetilde {\mathcal P}^\mu_\text{LF}(\uvec\Delta_\perp;P^+),\\
        M^\mu_\text{LF}(\uvec b_\perp;P^+)&=\int\frac{\ud^2\Delta_\perp}{(2\pi)^2}\,e^{-i\uvec\Delta_\perp\cdot\uvec b_\perp}\,\frac{1}{2P^+}\,\widetilde {M}^\mu_\text{LF}(\uvec\Delta_\perp;P^+).
    \end{aligned}
\end{equation}
Based on the expressions in Eq.~\eqref{LFmagpolampl}, we observe that the LF polarization distributions do not depend on $P^+$, while the longitudinal (transverse) LF magnetization distribution will be suppressed by one power (two powers) of $1/P^+$. In Fig.~\ref{Fig_2DLFMvTPvT}, we show the 2D LF transverse polarization and (scaled) magnetization distributions in the transverse plane for transversely polarized nucleons. To make the transverse magnetization distributions $P^+$-independent, a dimensionless factor $(P^+/M)^2$ has been introduced. While the BF polarization distribution vanishes, the transverse LF polarization distribution is nonzero even for $P_z=0$. This demonstrates once again that LF distributions provide distorted pictures of the system. A multipole decomposition of these distributions is discussed in Appendix~\ref{App-Multipole decomposition}.

Let us now compare the EF and LF distributions in the IMF. For the polarization distributions, we find that both sets coincide in that limit
\begin{equation}\label{spinhalf-IMFcoinLF-PvT}
    \lim_{P^+\to\infty}\mathcal P^\mu_\text{LF}(\uvec b_\perp;P^+)=\lim_{P_z\to\infty}\mathcal P^\mu_\text{EF}(\uvec b_\perp;P_z).
\end{equation}
Interestingly, while the longitudinal magnetization distribution vanishes in both cases
\begin{equation}
    \lim_{P^+\to\infty} M^+_\text{LF}(\uvec b_\perp;P^+)=\lim_{P_z\to\infty}M_{z,\text{EF}}(\uvec b_\perp;P_z)=0,
\end{equation}
it turns out that the scaled distributions do also coincide
\begin{equation}
\begin{aligned}
    \lim_{P^+\to\infty} \frac{P^+}{M}\,M^+_\text{LF}(\uvec b_\perp;P^+)&=\lim_{P_z\to\infty}\frac{P_z}{M}\,M_{z,\text{EF}}(\uvec b_\perp;P_z)\\
    &=\frac{e}{2M}\,\sigma_z\int\frac{\ud^2\Delta_\perp}{(2\pi)^2}\,e^{-i\uvec\Delta_\perp\cdot\uvec b_\perp}\,G_M(\uvec\Delta_\perp^2).
    \end{aligned}
\end{equation}
For the transverse magnetization, we find a relation for the scaled momentum amplitudes
\begin{equation}
\begin{aligned}\label{spinhalf-IMFcoinLF-MvT}
    \lim_{P^+\to\infty} \frac{M}{P^-}\,\widetilde{\uvec M}_{\perp,\text{LF}}(\uvec \Delta_\perp;P^+)&=\lim_{P_z\to\infty}\frac{M}{P_z}\,\widetilde{\uvec M}_{\perp,\text{EF}}(\uvec \Delta_\perp;P_z)\\
    &=e\left[\uvec\sigma_\perp+\frac{(\uvec e_z\times i\uvec\Delta)_\perp}{2M}\right]\frac{G_M(\uvec\Delta^2_\perp)}{1+\tau}.
    \end{aligned}
\end{equation}
Unfortunately, $P^-$ depends on the momentum transfer and therefore cannot be factored out of the Fourier transform, implying that the above relation does not hold in position space. A similar problem was observed for the $J^-$-component of the electromagnetic four-current in Ref.~\cite{Chen:2022smg}. This is of course not too surprising since the longitudinal LF polarization current reads $J^-_{P,\text{LF}}=-(\uvec\nabla_\perp\times\uvec M_{\perp,\text{LF}})_z$.

\subsection{Light-front electric and magnetic dipole moments}

Similarly to Eq.~\eqref{EFEDM}, the (transverse) LF EDM is defined as
\begin{equation}\label{LFEDM}
    \uvec d_{\perp,\text{LF}}(P^+) = \int\ud^2b_\perp\,\uvec b_\perp\,J^+_\text{LF}(\uvec b_\perp;P^+),
\end{equation}
and is given for a spin-$\frac{1}{2}$ target by~\cite{Burkardt:2002hr,Burkardt:2002ks}
\begin{equation}\label{EDM-LF-dy}
    \uvec d_{\perp,\text{LF}}(P^+)=(\uvec e_z\times\uvec\sigma)_\perp\,F_2(0)\,\frac{e}{2M}.
\end{equation}
This quantity does not depend on $P^+$ and is proportional to the anomalous MDM $\kappa=F_2(0)$. 

Since it is well known that objects with MDM in the rest frame display an EDM when viewed from a moving frame~\cite{Einstein:1908}, LF magnetization distributions were defined in Refs.~\cite{Miller:2007kt,Miller:2010nz,Venkat:2010by} directly in terms of 2D Fourier transforms of $F_2(Q^2)$, as suggested by Eq.~\eqref{EDM-LF-dy}. To explain why in the LF formalism $F_2(Q^2)$ appears instead of $G_M(Q^2)$, the authors invoked ``relativistic corrections caused by the transverse localization of the wave packet'' and referred to~\cite{Burkardt:2005hp} for more explanations. In the latter paper, it is argued that Melosh rotations~\eqref{Meloshrotation} cause a transverse shift\footnote{This shift is crucial for understanding the relation between transverse angular momentum and the dipole moment of the longitudinal LF momentum distributions~\cite{Burkardt:2005hp,Lorce:2018zpf}.} of the center of $P^+$ (identified with the origin within the LF formalism) relative to the center of mass of the system. The expression in Eq.~\eqref{EDM-LF-dy} represents therefore the EDM defined relative to the center of $P^+$. It coincides with the IMF limit of the EF EDM~\eqref{EDM-EF-dy}
\begin{equation}\label{EDM-IMF-dy}
    \uvec d_{\perp,\text{LF}}(P^+) = \lim_{P_z\to\infty}\uvec d_{\perp,\text{EF}}(P_z)=(\uvec e_z\times\uvec\sigma)_\perp\left[G_M(0)-G_E(0)\right]\frac{e}{2M},
\end{equation}
since $G_M(0)-G_E(0)=F_2(0)$. We have seen in Sec.~\ref{sec:Elastic frame electric and magnetic dipole moments} that the first term corresponds to the contribution associated with the rest-frame MDM. The second term arises from the sideways shift of the center of spin relative to the center of mass. In the IMF, the center of spin coincides with the center of $P^+$~\cite{Lorce:2018zpf}, see Fig.~\ref{Fig_RelativisticCenters} in Appendix~\ref{App-Relativistic centers of the nucleon}, and we can identify the second term with the shift pointed out in Ref.~\cite{Burkardt:2005hp} (equal to one half of the reduced Compton wavelength when the spin-$\frac{1}{2}$ system is transversely polarized) multiplied by the total charge $G_E(0)\,e$ of the system. Contrary to Refs.~\cite{Miller:2007kt,Miller:2010nz,Venkat:2010by}, we interpret this contribution as a relativistic artifact rather than a ``relativistic correction''. Genuine LF magnetization distributions should therefore be defined in terms of Fourier transforms of $G_M(Q^2)$ rather than $F_2(Q^2)$.

For a spin-$j$ target, the EF EDM~\eqref{spinjEDM} reduces in the IMF limit to
\begin{equation}
   \lim_{P_z\to \infty} \uvec d^{(j)}_{\perp,\text{EF}}(P_z)=(\uvec e_z\times\uvec\Sigma)_\perp\left[G_{M1}(0)-2j\,G_{E0}(0)\right]\frac{e}{2M}
\end{equation}
and coincides with the spin-$j$ LF EDM derived in Ref.~\cite{Lorce:2009bs}. Interestingly, this EDM vanishes when $G_{M1}(0)=2j\,G_{E0}(0)$, i.e.~when the Land\'e factor assumes the universal value $g=2$. The combination $\kappa\equiv G_{M1}(0)-2j\,G_{E0}(0)$ is then interpreted in general as the anomalous MDM for a spin-$j$ system. For $j=\frac{1}{2}$, we recover naturally $\kappa=F_2(0)$.

If we integrate the transverse LF polarization distribution~\eqref{spinhalf-LFPolMag} over the impact-parameter space, we will find
\begin{equation}
    \int\ud^2b_\perp\,\uvec{\mathcal P}_{\perp,\text{LF}}(\uvec b_\perp;P^+)=\frac{1}{2P^+}\,\widetilde{\uvec{\mathcal P}}_{\perp,\text{LF}}(\uvec 0_\perp;P^+)=(\uvec e_z\times\uvec\sigma)_\perp\,G_M(0)\,\frac{e}{2M}.
\end{equation}
This quantity corresponds to the first term in Eq.~\eqref{EDM-IMF-dy} since it is simply the LF EDM arising from the polarization part of the LF charge distribution $J^+_{\text{LF}}(\uvec b_\perp;P^+)$~\cite{Chen:2022smg}
\begin{equation}
    \uvec d_{P,\perp,\text{LF}}(P^+)=\int\ud^2b_\perp\,\uvec b_\perp \,\rho_{P,\text{LF}}(\uvec b_\perp;P^+)= \int\ud^2b_\perp\,\uvec{\mathcal P}_{\perp,\text{LF}}(\uvec b_\perp;P^+),
\end{equation}
where the LF polarization charge distribution $\rho_{P,\text{LF}}(\uvec b_\perp;P^+)$ coincides with the infinite-momentum limit of the corresponding EF polarization charge distribution~\eqref{2DEFrhoP}, namely
\begin{equation}
	\begin{aligned}\label{spinhalf-LFrhoP}
		\rho_{P,\text{LF}}(\uvec b_\perp;P^+)
		&= -\uvec\nabla_\perp \cdot\uvec{\mathcal P}_\text{LF}(\uvec b_\perp;P^+)=\lim_{P_z\to \infty}\,\rho_{P,\text{EF}}(\uvec b_\perp;P_z)\\
		&= e\int\frac{\ud^2\Delta_\perp}{(2\pi)^2}\,e^{-i\uvec\Delta_\perp\cdot\uvec b_\perp}\left[\tau + \frac{(\uvec\sigma\times i\uvec\Delta_\perp)_z}{2M}\right]\frac{G_M(\uvec\Delta_\perp^2)}{1+\tau}.
	\end{aligned}
\end{equation}

The LF magnetization distributions studied in the present work are directly defined in terms of the matrix elements of a polarization-magnetization tensor operator, see Eqs.~\eqref{LF-PolMagDef} and \eqref{spinhalf-LFPolMag}. These distributions therefore exclude from the beginning any contribution from the convective part of the electromagnetic four-current, and are naturally given by 2D Fourier transforms of $G_M(Q^2)$ rather than $F_2(Q^2)$. In particular, longitudinal and transverse LF MDMs are respectively defined as
\begin{equation}
    \begin{aligned}
        \mu_{z,\text{LF}}(P^+)&=\frac{1}{\sqrt{2}}\int\ud^2b_\perp\,M^+_\text{LF}(\uvec b_\perp;P^+)=\sigma_z\,\frac{M}{\sqrt{2}P^+}\,G_M(0)\,\frac{e}{2M},\\
        \uvec\mu_{\perp,\text{LF}}(P^+)&=\int\ud^2b_\perp\,\uvec M_{\perp,\text{LF}}(\uvec b_\perp;P^+)=\uvec\sigma_\perp\,\frac{M^2}{2(P^+)^2}\,G_M(0)\,\frac{e}{2M},
    \end{aligned}
\end{equation}
which agree in the rest frame (i.e.~when $P^+=M/\sqrt{2}$ with $\uvec\Delta_\perp=\uvec 0_\perp$ resulting from the integration over the impact-parameter space) with the BF results~\eqref{spinhalf-BFMDM}. It may seem a priori surprising that $\uvec\mu_{\perp,\text{LF}}(\infty)=0$ whereas $\uvec\mu_{\perp,\text{EF}}(\infty)=\uvec\sigma_\perp\,G_M(0)\,e/(2M)$. This can however be understood by the fact that $d^i_{P,\perp,\text{EF}}(\infty)=-\epsilon^{ij}_\perp \mu^j_{\perp,\text{EF}}(\infty)$, where $\uvec d_{P,\perp,\text{EF}}(P_z)= \int\ud^2b_\perp\,\uvec{\mathcal P}_\text{EF}(\uvec b_\perp;P_z)$ is the polarization part of the transverse EF EDM. Using Eq.~\eqref{spinhalf-LFMag0} we then find that $\mu^i_{\perp,\text{LF}}(\infty)\propto \mu^i_{\perp,\text{EF}}(\infty)-\epsilon^{ij}_\perp d^j_{P,\perp,\text{EF}}(\infty)=0$.

Following the spirit of the LF polarization charge distribution (\ref{spinhalf-LFrhoP}), we can likewise define the LF effective magnetic charge distribution via
\begin{equation}
	\begin{aligned}
		\rho_{M,\text{LF}}(\uvec b_\perp;P^+)
		&= -\uvec\nabla_\perp\cdot\uvec{M}_\text{LF}(\uvec b_\perp;P^+)\\ 
		&= \frac{e}{2M}\,\frac{M^2}{2(P^+)^2}\int\frac{\ud^2\Delta_\perp}{(2\pi)^2}\,e^{-i\uvec\Delta_\perp\cdot\uvec b_\perp}\,(i\uvec\Delta_\perp\cdot\uvec\sigma_\perp)\,G_M(\uvec\Delta_\perp^2),
	\end{aligned}
\end{equation}
and hence equivalently rewrite the transverse LF MDM as follows
\begin{equation}
    \uvec \mu_{\perp,\text{LF}}(P^+)=\int\ud^2b_\perp\,\uvec b_\perp \,\rho_{M,\text{LF}}(\uvec b_\perp;P^+).
\end{equation}

\section{Summary}
\label{sec:Summary}

In this paper, we extended our study of the relativistic electromagnetic four-current distributions inside a spin-$\tfrac{1}{2}$ system and applied the quantum phase-space formalism to the polarization-magnetization tensor operator. In doing so, relativistic polarization and magnetization distributions were for the first time systematically studied in the Breit frame, the elastic frame and on the light-front.

In the literature, the polarization-magnetization tensor is usually motivated by the Gordon decomposition of the electromagnetic four-current and is accordingly defined in terms of the tensor Dirac bilinear. However, we pointed out that a Sachs decomposition of the electromagnetic four-current suggests instead a definition in terms of the axial-vector Dirac bilinear. Axial-vector and tensor Dirac bilinears simply correspond to two natural ways of describing spin in a relativistic theory, differing by the reference point used in the definition of the internal angular momentum. Through our analysis of the polarization and magnetization distributions in the Breit frame (where the spin structure assumes its simplest form), we observed that the axial-vector description leads to the simplest and physically most natural picture of the polarization and magnetization content of the system.

Relativistic polarization and magnetization distributions are in general frame-dependent. We studied in detail their frame-dependence and compared them in the infinite-momentum frame with the corresponding light-front distributions. We explicitly showed that the genuine light-front magnetization distributions are defined in terms of 2D Fourier transforms of the Sachs magnetic form factor, rather than the Pauli form factor (as suggested earlier in the literature). We explained that the difference results from the transverse shift of the center of light-front momentum relative to the center of mass. 

For illustration, we finally applied our results to the case of a nucleon using the corresponding electromagnetic form factors extracted from experimental data. Our analytic expressions and physical interpretations of relativistic polarization and magnetization distributions hold in fact for any physical spin-$\frac{1}{2}$ targets and can be easily generalized to higher-spin targets. All that is required from the experimental side is an extraction of the corresponding electromagnetic form factors.

\begin{acknowledgements}

Y.~C. is grateful to Prof.~Qun Wang, Prof.~Shi Pu, and the Department of Modern Physics for their very kind hospitality and help during his visit to the University of Science and Technology of China. Y.~C. thanks Prof.~Qun Wang, Prof.~Yang Li, Prof.~Dao-Neng Gao and Prof.~Ming-Zhe Li for very insightful discussions at the early stage of this work. This work is supported in part by the National Natural Science Foundation of China (NSFC) under Grant Nos.~12135011, 11890713 (a sub-Grant of 11890710), and by the Strategic Priority Research Program of the Chinese Academy of Sciences (CAS) under Grant No.~XDB34030102. 
	
\end{acknowledgements}
\appendix
\section{Charge radii}
\label{App-Charge radii}

In this Appendix, we review the concept of relativistic mean square radii for spatial distributions, apply it to the case of the relativistic charge distribution for a spin-$\frac{1}{2}$ system, and study in particular the momentum dependence in the 2D case.

\subsection{In the 3D Breit frame}
The mean square radius of a 3D spatial distribution $O(\uvec r)$ is defined as
\begin{equation}\label{3DBFMSR-Definition}
		\langle \uvec r_{O}^2 \rangle \equiv \frac{\int\ud^3r\,\uvec r^2 O(\uvec r)}{\int\ud^3r\,O(\uvec r)}.
\end{equation}
Applying this definition to the 3D BF charge distribution leads to~\cite{Yennie:1957rmp,Ernst:1960zza}
\begin{equation}\label{totalchargeradius}
		\langle \uvec r^2_\text{ch} \rangle = \frac{\int\ud^3r\,\uvec r^2 J^0_B(\uvec r)}{\int\ud^3r\,J^0_B(\uvec r)} =\langle \uvec r^2_E \rangle+\frac{3}{4M^2},
\end{equation}
where the first term is the conventional Sachs mean square radius defined as~\cite{Gao:2021sml,Xiong:2023zih}\footnote{Since $G_E^n(0)=0$, the neutron Sachs mean square radius is defined with $G^p_E(0)=1$ in the denominator.}
\begin{equation}\label{Sachs-radiusE}
	\langle \uvec r^2_E \rangle \equiv -\frac{6}{G_E(0)}\,\frac{\ud G_E(Q^2)}{\ud Q^2}\bigg{|}_{Q=0} = \frac{1}{G_E(0)}\left[-\uvec\nabla_{\uvec\Delta}^2 G_E(\uvec\Delta^2) \right]_{\uvec\Delta=\uvec 0},
\end{equation}
and the second term is known as the Darwin-Foldy term~\cite{Foldy:1949wa,Foldy:1952csp,Foldy:1952zz}. For purely historical reasons, the Darwin-Foldy term is kept separate in the literature, and so the charge radius of a spin-$\frac{1}{2}$ system is traditionally defined by $r_E\equiv\sqrt{\langle \uvec r^2_E \rangle}$~\cite{Friar1975,Miller:2018ybm}. Similarly, one can consider the mean square radius of the effective magnetic charge distribution, but the result is trivial, viz.~$\int\ud^3r\,\uvec r^2\rho_{M,B}(\uvec r)=\int\ud^3r\,\rho_{M,B}(\uvec r) = 0$, because the expression for $\rho_{M,B}$ in momentum space is odd in $\uvec\Delta$, see Eq.~\eqref{effectiveMagCharge}. In the literature, the conventional magnetic mean square radius is in fact simply defined by analogy with Eq.~\eqref{Sachs-radiusE}
\begin{equation}
        \langle \uvec r^2_M \rangle \equiv -\frac{6}{G_M(0)} \frac{\ud G_M(Q^2)}{\ud Q^2}\bigg{|}_{Q=0} =\frac{1}{G_M(0)}[-\uvec\nabla_{\uvec\Delta}^2 G_M(\uvec\Delta^2)]_{\uvec\Delta=\uvec 0}.
\end{equation}

If one adopts a $T$-type decomposition of the charge density~\eqref{spinhalf-BFJ0Ttype}, the mean square charge radius can be split as follows
\begin{equation}
    	\langle \uvec r^2_\text{ch} \rangle = \langle \uvec r_{\text{ch},c'}^2 \rangle +
		\langle\uvec r_{\text{ch},P'}^2 \rangle,
\end{equation}
where the convection and polarization contributions are respectively given by
\begin{equation}
    \begin{aligned}
        \langle \uvec r_{\text{ch},c'}^2 \rangle&=\langle \uvec r^2_D \rangle-\frac{3}{4M^2},\\
        \langle\uvec r_{\text{ch},P'}^2 \rangle&=\frac{3}{2M^2}\,\frac{G_M(0)}{G_E(0)}.
    \end{aligned}
\end{equation}
Beside the Dirac mean square radius
\begin{equation}
	\langle \uvec r^2_D \rangle \equiv -\frac{6}{F_1(0)}\,\frac{\ud F_1(Q^2)}{\ud Q^2}\bigg{|}_{Q=0},
\end{equation}
we observe in the convection contribution a negative Darwin-Foldy term coming from the factor $P^0_B/M$ in Eq.~\eqref{spinhalf-BFJ0cTtype}, analogous to the positive Darwin-Foldy term in Eq.~\eqref{totalchargeradius} coming from the factor $M/P^0_B$ in Eq.~\eqref{BFdensities}. Interestingly, even if the $T$-type polarization does not contribute to the total charge of the system, it does contribute to the charge radius. This is reflected in momentum space by the global factor of $\tau=Q^2/(4M^2)$ in Eq.~\eqref{spinhalf-BFJ0PTtype}.

\subsection{In the 2D elastic and light-front frames}

The mean square transverse radius of a 2D spatial distribution $O(\uvec b_\perp)$ is defined similarly to its 3D counterpart~\eqref{3DBFMSR-Definition}
\begin{equation}\label{2DEFMSR-Definition}
		\langle \uvec b_{\perp,O}^2 \rangle \equiv \frac{\int\ud^2b_\perp\,\uvec b^2_\perp O(\uvec b_\perp)}{\int\ud^2b_\perp\,O(\uvec b_\perp)}.
\end{equation}
Applying this definition to the 2D EF charge distribution $J^0_\text{EF}(\uvec b_\perp;P_z)$ in Eq.~\eqref{spinhalfEFJ0Jv} leads to
\begin{equation}
	\begin{aligned}\label{2DEFMSR-J0}
		\langle \uvec b_{\perp,\text{ch}}^2 \rangle_\text{EF}(P_z) &= \frac{\int\ud^2b_\perp\, \uvec b_\perp^2 J^0_\text{EF}(\uvec b_\perp;P_z)}{\int\ud^2b_\perp\, J^0_\text{EF}(\uvec b_\perp;P_z)}\\
		&= \frac{2}{3}\,\langle \uvec r^2_E \rangle+\frac{1}{M^2}\left[\frac{E_P}{E_P+M} -  \frac{E_P-M}{E_P}\, \frac{G_M(0)}{G_E(0)} \right]
	\end{aligned}
\end{equation}
with $E_P=\sqrt{P^2_z+M^2}$. In particular, in the BF we have
\begin{equation}
	\begin{aligned}
		\lim_{P_z \to 0} \langle \uvec b_{\perp, \text{ch}}^2\rangle_\text{EF} (P_z) = \frac{2}{3}\,\langle \uvec r^2_E \rangle+\frac{1}{2M^2},
	\end{aligned}
\end{equation}
which is consistent with our expectation $\langle \uvec b_{\perp, \text{ch}}^2 \rangle_\text{EF}(0) = \frac{2}{3}\, \langle \uvec r^2_\text{ch} \rangle$ for a spherically symmetric BF charge distribution. In the IMF, we find
\begin{equation}
	\begin{aligned}
		\lim_{P_z \to \infty} \langle \uvec b_{\perp, \text{ch}}^2 \rangle_\text{EF}(P_z) 
		&= \frac{2}{3}\,\langle \uvec r^2_D \rangle,
	\end{aligned}
\end{equation}
where we used the relation  
\begin{equation}
    \langle \uvec r_E^2\rangle=\langle\uvec r^2_D\rangle+\frac{3}{2M^2}\,\frac{F_2(0)}{F_1(0)}
\end{equation}
between the Sachs and Dirac mean square radii.

\begin{figure}[tb!]
	\centering
	{\includegraphics[angle=0,scale=0.46]{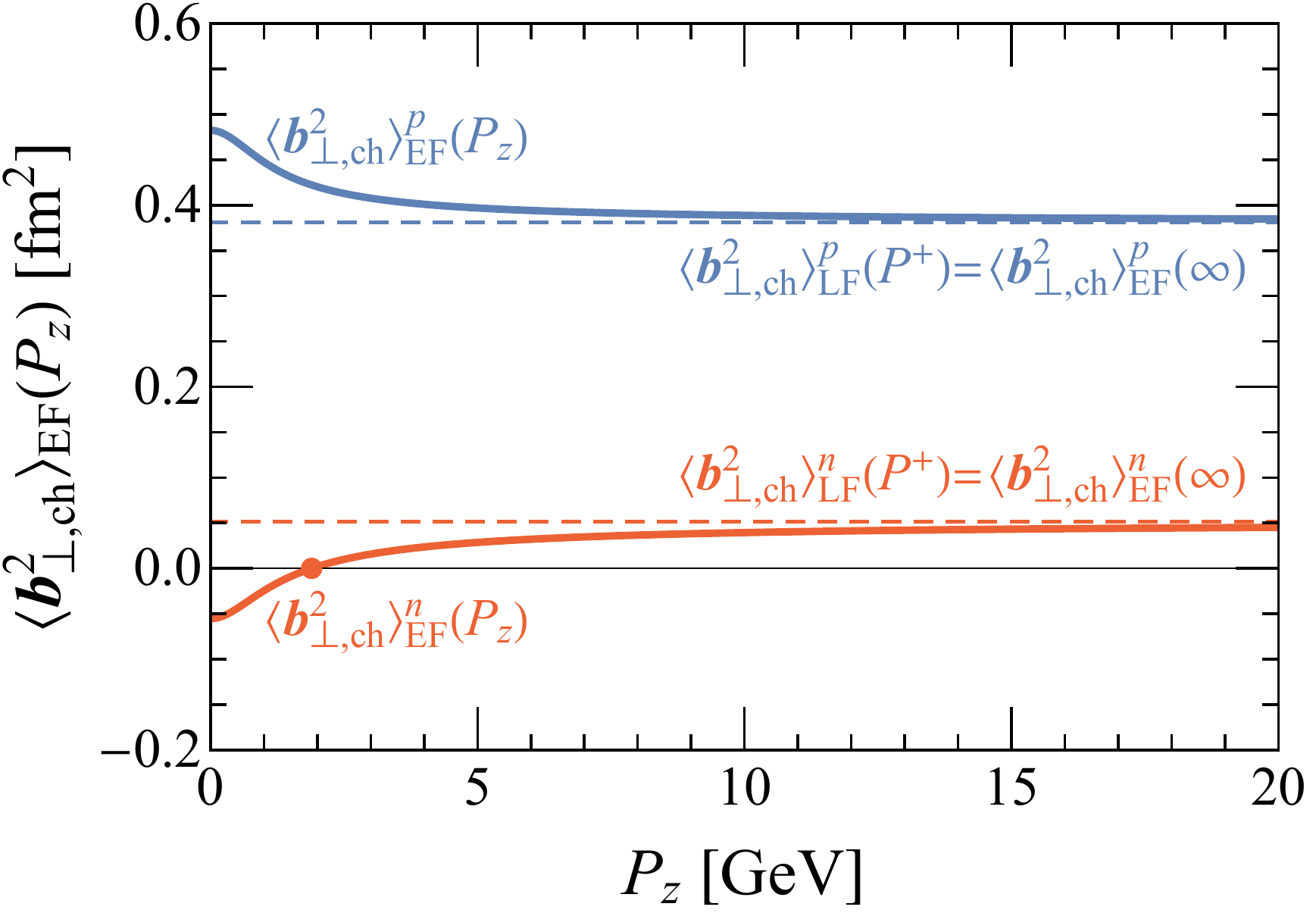}}
	\caption{Mean-square transverse charge radii $\langle \uvec b_{\perp,\text{ch}}^2\rangle_{\text{EF}}(P_z)$ of the nucleon in the elastic frame, see Eq.~\eqref{2DEFMSR-J0}, as functions of the nucleon average momentum $P_z$. The proton and neutron Sachs mean square radii $\langle \uvec r^2_E \rangle$ are taken from the recent measurements by the PRad Collaboration~\cite{Xiong:2019umf,PRad:2020oor} and the data tables by the Particle Data Group~\cite{ParticleDataGroup:2022pth}, respectively.}
	\label{Fig_2DEFJ0JzMSR}
\end{figure}

In Fig.~\ref{Fig_2DEFJ0JzMSR}, we show the EF mean square transverse charge radii $\langle \uvec b_{\perp,\text{ch}}^2\rangle(P_z)$ of the nucleon as a function of the average momentum $P_z$. The proton and neutron Sachs mean square radii 
\begin{equation}
	\begin{aligned}
	    \langle \uvec r^2_E \rangle^{p} &= (0.831 \pm 0.007_{\text{stat.}}\pm 0.012_{\text{syst.}})^2~\text{fm}^2,\\
	    \langle \uvec r^2_E \rangle^{n} &= (-0.1161 \pm 0.0022)~\text{fm}^2,
	\end{aligned}
\end{equation}
are taken from recent measurements by the PRad Collaboration~\cite{Xiong:2019umf,PRad:2020oor} and from the Particle Data Group~\cite{ParticleDataGroup:2022pth}, respectively. Interestingly, we observe that $\langle \uvec b_{\perp,\text{ch}}^2\rangle^n(P_z)$ switches sign from negative to positive around $P_z \approx 1.893~\text{GeV}$. 

Applying now the definition~\eqref{2DEFMSR-Definition} to the 2D LF charge distribution $J^+_\text{LF}(\uvec b_\perp;P^+)$ leads to~\cite{Miller:2018ybm}
\begin{equation}
		\langle \uvec b_{\perp, \text{ch}}^2 \rangle_\text{LF}(P^+) = \frac{\int\ud^2b_\perp\, \uvec b_\perp^2 J^+_\text{LF}(\uvec b_\perp;P^+)}{\int\ud^2b_\perp\, J^+_\text{LF}(\uvec b_\perp;P^+)}=\frac{2}{3}\,\langle \uvec r^2_D \rangle =\langle \uvec b_{\perp,\text{ch}}^2 \rangle_\text{EF}(\infty),
\end{equation}
which is consistent with the fact that $J^+_\text{LF}(\uvec b_\perp;P^+)=J^0_\text{EF}(\uvec b_\perp;\infty)$~\cite{Chen:2022smg}.

\section{Relativistic centers of the nucleon}
\label{App-Relativistic centers of the nucleon}

In this Appendix, we remind the relations between the positions of the various possible centers of a relativistic spin-$\tfrac{1}{2}$ system~\cite{Lorce:2018zpf,Lorce:2021gxs}. For a spin-$j$ system, it suffices to replace $\tfrac{1}{2}\uvec S$ by $j\uvec S$ in the following expressions.

The position of the center of canonical spin $\uvec R_c$ (the point about which the internal angular momentum takes the same value as in the rest frame) coincides with the average position $\uvec R$ appearing in the quantum phase-space formalism, namely
\begin{equation}
	\uvec R_c = \uvec R = \tfrac{1}{2}(\uvec r + \uvec r').
\end{equation}
Since in the literature one is usually interested only in the internal structure of the target, one often sets $\uvec R=\uvec 0$ for convenience.

The positions of the center of energy (or inertia) $\uvec R_E$ and the center of mass $\uvec R_M$ are respectively given by
\begin{equation}
\begin{aligned}
    \uvec R_E &= \uvec R + \frac{\uvec P \times \uvec S}{2E_P(E_P+M)} ,\\
    \uvec R_M &= \uvec R - \frac{\uvec P \times \uvec S}{2M(E_P+M)} ,
\end{aligned}
\end{equation}
where $\uvec S$ is the unit polarization vector. For a system at rest ($\uvec P=\uvec 0$) or longitudinally polarized ($\uvec P\times\uvec S=\uvec 0$), all these relativistic centers coincide
\begin{equation}
		\uvec R_M = \uvec R_E = \uvec R_c = \uvec R .
\end{equation}
The center of mass is the only one transforming as the spatial part of a Lorentz four-vector, and corresponds therefore to the \emph{true} center of the system. The shifts
\begin{equation}\label{shift}
	\begin{aligned}
	\uvec R_c-\uvec R_M &= \frac{\uvec P \times \uvec S}{2M(E_P+M)},\\
	\uvec R_E-\uvec R_M &= \frac{\uvec P \times \uvec S}{2M E_P},
	\end{aligned}
\end{equation}
are a pure relativistic effect. The set of all possible centers of energy forms a disk centered at $\uvec R_M$ and orthogonal to $\uvec S$, known as M\o ller's disk~\cite{Moller:1949bis,Moller:1949}. Its radius is equal to half the reduced Compton wavelength 
\begin{equation}\label{Rmoller}
		R_{\text{M\o ller}} = \frac{1}{2M},
\end{equation}
and corresponds to the maximum value of $|\uvec R_{c,E}-\uvec R_M|$ in Eq.~\eqref{shift}, reached in the IMF for a purely transverse polarization.

\begin{figure}[tb!]
	\centering
	{\includegraphics[angle=0,scale=0.270]{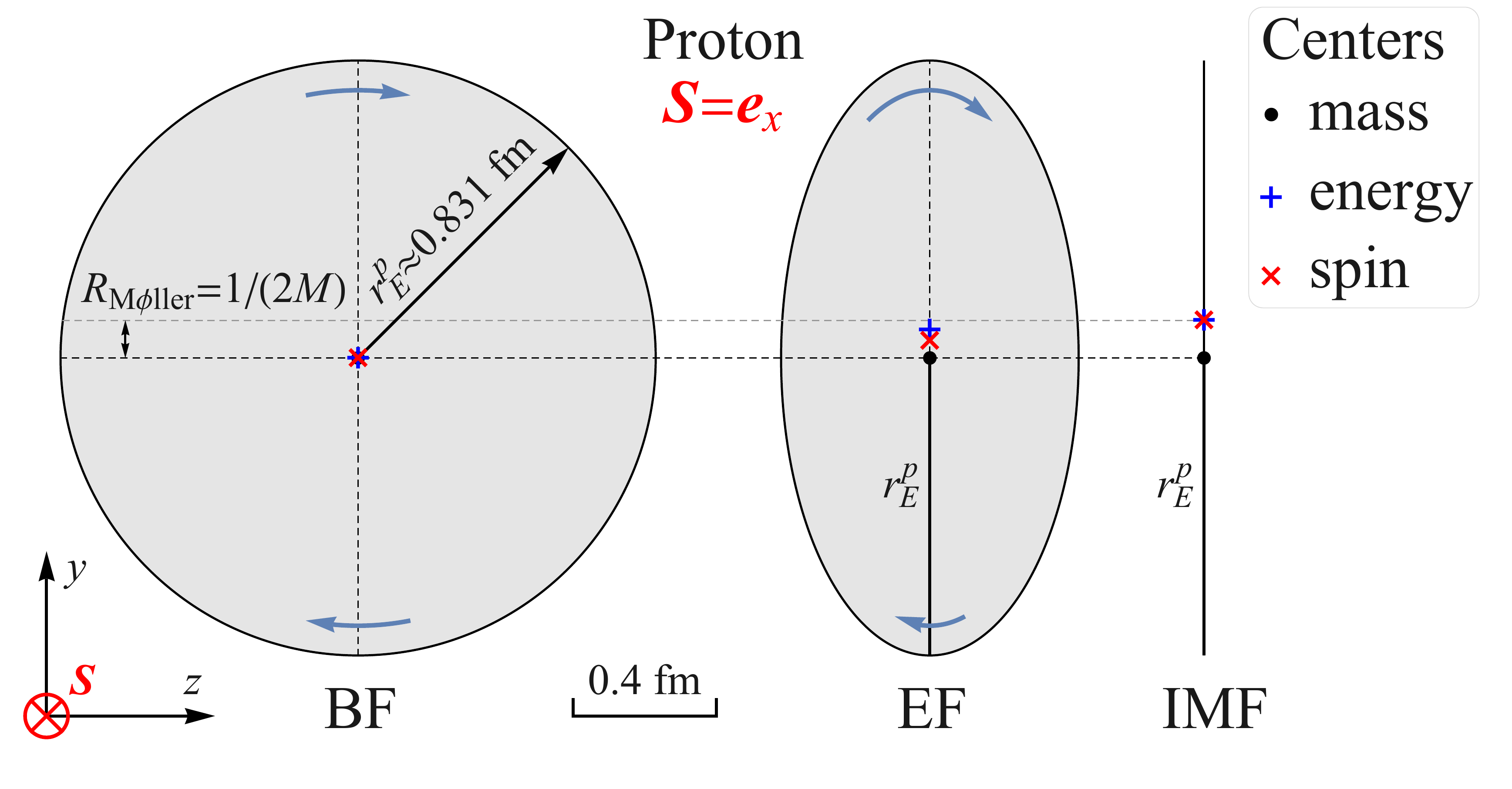}}
	{\includegraphics[angle=0,scale=0.390]{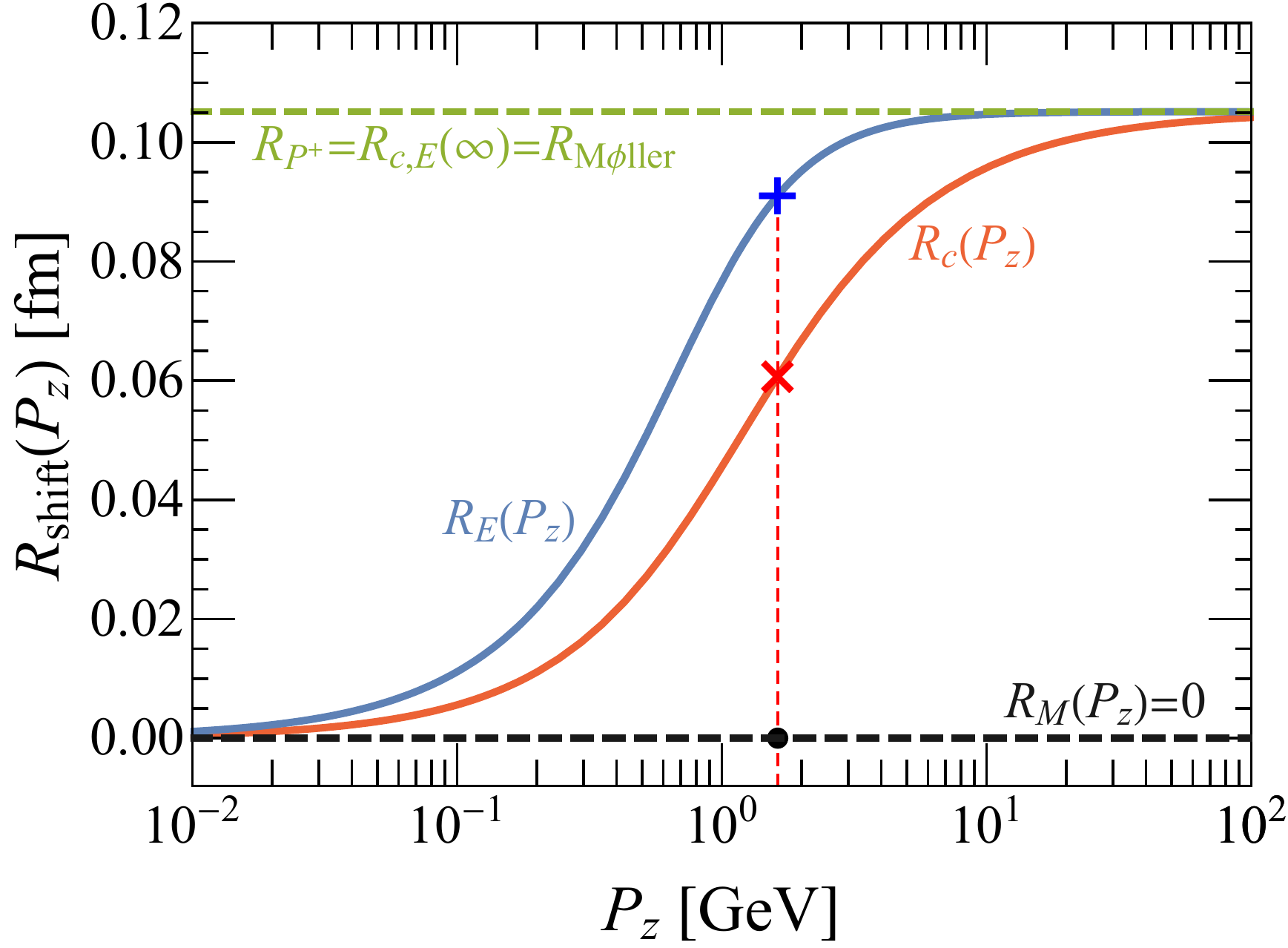}}
	\caption{Left panel: Illustration of the relative positions in the $r_x=0$ plane of the relativistic centers of mass $\uvec R_M$, energy $\uvec R_E$ and canonical spin $\uvec R_c$ inside a transversely polarized proton viewed in the Breit, elastic and infinite-momentum frames. The light-blue arrows represent the local momentum density. The proton charge radius $r_E^p\approx 0.831\,\text{fm}$ is taken from recent precision measurements by the PRad Collaboration~\cite{Xiong:2019umf,PRad:2020oor}. The horizontal gray-dashed line corresponds to the maximum shift given by the M\o ller radius~\eqref{Rmoller}. Right panel: Momentum dependence of sideways shifts along the $y$-axis of the relativistic centers inside a proton. As an example, the vertical dashed red line at $P_z=\sqrt{3}M \approx 1.625\,\text{GeV}$ corresponds to the elastic frame case (with Lorentz factor $\gamma_P=2$) in the left panel.}
	\label{Fig_RelativisticCenters}
\end{figure}

In the LF formalism one identifies the center of the target with the center of $P^+$~\cite{Burkardt:2002hr}, whose transverse position is given by~\cite{Burkardt:2005hp,Lorce:2018zpf}
\begin{equation}
    \uvec R_{P^+,\perp}=\uvec R_{M,\perp}+\frac{(\uvec e_z\times\uvec S)_\perp}{2M}.
\end{equation}
The center of $P^+$ can therefore be identified with the IMF center of energy (or equivalently the IMF center of spin). 

The relative positions of the various relativistic centers are illustrated in Fig.~\ref{Fig_RelativisticCenters}. The left panel shows a representation of the proton viewed from different Lorentz frames. The right panel shows the momentum dependence of the transverse position of the center of energy, spin and $P^+$ relative to the center of mass. The EF situation represented in the left panel corresponds to the Lorentz factor $\gamma_P=2$ (i.e. $P_z=\sqrt 3 M\approx 1.625$ GeV for a proton) and is represented by the vertical dashed line in the right panel.

\section{Multipole decomposition of the relativistic polarization and magnetization distributions}
\label{App-Multipole decomposition}

In this Appendix, we discuss the multipole decomposition of the relativistic polarization and magnetization distributions in both 3D and 2D cases. Since polarization and magnetization transform as vectors under rotations, their matrix elements for a spin-$\frac{1}{2}$ system can only consist in monopole, dipole and quadrupole contributions.

\subsection{In the 3D Breit frame}

In the $n$D Euclidean space, the Fourier transform of a quadrupole in $\uvec\Delta$ can conveniently be expressed as follows
\begin{equation}
	\begin{aligned}\label{3DBF-Multipole}
		\int\frac{\ud^n\Delta}{(2\pi)^n}\,&e^{-i\uvec\Delta\cdot\uvec r}\left(\Delta^i\Delta^j-\tfrac{1}{n}\,\delta^{ij}\uvec\Delta^2\right) f(\uvec\Delta^2)\\
		&=\frac{r^ir^j-\frac{1}{n}\,\delta^{ij}\uvec r^2}{r^2}\left(\frac{1}{r}\frac{\ud}{\ud r} - \frac{\ud^2}{\ud r^2} \right)\int\frac{\ud^n\Delta}{(2\pi)^n}\,e^{-i\uvec\Delta\cdot\uvec r}\,f(\uvec\Delta^2)
	\end{aligned}
\end{equation}
with $r=|\uvec r|$. In the 3D Euclidean space, we have in particular
\begin{equation}
	\int\frac{\ud^3\Delta}{(2\pi)^3}\,e^{-i\uvec\Delta\cdot\uvec r}\left(\Delta^i\Delta^j-\tfrac{1}{3}\,\delta^{ij}\uvec\Delta^2\right) f(\uvec\Delta^2)
	=-\frac{r^ir^j-\frac{1}{3}\,\delta^{ij}\uvec r^2}{r^2}\int\frac{\ud Q}{2\pi^2}\,Q^4j_2(Qr)\,f(Q^2),
\end{equation}
where the $n$th order spherical Bessel function $j_n(x)$ is given by
\begin{equation}\label{spherical-Bessel-jn}
	j_n(x)=(-1)^n x^n\left(\frac{1}{x}\frac{\ud }{\ud x}\right)^{\!n} j_0(x)
\end{equation}
with $j_0(x)=\sin x/x$ the zeroth order spherical Bessel function.

It is then straightforward to decompose the BF magnetization distribution in Eq.~(\ref{simplepicture}) into two terms $\uvec M_B = \uvec M^{(M)}_B+ \uvec M^{(Q)}_B$, where $\uvec M^{(M)}_B(\uvec r)$ corresponds to the monopole contribution
\begin{equation}
	\begin{aligned}\label{3DBF-monopole-Mv}
		\uvec M^{(M)}_B(\uvec r) 
		&= \frac{e}{2M}\,\uvec\sigma\int\frac{\ud Q}{2\pi^2}\,Q^2j_0(Qr)\, \frac{1}{3}\left[2+\frac{M}{P^0_B} \right]\frac{M}{P^0_B}\, G_M(Q^2),
	\end{aligned}
\end{equation}
and $\uvec M^{(Q)}_B(\uvec r)$ corresponds to the quadrupole contribution
\begin{equation}
	\label{3DBF-quadrupole-Mv}
		\uvec M^{(Q)}_B(\uvec r) 
		=\frac{e}{2M}\left[\hat{\uvec r} (\hat{\uvec r}\cdot\uvec \sigma)-\tfrac{1}{3}\,\uvec\sigma\right] \int \frac{\ud Q}{2\pi^2} \, Q^4 j_2(Q r)\,  \frac{1}{4P_B^0(P_B^0+M)}\, \frac{M}{P^0_B}\,G_M(Q^2)
\end{equation}
with $\hat{\uvec r}\equiv \uvec r/|\uvec r|$ the unit vector along $\uvec r$. 
 
\begin{figure}[tb!]
	\centering
	{\includegraphics[angle=0,scale=0.38]{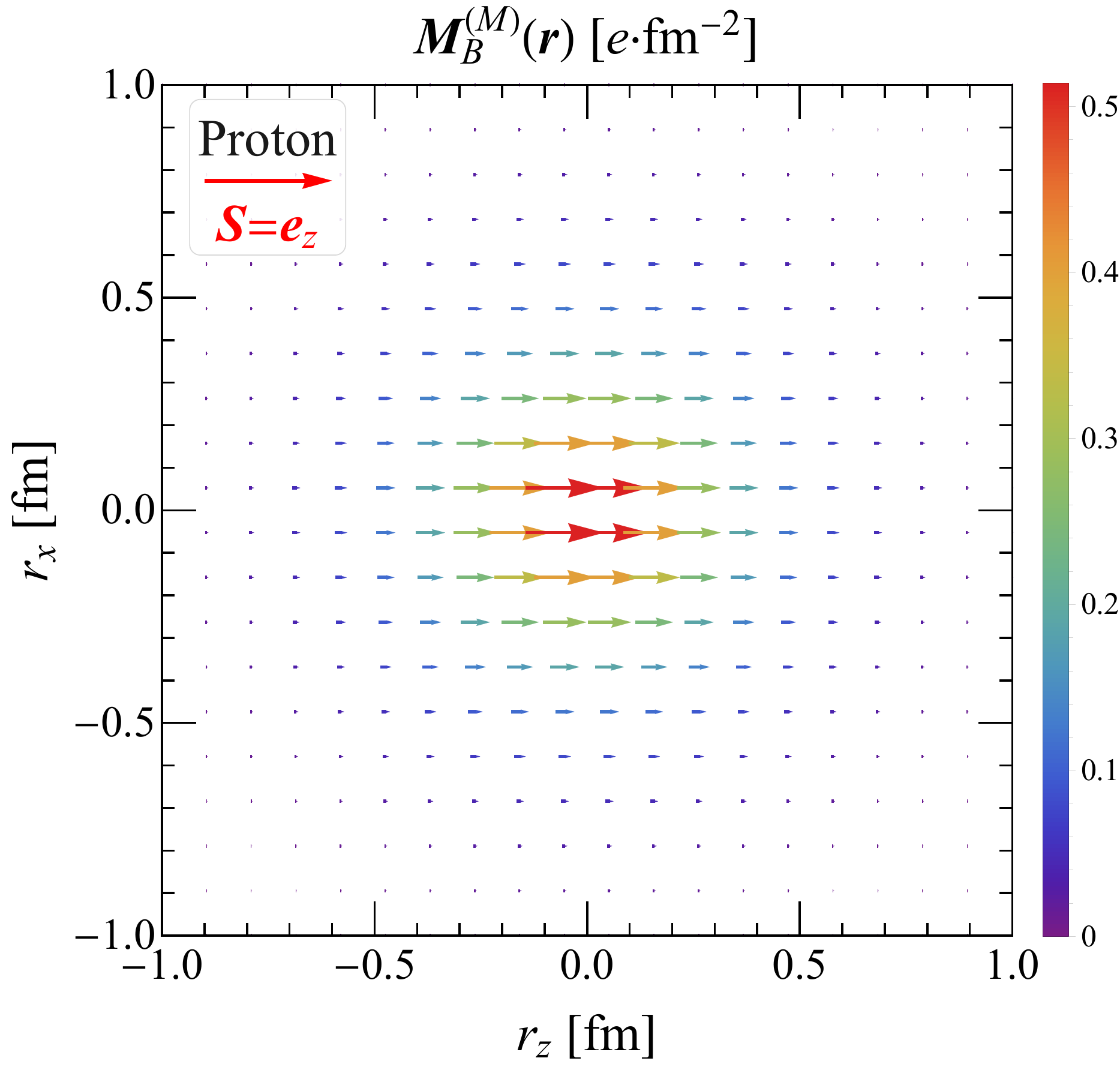}\hspace{0.2cm}}
	{\includegraphics[angle=0,scale=0.38]{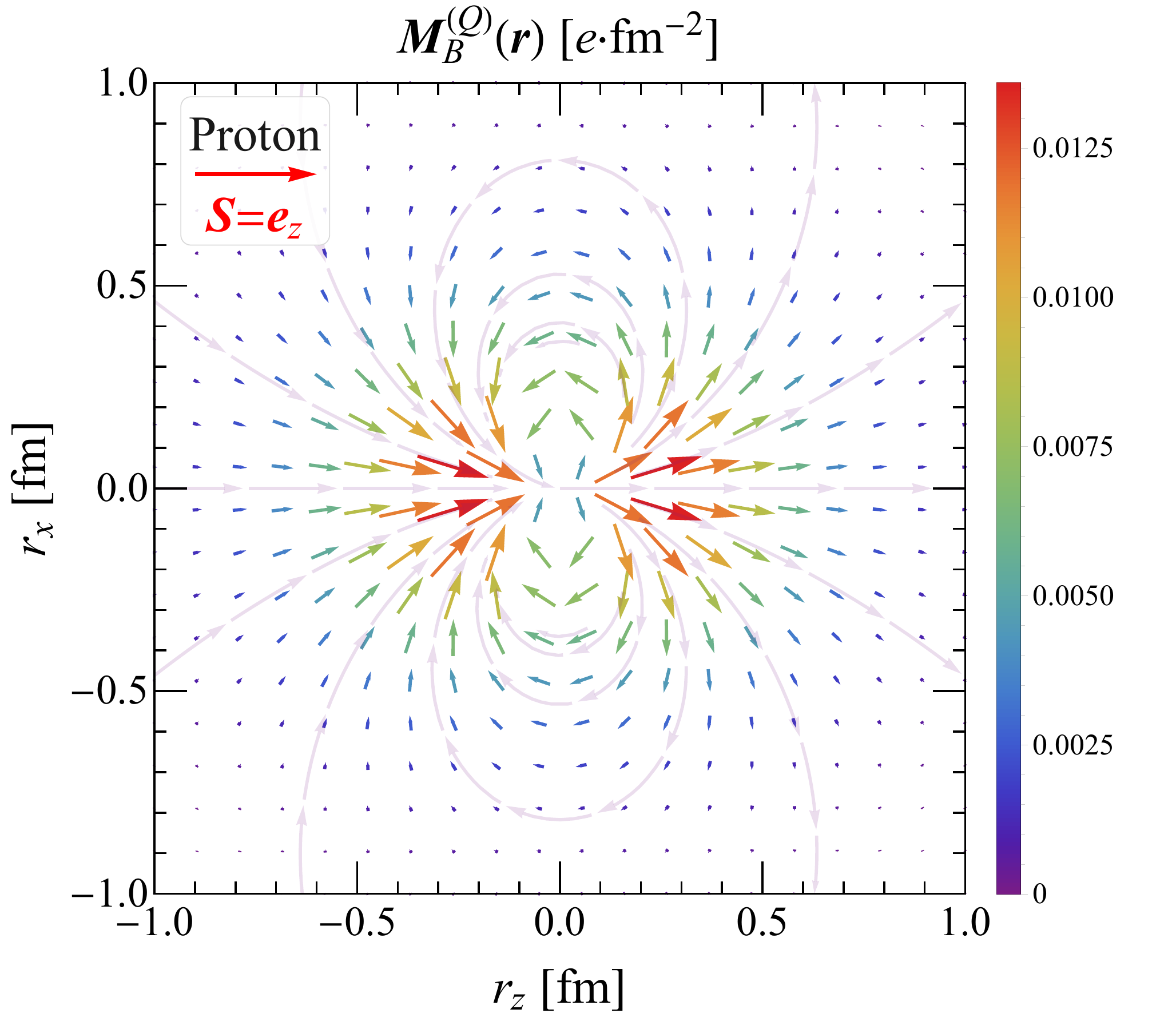}}
	\caption{Monopole (left panel) and quadrupole (right panel) contributions, see Eqs.~(\ref{3DBF-monopole-Mv}) and (\ref{3DBF-quadrupole-Mv}), to the Breit frame magnetization distribution inside a proton polarized along the $z$-direction in the $r_y=0$ plane. Based on the parametrization for the nucleon electromagnetic form factors given in Ref.~\cite{Bradford:2006yz}.}
	\label{Fig_3DBFMvMonoQuad}
\end{figure}

In Fig.~\ref{Fig_3DBFMvMonoQuad}, we show the monopole and quadrupole contributions to the BF magnetization distribution of a proton presented in the upper left panel of Fig.~\ref{Fig_Mv3DBF2D}. The quadrupole contributions have an interesting structure which we highlighted with streamlines. These contributions are however small, explaining why the BF magnetization distributions presented in the first row of Fig.~\ref{Fig_Mv3DBF2D} look essentially like monopoles.

\subsection{In the 2D elastic and light-front frames}

\begin{figure}[tb!]
	\centering
	{\includegraphics[angle=0,scale=0.38]{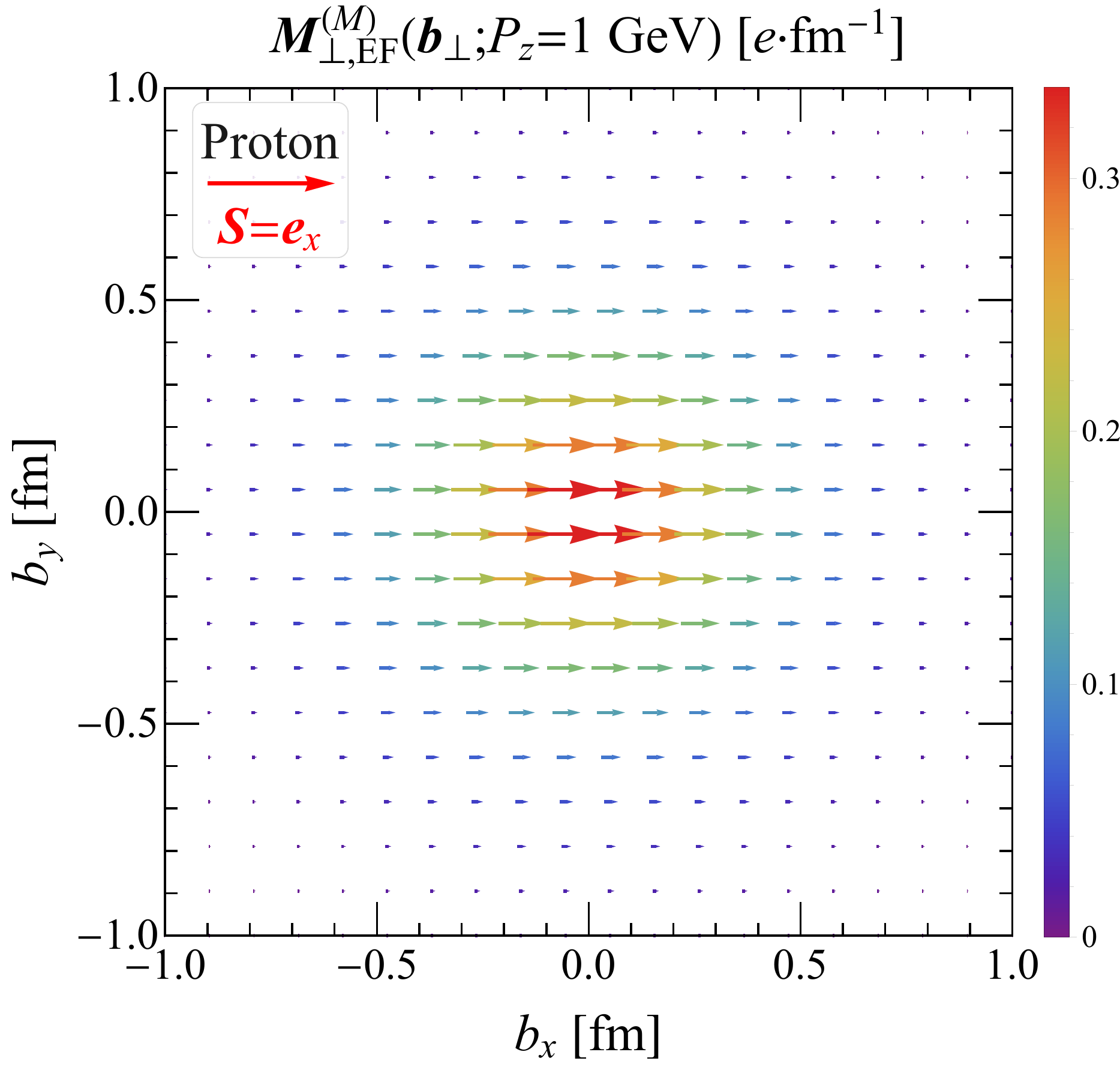}\ \ }
	{\includegraphics[angle=0,scale=0.38]{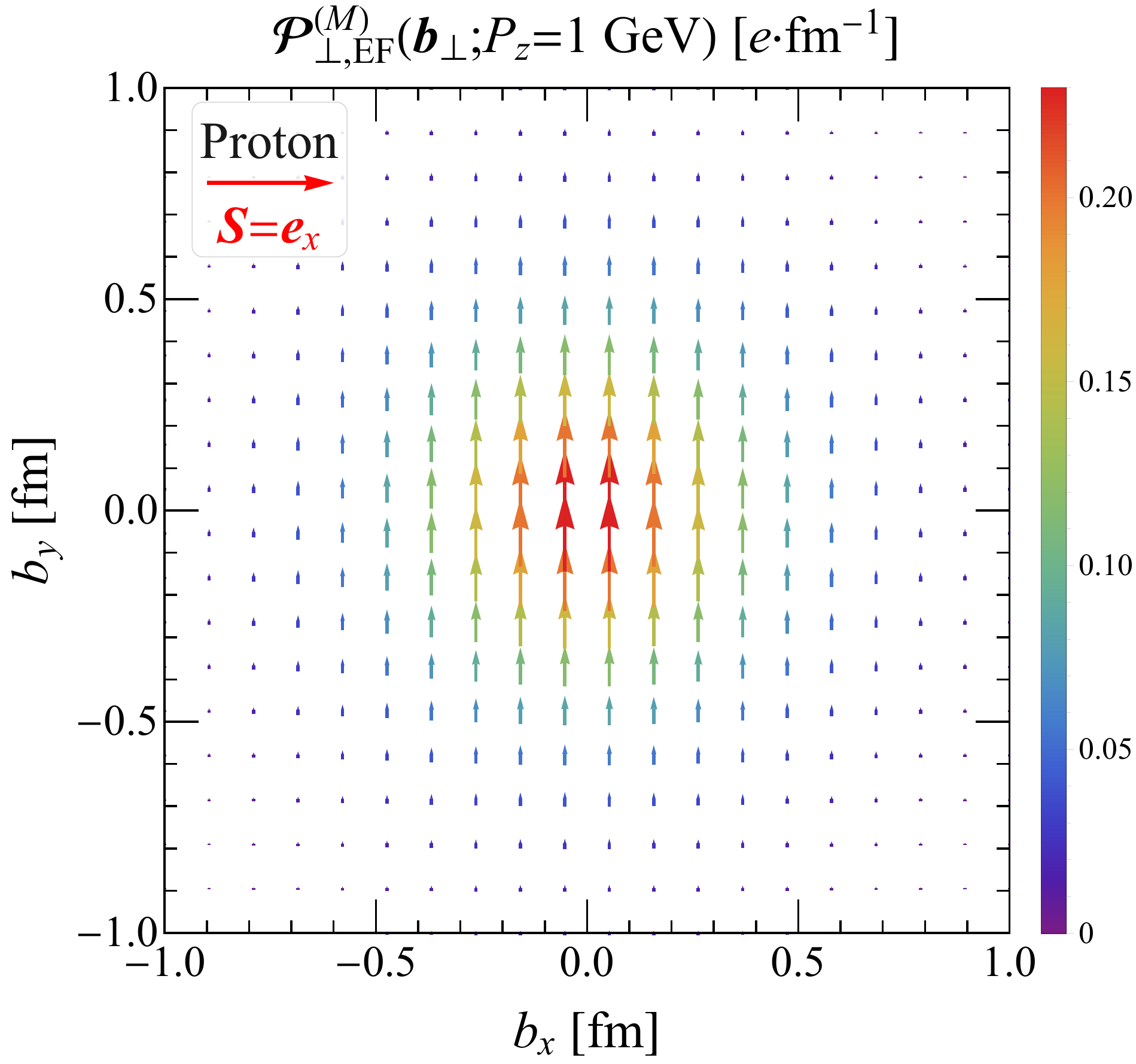}} 
	{\includegraphics[angle=0,scale=0.38]{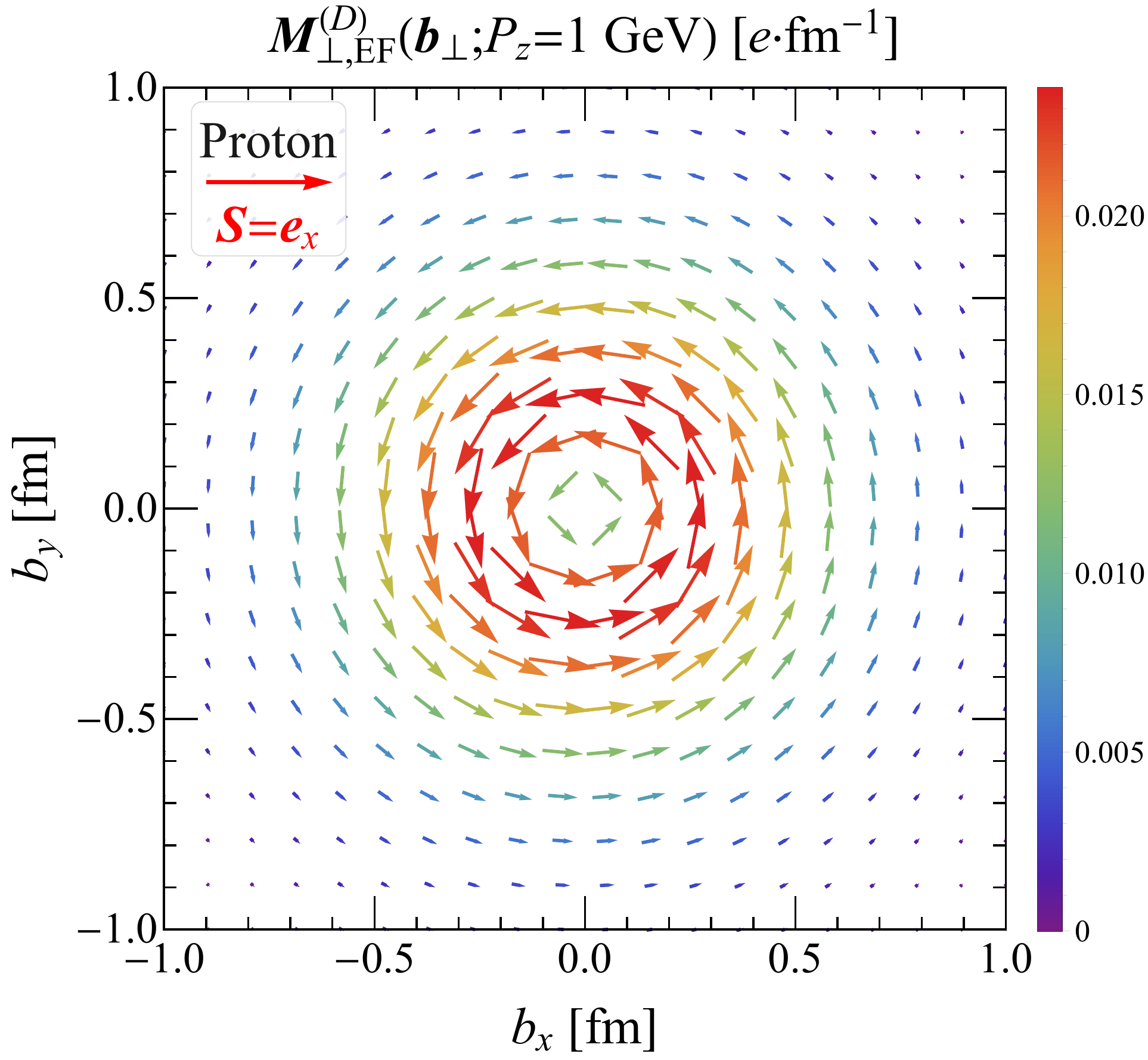}}
	{\ \includegraphics[angle=0,scale=0.38]{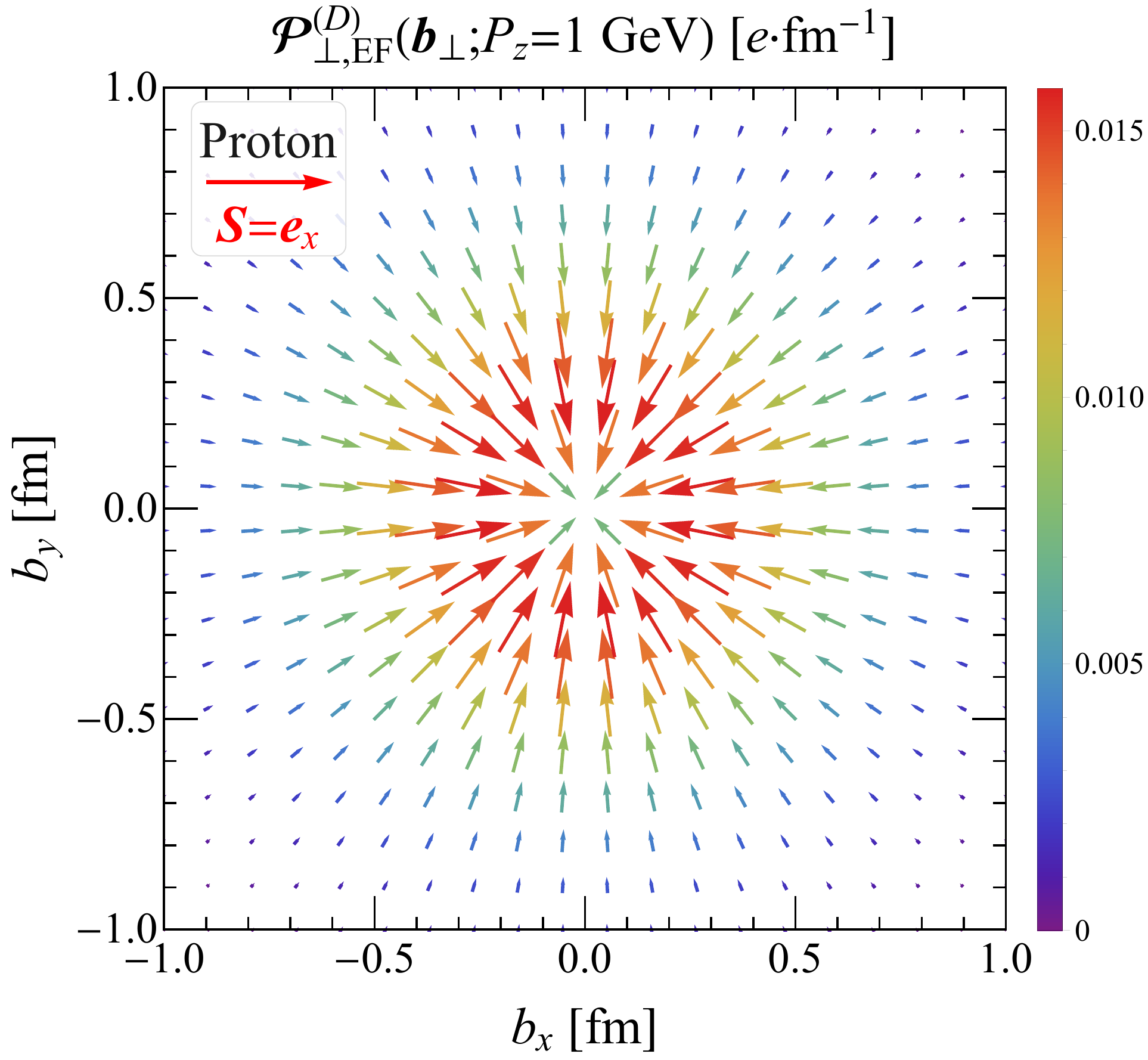}}
	{\includegraphics[angle=0,scale=0.38]{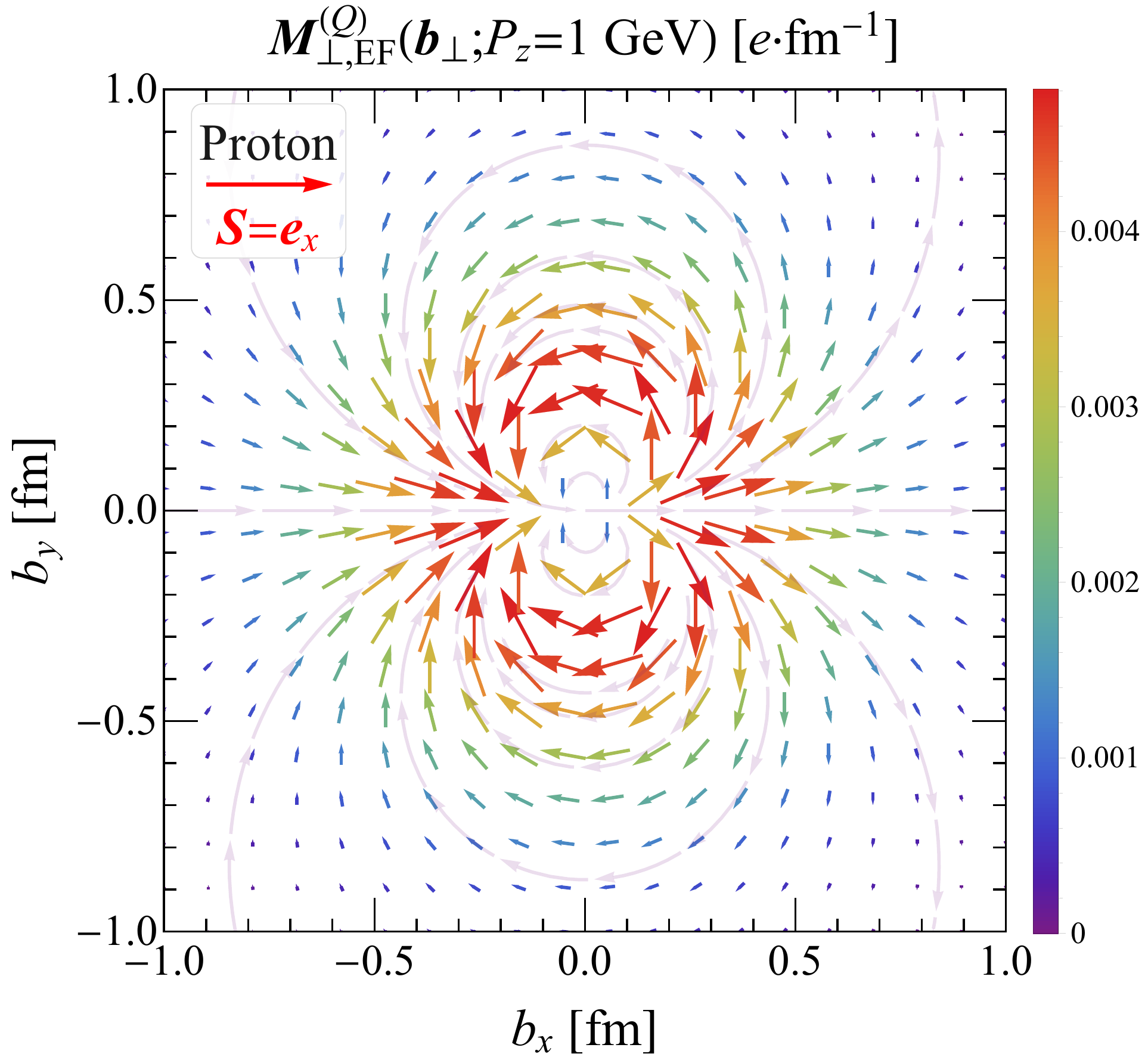}}
	{\includegraphics[angle=0,scale=0.38]{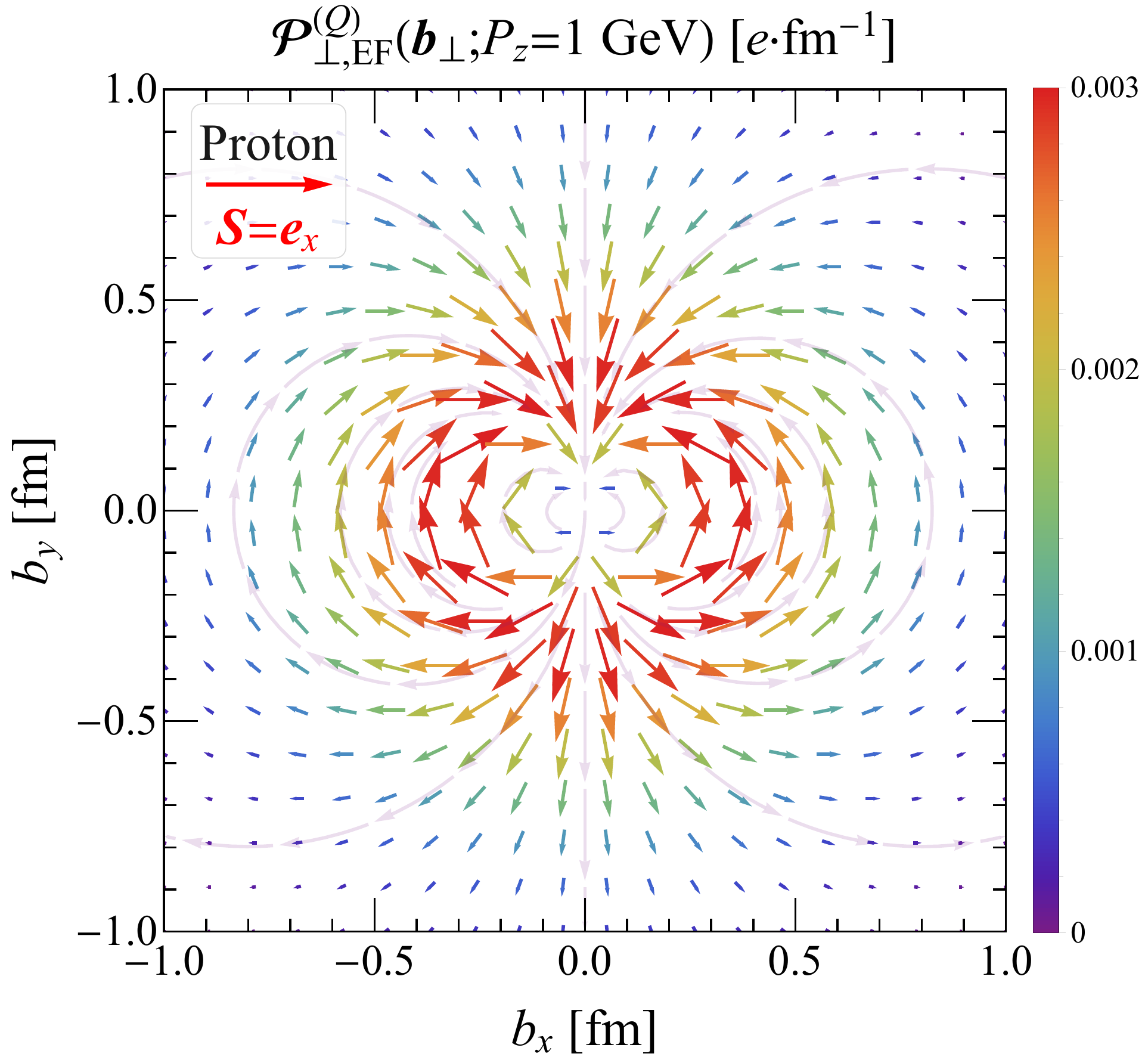}}
	\caption{Monopole (upper panels), dipole (middle panels) and quadrupole (lower panels) contributions, see Eqs.~(\ref{2DEFMv-MultipoleDecom}) and (\ref{2DEFPv-MultipoleDecom}), to the elastic frame magnetization (left panels) and polarization (right panels) distributions inside a proton polarized along the $x$-direction and with average momentum $P_z=1$ GeV. Based on the parametrization for the nucleon electromagnetic form factors given in Ref.~\cite{Bradford:2006yz}.}
	\label{Fig_2DEFMvTPvTMDQ}
\end{figure}

In the 2D transverse Euclidean plane, the relation~\eqref{3DBF-Multipole} for the Fourier transform of a quadrupole in $\uvec\Delta$ reduces to
\begin{equation}
	\begin{aligned}
		\int\frac{\ud^2\Delta_\perp}{(2\pi)^2}\,&e^{-i\uvec\Delta_\perp\cdot\uvec b_\perp}\left(\Delta_\perp^i\Delta_\perp^j-\tfrac{1}{2}\,\delta_\perp^{ij}\uvec\Delta_\perp^2\right) f(\uvec\Delta_\perp^2)= -\frac{b_\perp^ib_\perp^j-\frac{1}{2}\,\delta_\perp^{ij}\uvec b_\perp^2}{b^2} \int\frac{\ud Q}{2\pi} \,Q^3 J_2(Qb) f(Q^2),
	\end{aligned}	
\end{equation}
where $b=|\uvec b_\perp|$ and the $n$th order cylindrical Bessel function $J_n(x)$ is given by
\begin{equation}
	J_n(x)=(-1)^n x^n\left(\frac{1}{x} \frac{\ud }{\ud x}\right)^{\!n} J_0(x),
\end{equation}
with $J_0(x)=\frac{1}{2\pi}\int_{-\pi}^\pi\,\ud\theta\,e^{-ix\cos\theta}$ the zeroth order cylindrical Bessel function.

It is then straightforward to decompose the transverse EF magnetization distribution in Eq.~(\ref{spinhalfPvMv}) into three terms $\uvec M_{\perp,\text{EF}}= \uvec M_{\perp,\text{EF}}^{(M)} + \uvec M_{\perp,\text{EF}}^{(D)} + \uvec M_{\perp,\text{EF}}^{(Q)}$, where the monopole, dipole and quadrupole contributions are respectively given by
\begin{equation}
	\begin{aligned}\label{2DEFMv-MultipoleDecom}
		\uvec M_{\perp,\text{EF}}^{(M)}(\uvec b_\perp;P_z)
		&= \frac{e}{2M} \,\uvec \sigma_\perp \int\frac{\ud Q}{2\pi}\,QJ_0(Qb)\, \frac{P^0+M(1+\tau/2)}{(P^0+M)(1+\tau)}\, G_M(Q^2),\\
		\uvec M_{\perp,\text{EF}}^{(D)}(\uvec b_\perp;P_z)
		&=\frac{e}{2M} \,\frac{P_z}{2M}\,(\uvec e_z\times\hat{\uvec b}_\perp) \int\frac{\ud Q}{2\pi}\,Q^2J_1(Qb)\, \frac{G_M(Q^2)}{(P^0+M)(1+\tau)},\\
		\uvec M_{\perp,\text{EF}}^{(Q)}(\uvec b_\perp;P_z)
		&= \frac{e}{2M} \left[\hat{\uvec b}_\perp(\hat{\uvec b}_\perp\cdot\uvec\sigma_\perp)-\tfrac{1}{2}\,  \uvec\sigma_\perp\right] \int\frac{\ud Q}{2\pi}\, Q^3 J_2(Qb) \,\frac{ G_M(Q^2) }{4M(P^0+M)(1+\tau)}
	\end{aligned}
\end{equation}
with $\hat{\uvec b}_\perp\equiv \uvec b_\perp/|\uvec b_\perp|$ the unit vector along $\uvec b_\perp$.  Similarly, the transverse EF polarization distribution can also be decomposed into three terms $\uvec {\mathcal P}_{\perp,\text{EF}}= \uvec {\mathcal P}_{\perp,\text{EF}}^{(M)} + \uvec {\mathcal P}_{\perp,\text{EF}}^{(D)} + \uvec {\mathcal P}_{\perp,\text{EF}}^{(Q)}$, where the monopole, dipole and quadrupole contributions are respectively given by
\begin{equation}
	\begin{aligned}\label{2DEFPv-MultipoleDecom}
	    \uvec{\mathcal P}_{\perp,\text{EF}}^{(M)}(\uvec b_\perp;P_z)
	    &=\frac{e}{2M}\,(\uvec e_z\times\uvec \sigma_\perp)\int\frac{\ud Q}{2\pi}\,QJ_0(Qb)\,\frac{P_z}{P^0} \,   \frac{P^0+M(1+\tau/2)}{(P^0+M)(1+\tau)}\, G_M(Q^2),\\
	    \uvec{\mathcal P}_{\perp,\text{EF}}^{(D)}(\uvec b_\perp;P_z) &= -\frac{e}{2M}\,\frac{ P_z }{2M}\,\hat{\uvec b}_\perp\int\frac{\ud Q}{2\pi}\,Q^2J_1(Qb)\,\frac{P_z}{P^0}  \,\frac{G_M(Q^2) }{(P^0+M)(1+\tau)}  ,\\
	    \uvec{\mathcal P}_{\perp,\text{EF}}^{(Q)}(\uvec b_\perp;P_z)
	    &=\frac{e}{2M} \left[(\uvec e_z\times\hat{\uvec b}_\perp)(\hat{\uvec b}_\perp\cdot\uvec\sigma_\perp)-\tfrac{1}{2}\,  (\uvec e_z\times\uvec\sigma_\perp)\right] \\
     &\qquad\qquad\qquad\quad \int\frac{\ud Q}{2\pi} \,Q^3 J_2(Qb) \,\frac{P_z}{P^0}\,\frac{ G_M(Q^2) }{4M(P^0+M)(1+\tau)}.
	\end{aligned}
\end{equation}
Expressions in Eqs.~(\ref{2DEFPv-MultipoleDecom}) and (\ref{2DEFMv-MultipoleDecom}) are very similar and follow simply from the relation between the momentum-space amplitudes $\widetilde{\uvec{\mathcal P}}_\text{EF} =\uvec\beta\times \widetilde{\uvec M}_\text{EF}$ in Eq.~(\ref{polmagEF}), which obviously holds also for the individual multipole contributions. In Fig.~\ref{Fig_2DEFMvTPvTMDQ}, we show the multipole decomposition of the transverse EF magnetization and polarization distributions inside a transversely polarized proton with average momentum $P_z=1~\text{GeV}$. As the result of the non-vanishing average momentum which breaks the $z\mapsto -z$ symmetry, Wigner rotations generate a dipole contribution on top of the quadrupole contribution. However, discrete spacetime symmetries prevent the appearance of $\uvec\sigma_\perp$ in the dipole contribution, explaining why the latter does not depend on the target polarization.

\begin{figure}[tb!]
	\centering
	{\includegraphics[angle=0,scale=0.38]{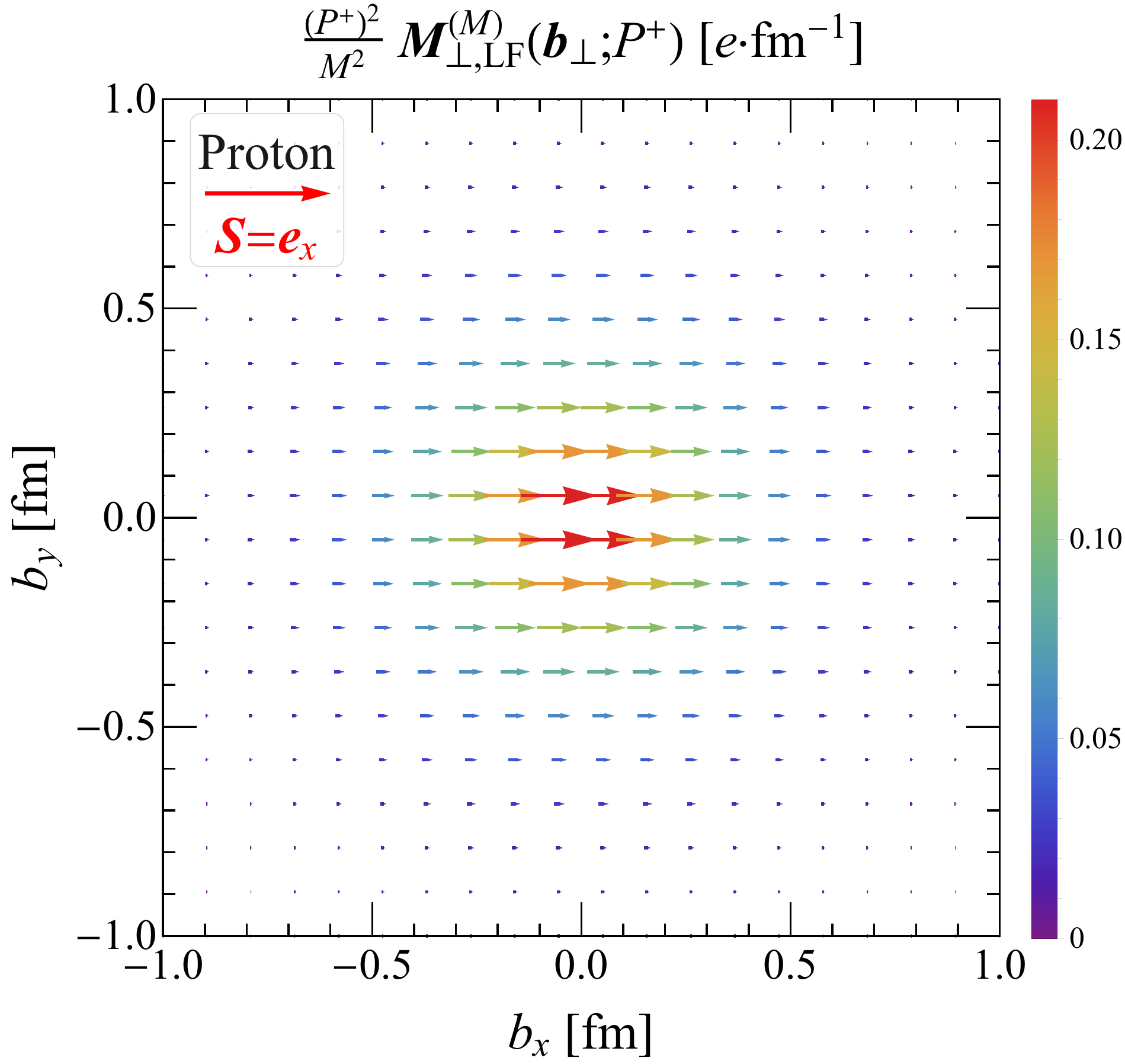}}
	{\includegraphics[angle=0,scale=0.38]{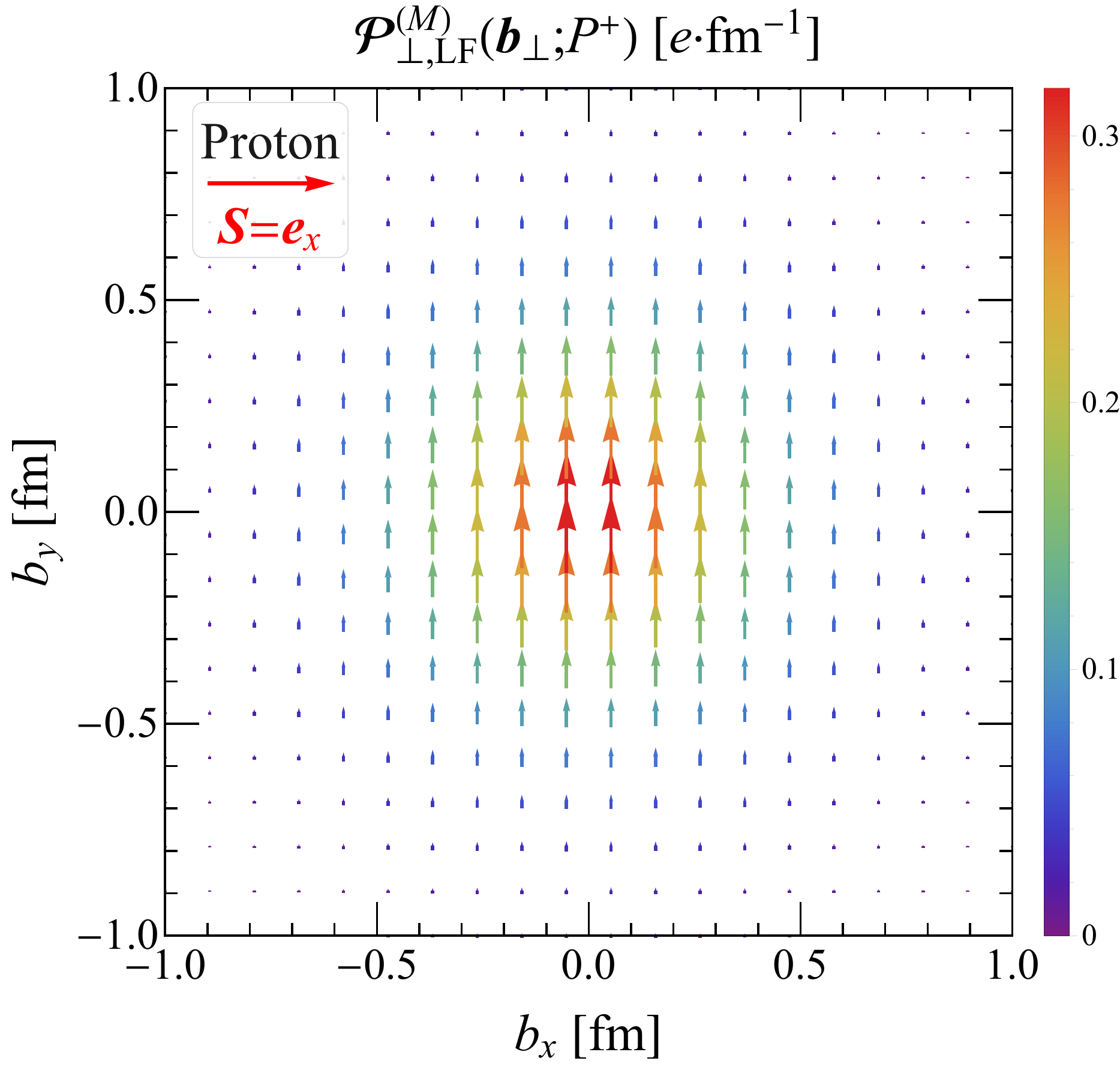}}
	{\includegraphics[angle=0,scale=0.38]{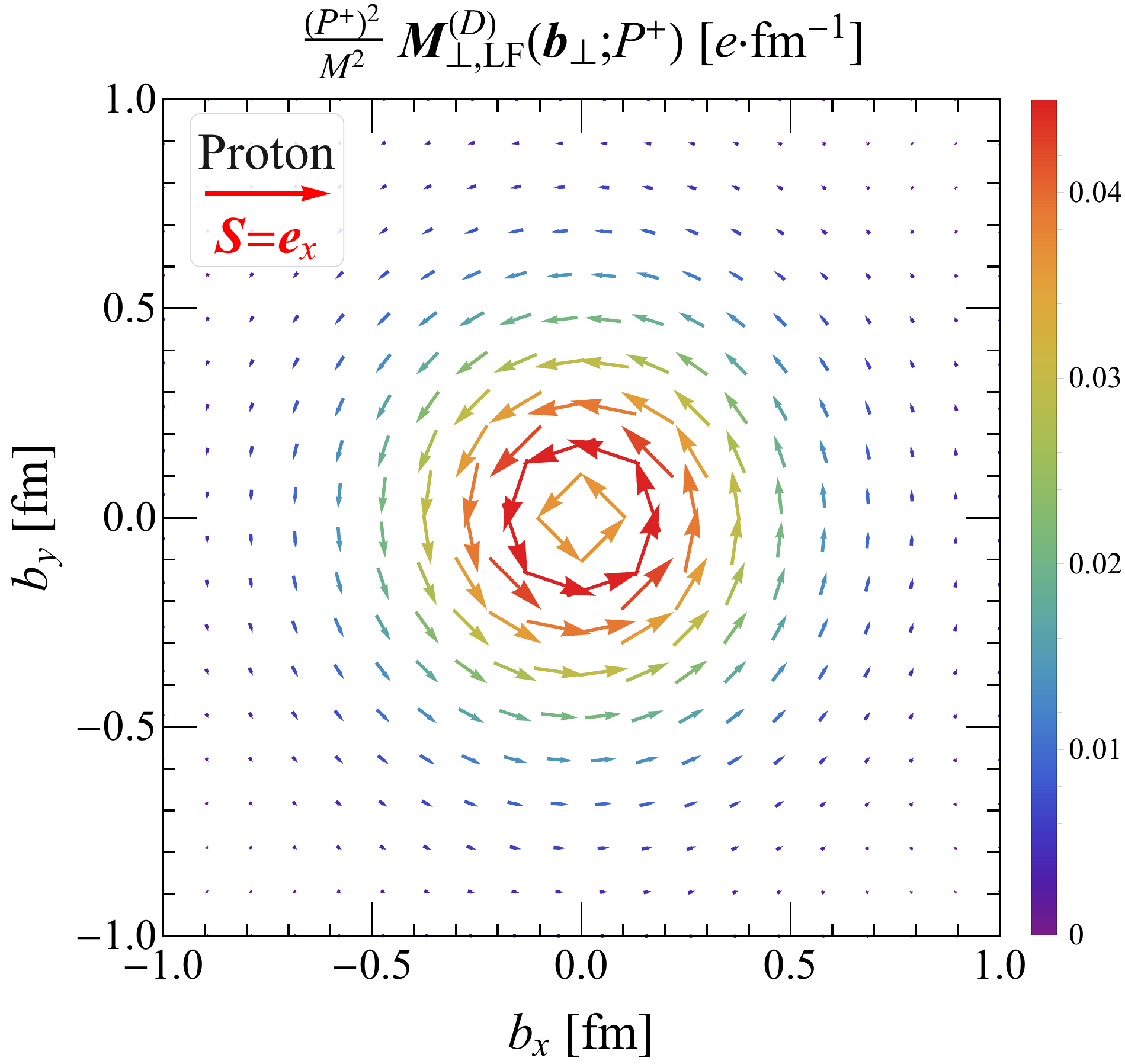}}
	{\includegraphics[angle=0,scale=0.38]{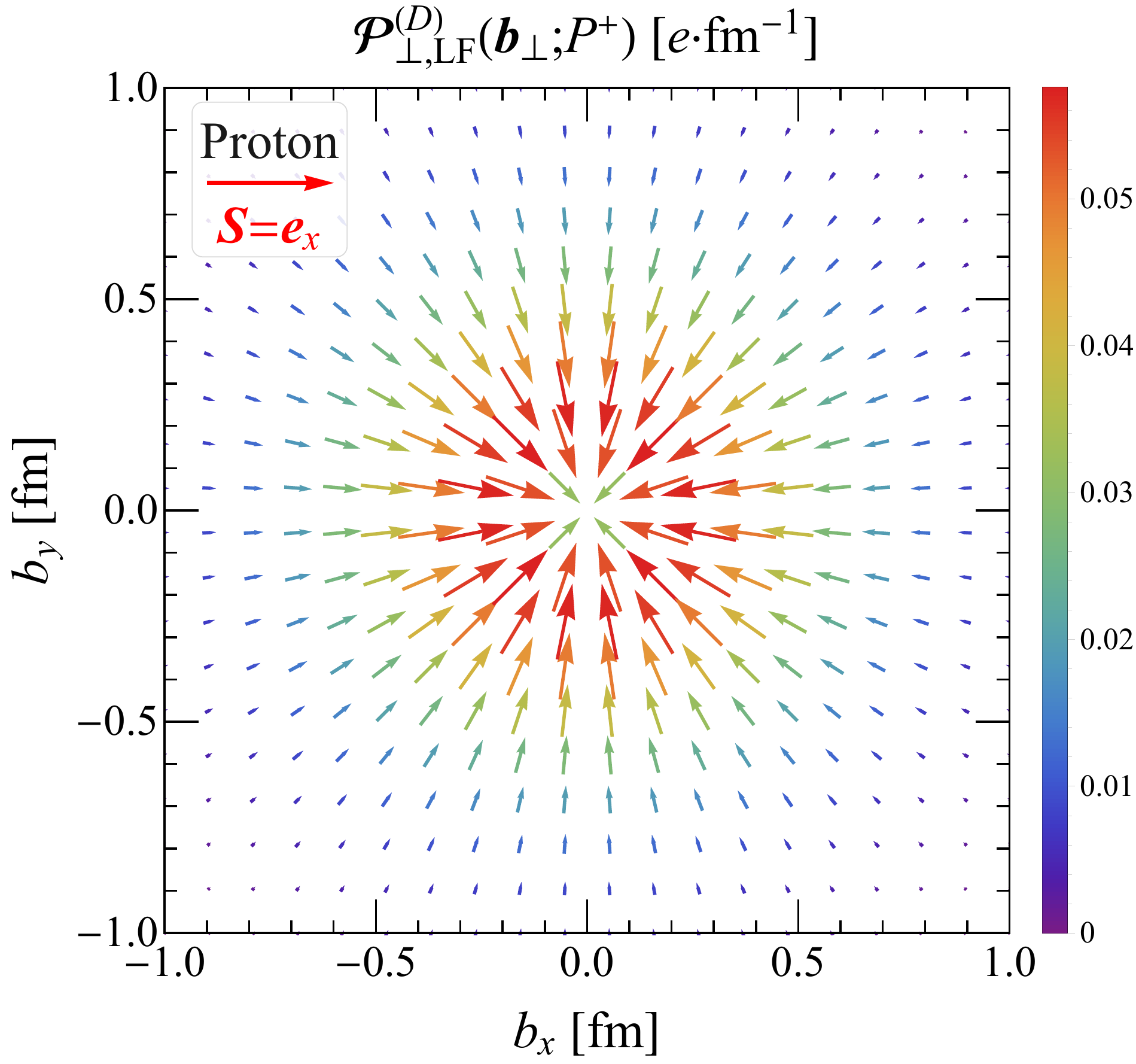}}
	\caption{Monopole (upper panels) and dipole (lower panels) contributions, see Eqs.~(\ref{2DLFMvT-MultipoleDecom}) and (\ref{2DLFPvT-MultipoleDecom}), to the (scaled) light-front magnetization (left panels) and polarization (right panels) distributions inside a proton polarized along the $x$-direction. Based on the parametrization for the nucleon electromagnetic form factors given in Ref.~\cite{Bradford:2006yz}.}
	\label{Fig_2DLFMvTPvTMD}
\end{figure}

According to the LF expressions in Eqs.~(\ref{spinhalf-LFPol0}-\ref{LFmagpolampl}),
\begin{equation}
    \mathcal P^+_\text{LF}(\uvec b_\perp;P^+)=\mathcal P^-_\text{LF}(\uvec b_\perp;P^+)=M^-_\text{LF}(\uvec b_\perp;P^+) = 0.
\end{equation}
For the transverse LF magnetization distribution in Eq.~(\ref{spinhalf-LFPolMag}), we apply the same procedure as in the EF and decompose it into two terms $\uvec M_{\perp,\text{LF}}= \uvec M_{\perp,\text{LF}}^{(M)} + \uvec M_{\perp,\text{LF}}^{(D)}$, where the monopole and dipole contributions are respectively given by
\begin{equation}
	\begin{aligned}\label{2DLFMvT-MultipoleDecom}
		\uvec M_{\perp,\text{LF}}^{(M)}(\uvec b_\perp;P^+) 
		&= \frac{e}{2M}\,\frac{M^2}{2(P^+)^2}\,\uvec\sigma_\perp\int\frac{\ud Q}{2\pi}\,QJ_0(Qb)\,  G_M(Q^2),\\
		\uvec M_{\perp,\text{LF}}^{(D)}(\uvec b_\perp;P^+) 
		&= \frac{e}{2M}\,\frac{M^2}{2(P^+)^2}\,(\uvec e_z \times \hat{\uvec b}_\perp)\int\frac{\ud Q}{2\pi}\,\frac{Q^2}{2M}\, J_1(Qb)\, G_M(Q^2). 
	\end{aligned}
\end{equation}

Likewise, the transverse LF polarization distributions from Eq.~(\ref{spinhalf-LFPolMag}) can be decomposed into two terms $\uvec{\mathcal P}_{\perp,\text{LF}}= \uvec{\mathcal P}^{(M)}_{\perp,\text{LF}} + \uvec{\mathcal P}^{(D)}_{\perp,\text{LF}}$, where the monopole and dipole contributions are respectively given by
\begin{equation}
	\begin{aligned}\label{2DLFPvT-MultipoleDecom}
		\uvec{\mathcal P}^{(M)}_{\perp,\text{LF}}(\uvec b_\perp;P^+) 
		&= \frac{e}{2M}\,(\uvec e_z \times\uvec\sigma_\perp) \int\frac{\ud Q}{2\pi}\,QJ_0(Qb)\,\frac{G_M(Q^2)}{1+\tau},\\
		\uvec{\mathcal P}^{(D)}_{\perp,\text{LF}}(\uvec b_\perp;P^+)
		&= -\frac{e}{2M}\,\hat{\uvec b}_\perp \int\frac{\ud Q}{2\pi}\, \frac{Q^2}{2M}\, J_1(Qb)\,\frac{G_M(Q^2)}{1+\tau}.
	\end{aligned}
\end{equation}
Like in the EF case, we observe similar structures in the LF magnetization and polarization distributions which follow this time from $\widetilde{\mathcal P}^i_{\perp,\text{LF}}=-\frac{P^+}{P^-}\,\epsilon^{ij}_\perp \widetilde{M}^j_{\perp,\text{LF}}$ in Eq.~(\ref{LFmagpolampl}). The multipole contributions to the transverse LF polarization distribution in Eq.~\eqref{2DLFPvT-MultipoleDecom} are $P^+$-independent and coincide with the IMF limit of the corresponding EF contributions in Eq.~\eqref{2DEFPv-MultipoleDecom}. In particular, the EF quadrupole contribution vanishes in the IMF in agreement with the absence of LF quadrupole contribution. By contrast, the multipole contributions to the transverse LF magnetization distribution in Eq.~\eqref{2DLFMvT-MultipoleDecom} differ from the IMF limit of the corresponding EF contributions in Eq.~\eqref{2DEFMv-MultipoleDecom} by a factor $P^-/P^+$. In Fig.~\ref{Fig_2DLFMvTPvTMD}, we show the multipole contributions to the transverse LF (scaled) magnetization and polarization distributions inside a transversely polarized proton. As expected, they look similar to the corresponding EF distributions in Fig.~\ref{Fig_2DEFMvTPvTMDQ}, albeit with differences in magnitude.

%
\bibliography{Nucleon_relativistic_polarization_and_magnetization.bbl}
\end{document}